\documentclass[twocolumn,tightenlines,aps,preprintnumbers,superscriptaddress,showpacs,floatfix]{revtex4-1}
\usepackage{graphicx}
\usepackage{epsf}
\usepackage{amssymb}
\usepackage{amsmath,amssymb,textcomp,array}

\begin{document}

\preprint{LA-UR-11-01798}

\title{Nonparametric Reconstruction of the Dark Energy Equation of State from Diverse Data Sets}
\author{Tracy Holsclaw}\affiliation{Department of Applied Mathematics and Statistics, 
 University of California, Santa Cruz, CA 95064}
\author{Ujjaini Alam}\affiliation{ISR-1, MS D466, Los Alamos National Laboratory, Los
Alamos, NM 87545}
\author{Bruno Sans\'o}\affiliation{Department of Applied Mathematics and Statistics, 
University of California, Santa Cruz, CA 95064}
\author{Herbie Lee}\affiliation{Department of Applied Mathematics and Statistics, 
 University of California, Santa Cruz, CA 95064}
\author{Katrin Heitmann}\affiliation{ISR-1, MS D466, Los Alamos National Laboratory, Los
Alamos, NM 87545}
\author{Salman Habib}\affiliation{T-2, MS B285, Los Alamos National Laboratory, Los
Alamos, NM 87545}
\author{David Higdon}\affiliation{CCS-6, MS F600, Los
Alamos National Laboratory, Los Alamos, NM 87545}

\date{\today}

\begin{abstract}
  The cause of the accelerated expansion of the Universe poses one of
  the most fundamental questions in physics today. In the absence of a
  compelling theory to explain the observations, a first task is to
  develop a robust phenomenology. If the acceleration is driven by
  some form of dark energy, then, the phenomenology is determined by
  the dark energy equation of state $w$. A major aim of ongoing and
  upcoming cosmological surveys is to measure $w$ and its time
  dependence at high accuracy. Since $w(z)$ is not directly accessible
  to measurement, powerful reconstruction methods are needed to
  extract it reliably from observations. We have recently introduced a
  new reconstruction method for $w(z)$ based on Gaussian process
  modeling. This method can capture nontrivial time-dependences in
  $w(z)$ and, most importantly, it yields controlled and unbaised
  error estimates. In this paper we extend the method to include a
  diverse set of measurements: baryon acoustic oscillations, cosmic
  microwave background measurements, and supernova data. We analyze
  currently available datasets and present the resulting constraints
  on $w(z)$, finding that current observations are in very good
  agreement with a cosmological constant. In addition we explore how
  well our method captures nontrivial behavior of $w(z)$ by analyzing
  simulated data assuming high-quality observations from future
  surveys. We find that the baryon acoustic oscillation measurements
  by themselves already lead to remarkably good reconstruction results
  and that the combination of different high-quality probes allows us
  to reconstruct $w(z)$ very reliably with small error bounds.
\end{abstract}


\maketitle

\section{Introduction}

The discovery of the accelerated expansion of the Universe little more
than a decade ago~\cite{riess,perlmutter} was a major surprise. Since
then, many observational efforts to understand the underlying cause
have been initiated (e.g. the Baryon Oscillation Spectroscopic Survey
(BOSS)~\cite{boss}, WiggleZ~\cite{wigglez}, the Dark Energy Survey
(DES, https://www.darkenergysurvey.org/), the Large Synoptic Survey
Telescope (LSST)~\cite{lsst}) and proposed (e.g.
BigBOSS~\cite{bigboss}, the Wide Field Infrared Survey Telescope
(WFIRST)~\cite{wfirst}, Euclid~\cite{euclid}). These efforts focus on a
set of diverse cosmological probes (supernovae, baryon acoustic
oscillations, clusters of galaxies, weak lensing, etc.) to combine the
best possible observations in order to help solve this puzzle.

The two currently most popular explanations are a form of dark energy
or a modification of Einstein's theory of gravity on the largest
observable scales. We will focus in this paper on dark energy as the
cause for the accelerated expansion. The simplest way to realize a
dark energy is via a cosmological constant with a dark energy equation
of state $w=p/\rho=-1$. A cosmological constant, however, is not
theoretically well-motivated. If we assume that the origin is due to a
vacuum energy, the predicted value is incorrect at the order of
$10^{60}$. Therefore, a more natural realization of dark energy might
be a dynamical field, similar to the inflaton that is believed to
drive the very early rapid expansion of the Universe. Such a dynamical
field, described for example by quintessence models~\cite{quint},
would lead to a non-constant dark energy equation of state $w(z)$. It
is therefore one of the major aims of ongoing and upcoming dark energy
missions to measure $w(z)$ and its time variation with high accuracy.
If $w(z)$ is modeled via a simple parametrization
$w(z)=w_0-w_az/(1+z)$~\cite{chev01,linder03}, current predictions for
future surveys promise measurements of the constant part at the 1\%
level accuracy and of the leading time varying part at the 10\% level.
At present, the best measurements are accurate to 10\% with respect to
$w_0$ with no strong constraints on the time
variation~\cite{hicken09,amanullah}.

With the prospect of high-accuracy measurements from supernova (SN)
surveys and complementary large-scale structure probes such as baryon
acoustic oscillation (BAO) surveys, it is desirable to develop an
accurate reconstruction method with reliable error bars that allows us
to extract the dark energy equation of state from different
measurements. While the earlier focus in the field was on parametric
methods~\cite{chev01,linder03,params}, non-parametric methods are
becoming more popular~\cite{nonparams}. The major advantage of
non-parametric models is that they are not biased (no assumptions are
made regarding the functional form for $w(z)$). A possible
disadvantage might be that if the data quality is insufficient,
non-parametric approaches might not provide much information about
$w(z)$. In principle this is also an advantage: if the data does not
have enough information it is better to obtain uncertain results with
large error bars than a prediction which might be biased (since the
functional form assumed for $w(z)$ is incorrect) without this bias
being reflected in the error bars.

In this paper, we discuss a recently-introduced reconstruction method
based on Gaussian process (GP) modeling~\cite{holsclaw1,holsclaw2}. A
GP is a stochastic process and each realization is a random draw from
a multivariate Normal distribution. It is characterized by a mean and
a covariance function, that are defined by a small number of
parameters. Bayesian estimation methods are used to determine the
parameters of the GP model together with any other physics parameters.
Therefore, the final form of the GP model is informed by the data
itself. The form of the covariance function is general enough to
accommodate a large variety of possible outcomes for $w(z)$. The only
assumption made is that $w(z)$ is somewhat smooth and continuous. If the underlying
cause for the accelerated expansion is due to a physically well
motivated reason this assumption is justified. We extend the approach
described in detail in Ref. \cite{holsclaw1} to include different
observational probes of $w(z)$, namely supernova measurements, cosmic
microwave background (CMB) observations, and BAO results. We begin
with an analysis of currently available data. We find, not
surprisingly, that our predictions are in good agreement with a
cosmological constant. Using simulated data, we then explore the
ability to extract variations of $w(z)$ away from a cosmological
constant with improving accuracy and statistics of the data. The
inclusion of the additional BAO and CMB measurements greatly help to
improve these predictions.

The paper is organized as follows. In Section~\ref{diverse} we
describe the different data sources included in our analysis, namely,
supernova, CMB, and BAO measurements. In Section~\ref{gpmodel} we
describe our GP model based reconstruction method. We carry out an
analysis of currently available data in Section~\ref{realdat}. We
demonstrate that the method will allow us to extract variations in the
dark energy equation of state by using simulated data in
Section~\ref{simdat}. Finally, we conclude in
Section~\ref{conclusion}.

\section{Dark Energy Equation of State from Diverse Data Sets}
\label{diverse}

Type Ia supernova measurements are currently the best source of
information regarding possible deviations of $w(z)$ from a constant
value. In the future, BAO (and other) measurements will be a strong
competitor and in combination will lead to the best possible
constraints on $w(z)$. The complementarity of the different probes is
important to break degeneracies and decrease the overall errors. In
our previous paper ~\cite{holsclaw1} we showed that our non-parametric
reconstruction method is able to capture even rather sharp transitions
in $w(z)$ well if we have very good knowledge about $\Omega_m$.
Supernova data alone does not provide this information and one needs a
strong prior on $\Omega_m$ to obtain good results. This strong prior
can be justified by the existence of complementary probes. In a more
direct and complete implementation, multiple probes are included in
the analysis and a joint analysis is performed. This allows us to
relax our prior assumptions on $\Omega_m$ and to tighten our final
constraints on the behavior of $w(z)$.

In the following, we provide a brief review of the different dark
energy probes employed in this paper -- supernovae, BAO, and CMB --
and how to extract information about $w(z)$ from these probes. We
focus in this paper on the geometric probes for $w(z)$. The GP
analysis is in this case very similar for all methods -- $w(z)$ is
connected via two derivatives with the different distance measures. In
the next section, we explain in detail how to set up a joint GP model
for the three different observations.

\subsection{Supernova Measurements}

In this paper we retain the notation from our previous
work~\cite{holsclaw1}. For completeness, we summarize the important
equations here.  The luminosity distance $d_L$ as measured by
supernovae is directly connected to the expansion history of the
Universe described by the Hubble parameter $H(z)$. For a spatially
flat Universe, the relation is given by
\begin{equation}\label{dl}
d_L(z)=(1+z)\frac{c}{H_0}\int_0^z\frac{ds}{h(s)},
\end{equation}
where $c$ is the speed of light, $H_0$, the current value of the
Hubble parameter ($H(z)=\dot{a}/a$, where $a$ is the scale factor and
the overdot represents a derivative with respect to cosmic time), and
$h(z)=H(z)/H_0$. The assumption of spatial flatness is in effect an
``inflation prior'', although there do exist strong constraints on
spatial flatness when CMB and BAO observations are combined (see,
e.g., Ref.~\cite{komatsu}). In principle, we can relax this
assumption, but enforce it here to simplify the analysis.

Instead of $d_L(z)$, supernova data are usually specified in terms of the
distance modulus $\mu$ as a function of  redshift. The relation between
$\mu$ and the luminosity distance is 
\begin{eqnarray}\label{muh}
\mu_B(z)&=&m_B-M_B=5\log_{10} \left(\frac{d_L (z)}{1~{\rm Mpc}}\right)+25\\
&=&5\log_{10} \left[(1+z)
  c\int_0^z\frac{ds}{h(s)}\right]-5\log_{10}(H_0)+25,\nonumber
\end{eqnarray} 
where we used Eq.~(\ref{dl}). $M_B$ is the absolute magnitude of the
object and $m_B$ the ($B$-band) apparent magnitude.
Writing out the expression for the reduced Hubble parameter $h(z)$ in
Eq.~(\ref{muh}) explicitly in terms of a general dark energy equation
of state for a spatially flat FRW Universe: 
\begin{eqnarray}\label{hubble}
h^2(z)&=&\Omega_r(1+z)^4+\Omega_m(1+z)^{3}\\
&&+(1-\Omega_r-\Omega_m)(1+z)^3
    \exp\left(3\int_0^{z}\frac{w(u)}{1+u}du\right)\nonumber
\end{eqnarray}
leads to the relation
\begin{eqnarray}
&&\mu_B(z)=25-5\log_{10}(H_0)\label{mu}\\
&&+5\log_{10}\left\{(1+z){c}
\int^z_0 ds\left[\Omega_r(1+s)^4+\Omega_m(1+s)^{3}\right.\right.\nonumber\\
&&+\left.\left.(1-\Omega_r-\Omega_m)(1+s)^3
    \exp\left(3\int_0^{s}\frac{w(u)}{1+u}du\right)\right]^{-1/2}\right\}.
\nonumber
\end{eqnarray}
While the term proportional to $\Omega_r$ (the radiation density for
photons and neutrinos) is negligible at low redshift, we include it in
the equations for completeness -- it will become important for the CMB
and BAO measurements. We use the following relation for  
$\Omega_r(z)$ when CMB is added to the analysis:
\begin{equation}
\Omega_r(z)=\Omega_r(0)[1+0.227 N_{\rm eff}f(m_\nu a/T_\nu)],
\end{equation}
where for the standard three neutrino species, $N_{\rm eff}=3.04$, we have
$m_\nu a/T_\nu=187/(1+z)\Omega_r h^2\cdot10^3$ and $f(y)\simeq[1+(0.3173y)^{1.83}]^{1/1.83}$ (In the
following, wherever we quote values for $\Omega_r$ they are quoted at $z=0$.)

Note that $H_0$ in Eq.~(\ref{mu}) cannot be
determined from supernova measurements in the absence of an
independent distance measurement.  Thus $H_0$ can be treated as an
unknown and absorbed in a re-definition of the absolute magnitude
${\cal M}_B=M_B-5\log_{10}H_0+25$, which accounts for the combined
uncertainty in the absolute calibration of the supernova data, as well
as in $H_0$. Using this, the $B$-band magnitude can be expressed as
$m_B=5 \log_{10}{D}_L(z) + {\cal M}_B$ where ${D}_L(z) = H_0 d_L (z)$
is the ``Hubble-constant-free'' luminosity distance (throughout this
paper we will follow the convention to use capital letters for
Hubble-constant-free distances and small letters for distances
measured in Mpc. Different papers use different conventions.). The
measurement of $\mu_B$ is only a relative measurement and ${\cal M}_B$
allows for an additive uncertainty which can be left as a nuisance
parameter. To simplify our notation, we absorb $5\log_{10}(H_0)-25$
into our definition of the distance modulus, leading to:
\begin{equation}\label{tildemu}
\tilde \mu_B=\mu_B+5\log_{10}(H_0)-25=5\log_{10}[D_L(z)].
\end{equation}
With this definition of the distance modulus we have calibrated the
overall offset of the data to be zero. To account for uncertainties in
this calibration, we introduce a shift parameter $\Delta_\mu$ with a
broad uniform prior. The expected value for $\Delta_\mu=0$.

\subsection{BAO Measurements}

Baryon acoustic oscillations provide another powerful measurement of
the expansion history of the Universe. Similar to supernovae they yield a
geometric probe of dark energy. By carrying out measurements of the
clustering along the transverse BAO scale one can obtain the angular
diameter distance $d_A(z)$, defined as
\begin{equation}
d_A(z)=\frac{1}{1+z}\frac{c}{H_0}\int_0^z\frac{ds}{h(s)}\label{da},
\end{equation}
and by measuring the BAO scale along the line of sight, one obtains
information on the Hubble parameter $H(z)$ itself (for details on
future measurements, see, e.g. Ref.~\cite{bigboss}). Both of these
measurements will be carried out in terms of the sound horizon at the
epoch of baryon drag, $r_s(z_d)$, given by:
\begin{equation}
  r_s(z_d)=\frac{c}{\sqrt{3}}\int_{z_d}^\infty \frac{ds}{H(s)\sqrt{1+
    \frac{3\Omega_b}{4\Omega_r(1+s)}}}, 
\end{equation}
and the final measurements will be in terms of $d_A(z)/r_s$ and
$H(z)r_s$. Current data provide information only on the angular
diameter distance. The structure of Eq.~(\ref{da}) with respect to its
$w$-dependence via two integrals is exactly the same as for $d_L(z)$
given in Eq.~(\ref{dl}). This makes it very easy to carry through a
reconstruction approach combining both probes.

\subsection{CMB Measurements}

For the CMB measurements we employ the so-called shift parameter
$R(z_\star)$ first introduced by Bond et al.~\cite{bond97}:
\begin{eqnarray}
R(z_\star)&=&\frac{\sqrt{\Omega_mH_0^2}}{c}(1+z_\star)d_A(z_\star)\nonumber\\
&=&\sqrt{\Omega_m}\int_0^{z_\star}\frac{ds}{h(s)},
\end{eqnarray}
where $z_\star$ is the redshift of decoupling ($z_\star\sim 1090$) and
the angular diameter distance $d_A$ is given in Eq.~(\ref{da}). The
shift parameter is related to the peak heights and the locations of
the peaks in the temperature power spectrum of the CMB. As we will
show in our analysis below, the shift parameter is very helpful in
breaking the degeneracy between $\Omega_m$ and $w(z)$ when used in a
combined analysis with supernova data. As an alternative to using the
full CMB power spectra, the shift parameter provides a good way to
summarize (see, e.g., Ref.~\cite{wang07}) CMB measurements, hence
simplifying dark energy investigations.

One caveat of using $R$ as pointed out in, e.g., Ref.~\cite{komatsu}
is the fact that $R$ is a derived quantity from fitting to the CMB
power spectrum and therefore assumes a certain cosmology. It is
therefore important to state explicitly the assumption made under
which the best-fit value for $R$ was derived. Several groups including
Refs.~\cite{elgaroy07,wang07,corasaniti08} have studied this point in
more detail and found that the constraints on $R$ are relatively
stable under minor modifications of the dark energy parameters
underlying the analysis, including dark energy
clustering~\cite{corasaniti08}. It was found that massive neutrinos
had a larger effect on $R$ (few percent level)~\cite{corasaniti08}. In
Ref.~\cite{wang07} an analysis of WMAP-3 data was carried out and it
was found that for non-flat cosmologies, the value for $R$ was very
similar for different dark energy models, including constant $w$ and
time-varying $w$ parametrized via $w_0+w_a(1-a)$. In addition, the
best-fit values for $R$ in the current WMAP-7 analysis are the same
within error bars for different underlying cosmologies, including
$w$CDM and open $w$CDM models.

In our analysis of currently available data it should be kept in mind
that we use the best-fit value for $R$ derived under the assumptions
of a flat FRW universe with $w=-1$, an effective number of neutrinos
of $N_{\rm eff}=3.04$ and a primordial power spectrum close to a power
law. As we show below, the inclusion of $R$ in the analysis in
addition to the supernova data does not alter the result for $w(z)$
itself, its main contribution is to help relax the assumption on
$\Omega_m$. For this reason, the fact that the value we use for $R$ is
derived for a specific model is of much less consequence. In the case
of our simulated data, the value of $R$ is obtained for the correct
underlying cosmology, in which case the above discussion does not
apply.

Another issue arises with the CMB measurement point due to its origin
at high redshift. The SNe and BAO data points occupy a redshift range
between $z\in (0,2)$ making it easier to set up a coherent
non-parametric reconstruction approach. The CMB data point on the
other hand is a single point around $z\sim 1000$, so far away that it
is bound to cause problems for any non-parametric method.
Consequently, we have to make some assumptions about the behavior of
$w(z)$ in the range $z\in (2,\infty)$ -- the simplest choice is
$w=const$.


\section{Reconstruction with Gaussian Process Modeling}
\label{gpmodel}
\subsection{Overview}

We have recently introduced a nonparametric reconstruction method
based on GP modeling and Markov chain Monte Carlo (MCMC), and applied
it to supernova data~\cite{holsclaw1,holsclaw2}. We refer the reader
to these papers for details on the implementation of the GP model.
Here, we provide a general introduction and explanation of the idea
behind the reconstruction process with GP models and then focus on how
to extend the method to include multiple data sources.

Gaussian processes extend the multivariate Gaussian distribution to
function spaces, with inference taking place in the space of
functions. The defining property of a GP is that the vector that
corresponds to the process at any finite collection of points follows
a multivariate Gaussian distribution.  Gaussian processes are elements
of an infinite dimensional space, and can be used as the basis for a
nonparametric reconstruction method.  Gaussian processes are
characterized by mean and covariance functions, defined by a small
number of hyperparameters~\cite{gprefs}. The covariance function
controls aspects such as roughness of the candidate functions and the
length scales on which they can change, aside from this, their shapes
are arbitrary. The use of Bayesian estimation methods (including the
MCMC algorithm) allows us to estimate the hyperparameters of the GP
correlation function together with any other parameters,
comprehensively propagating all estimation
uncertainties~\cite{gamerman}. Using the definition of a GP, we assume
that, for any collection $z_1,...,z_n$, $w(z_1),..., w(z_n)$ follow a
multivariate Gaussian distribution with a constant negative mean and
exponential covariance function written as
\begin{equation}
K(z,z')=\kappa^2\rho^{|z-z'|^\alpha}.  
\end{equation}
Here $\rho \in (0,1)$ is a free parameter that, together with $\kappa$
and the parameters defining the likelihood, are fit from the data
($\rho$ and $\kappa$ are the hyperparameters of the GP model). The
form of the assumed correlation function implies that, theoretically,
there is non-zero correlation between any two points. The parameter
$\rho$ controls the exponential decay of the correlation as a function
of distance in redshift, but it does not provide a bound for the
correlation between two points.

The value of $\alpha \in (0,2]$ influences the smoothness of the GP 
realizations: for $\alpha=2$, the realizations are smooth with 
infinitely many derivatives, while $\alpha=1$ leads to rougher 
realizations suited to modeling continuous non-differentiable 
functions. Here we use $\alpha=1$ to allow for maximum flexibility 
in reconstructing $w$. (For a comprehensive discussion of different 
choices for covariance functions and their properties, see 
Ref.~\cite{gprefs}.)  We set up the following GP for $w$:
\begin{equation}
w(u)\sim {\rm GP}(\vartheta,K(u,u')).
\end{equation}
The process is started using a mean value of $\vartheta=-1$; given
current observational constraints on $w$ this is a natural choice. Even
though the mean is fixed, each GP realization actually has a different
mean with a spread controlled by $\kappa$ and the means are adjusted
during the analysis to slightly different values suggested by
preliminary runs (we use this strategy for some of the simulated data
sets below).  This adjustment is purely informed by the data and
demonstrates the flexibility of the approach. In principle, the mean
could also be left as a free parameter. After the adjustment we
measure the posterior mean and ensure that it is close to the prior
mean.

\subsection{Combining Multiple Data Sources}

In order to determine the optimal values for the GP modeling
parameters and the cosmological parameters, we follow a Bayesian
analysis approach~\cite{gelman}. We use MCMC algorithms to fit for the
parameters~\cite{gamerman}, resulting in posterior estimates and
probability intervals for $\Omega_m$ and $\Delta_\mu$, and the
hyperparameters that specify the GP model, $\kappa$ and $\rho$.  We
choose the following priors for the hyperparameters:
\begin{eqnarray}
\pi(\kappa)&\sim&IG(6,2),\\
\pi(\rho)&\sim&Beta(6,1).
\end{eqnarray}
Here the notation ``$\sim$'' means ``distributed according
to'', and $IG$ is an inverse Gamma distribution prior, with the probability
density function $f(x;\alpha,\beta)=\beta^{\alpha}x^{-\alpha-1}
\Gamma(\alpha)^{-1}\exp(-\beta/x)$, with $x>0$. The probability
distribution of the $Beta$ prior is given by
$f(x;\alpha,\beta)=\Gamma(\alpha+\beta)x^{\alpha-1}(1-x)^{\beta-1}/
[\Gamma(\alpha)\Gamma(\beta)]$ (for examples of these distributions,
see, e.g. \cite{holsclaw2}).  

Turning to the cosmological parameters, we choose: 
\begin{eqnarray}
  \pi(\Omega_{m})& \sim& N(0.27,0.04^2) {\rm ~~~~SN~data~only},\label{om}\\
  \pi(\Omega_{m})& \sim&U(0,1) {\rm ~~~~~~combined~analyses},\label{om_u}\\
  \pi(\Delta_\mu) &\sim& U(-0.5,0.5),\label{H0}\\
  \pi(\sigma^2) &\propto& \sigma^{-2}{\rm ~~~~~~~~~~~~~~~~~~~~~~~SN~data},\\
   \pi(\sigma_B^2) &\propto& \sigma_B^{-2}{\rm~~~~~~~~~~~~~~~~~~~~BAO~data},
\end{eqnarray}  
where $U$ is a uniform prior, with the probability density function
$f(x;a,b)=1/(b-a)$ for $x\in [a,b]$ and $0$ otherwise. $N$ is a
Gaussian (or Normal distributed) prior with the probability density
function
$f(x;\mu,\sigma^2)=\exp[-(x-\mu)^2/(2\sigma^2)]/\sqrt{2\pi\sigma^2}$.
The squared notation for the second parameter in $N(\mu,\sigma^2)$ is
used to indicate that $\sigma$ is the standard deviation (to prevent
possible confusion with the variance $\sigma^2$). (The parameters in
the $U$ and $IG$ distributions do not have this same meaning of mean
and standard deviation as in the Normal distribution.)

\begin{figure*}[t]	
\centerline{
  \includegraphics[width=2.1in]{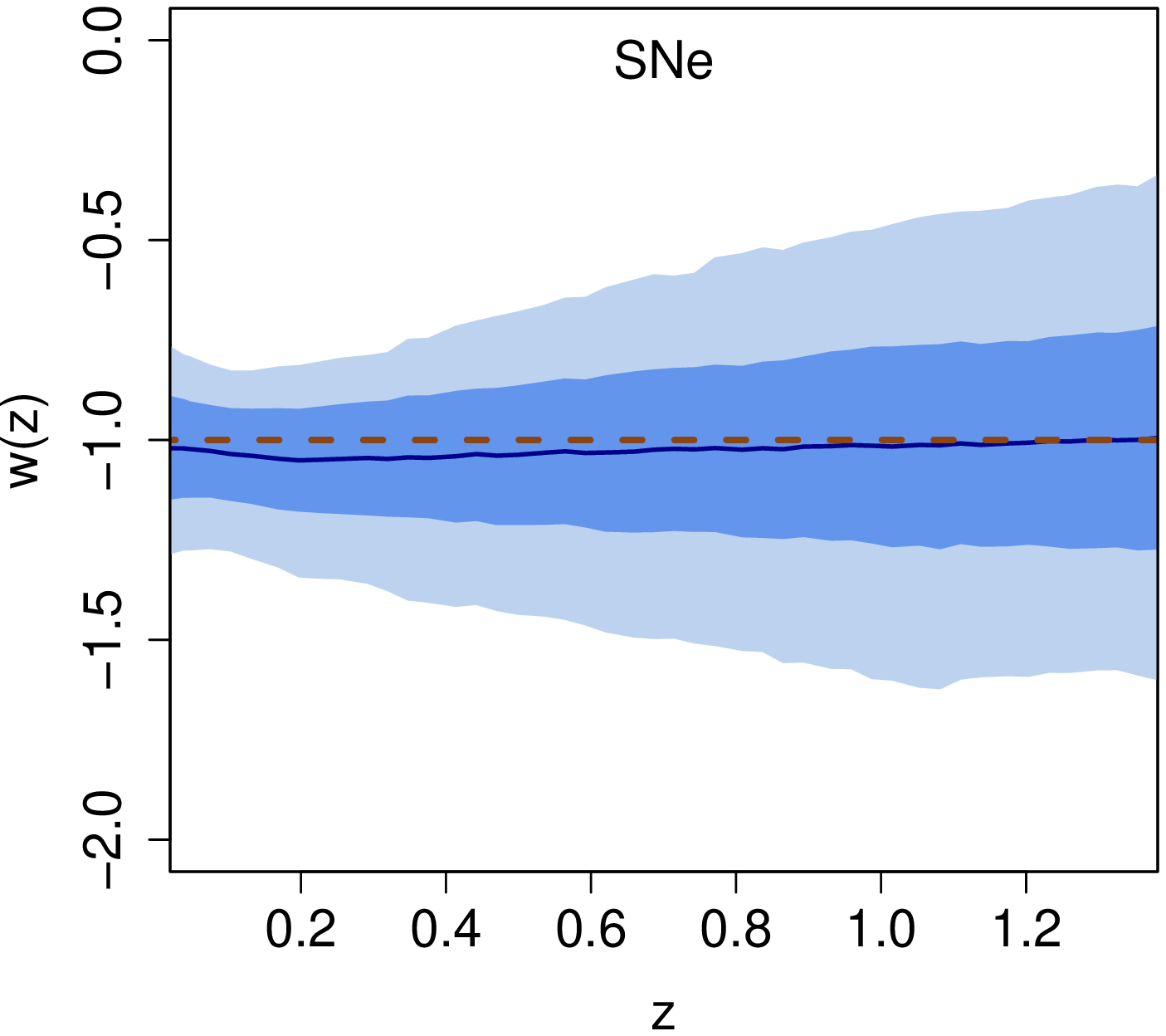}
  \hspace{-1.2cm}\includegraphics[width=2.1in]{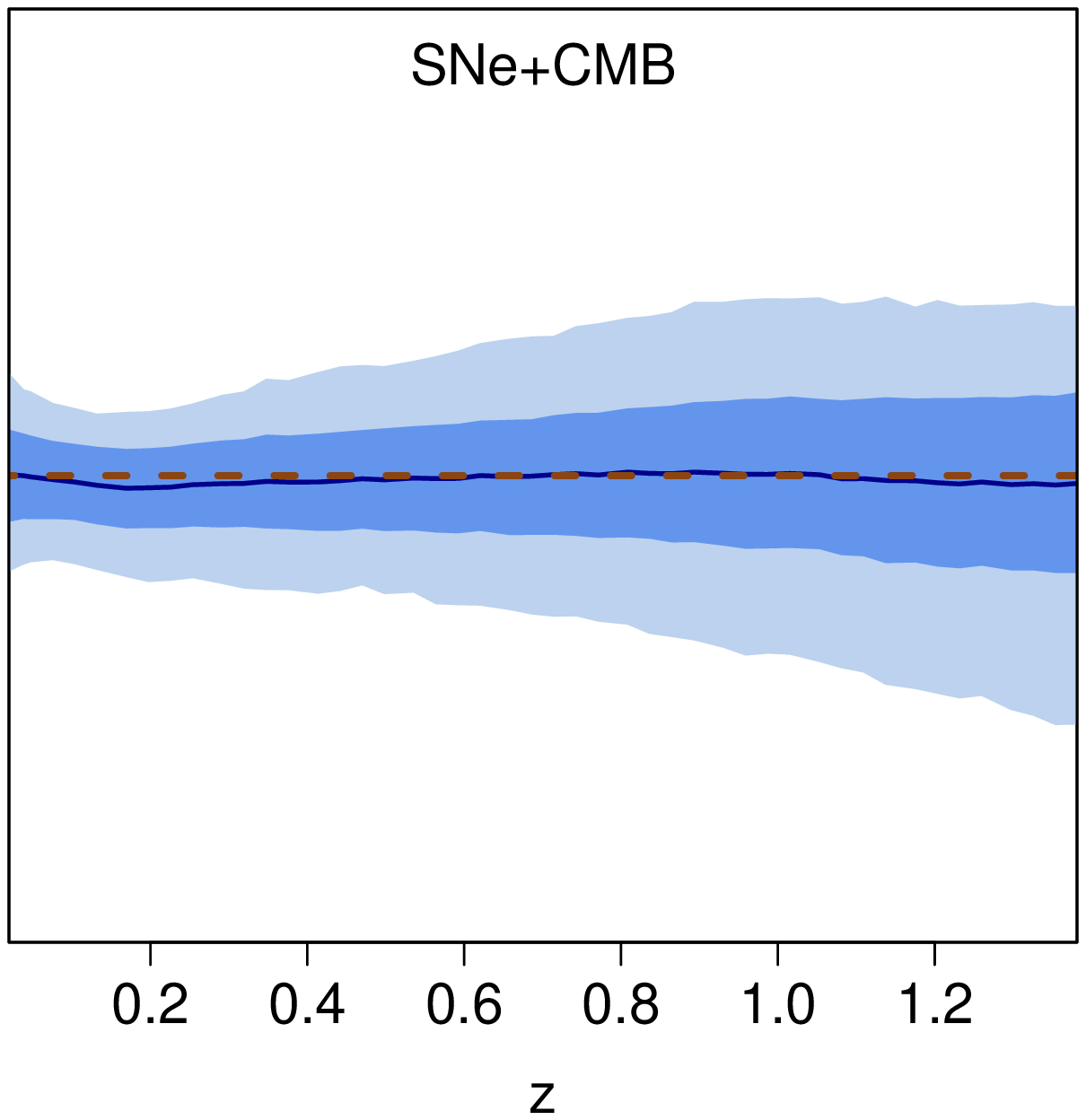}
  \hspace{-1.2cm}\includegraphics[width=2.1in]{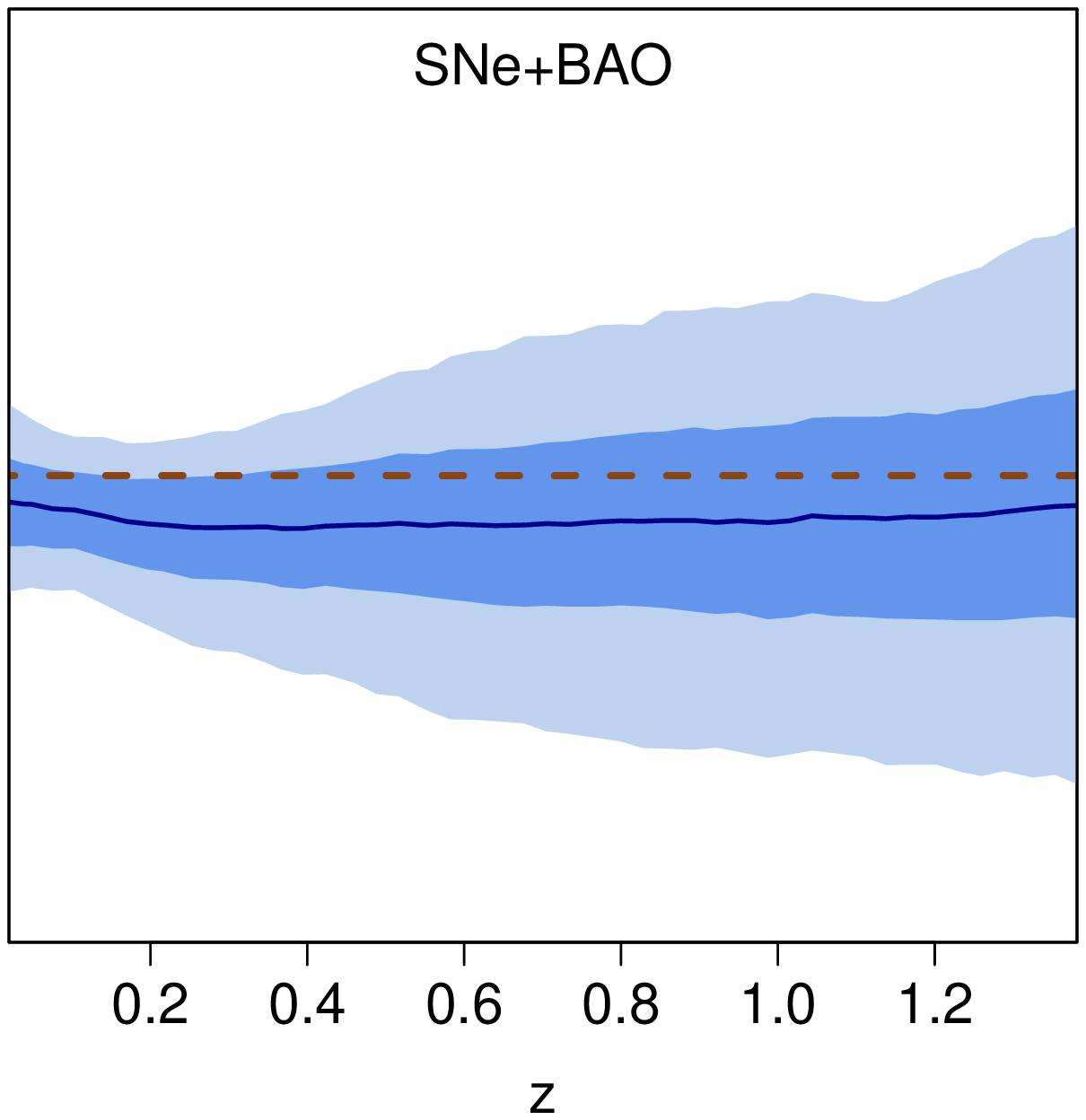}
  \hspace{-1.2cm}\includegraphics[width=2.1in]{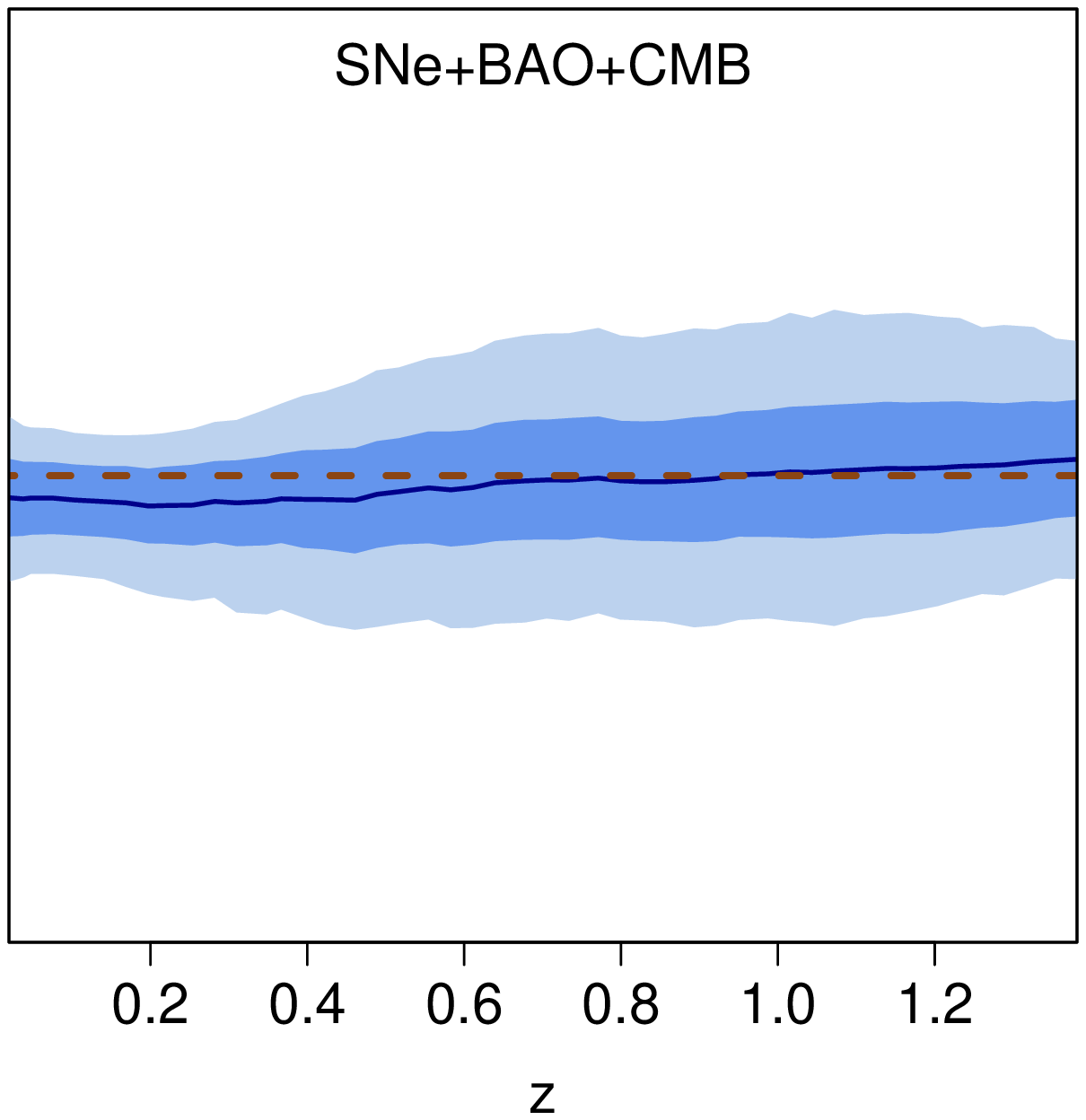}} 

  \vspace{-0.53cm} 

  \centerline{
   \includegraphics[width=2.1in]{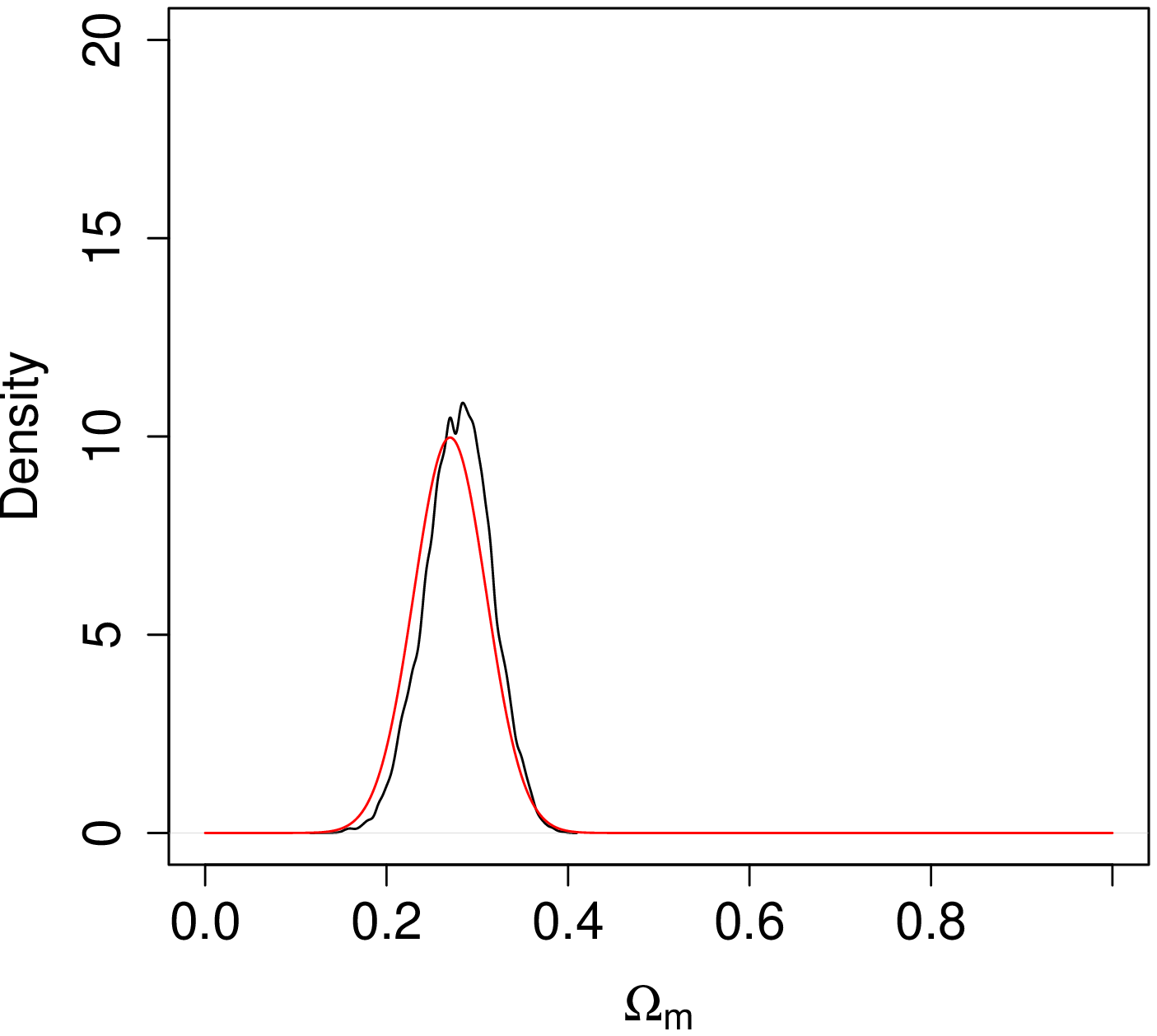}
  \hspace{-1.2cm}\includegraphics[width=2.1in]{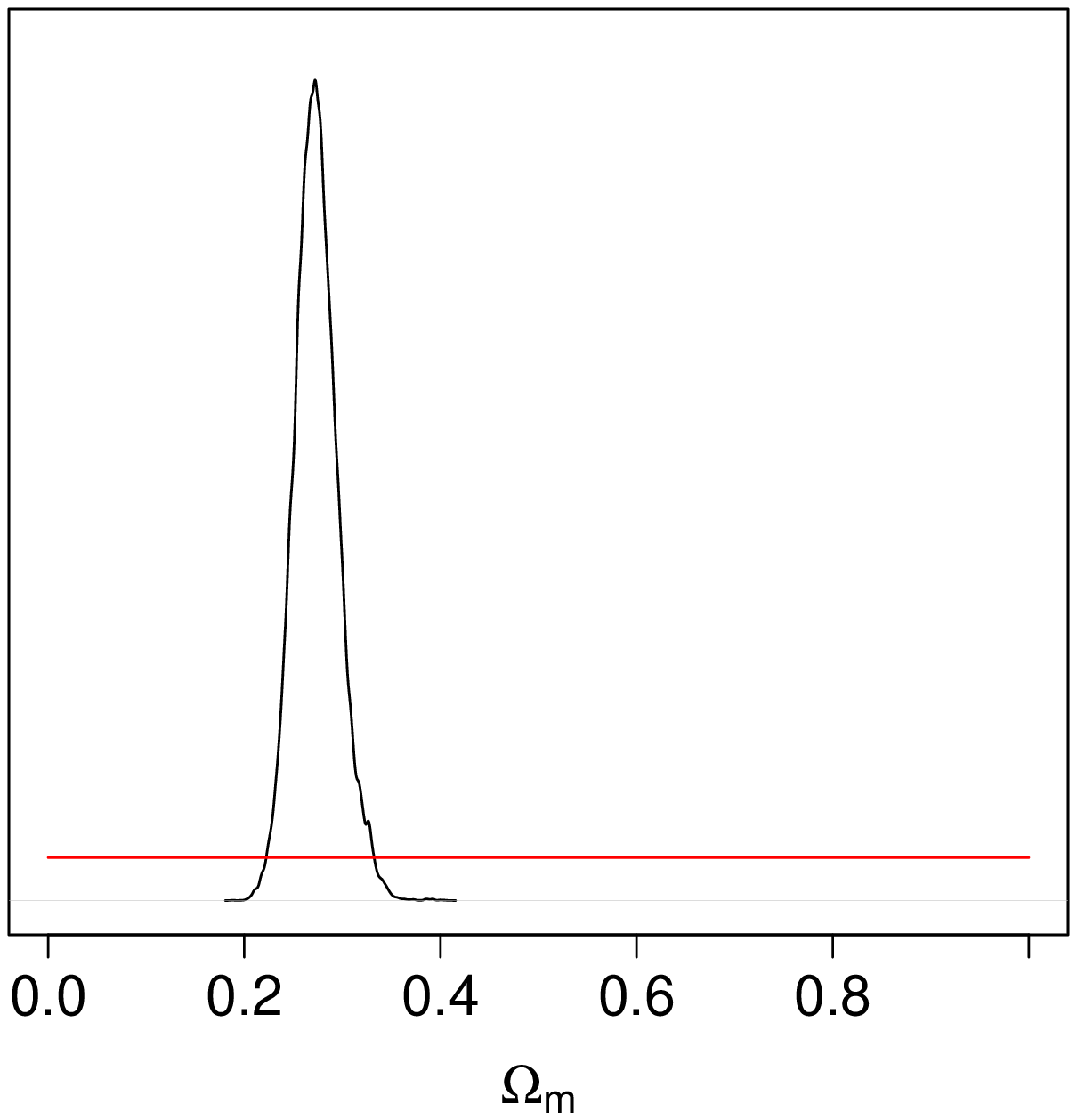}
  \hspace{-1.2cm}\includegraphics[width=2.1in]{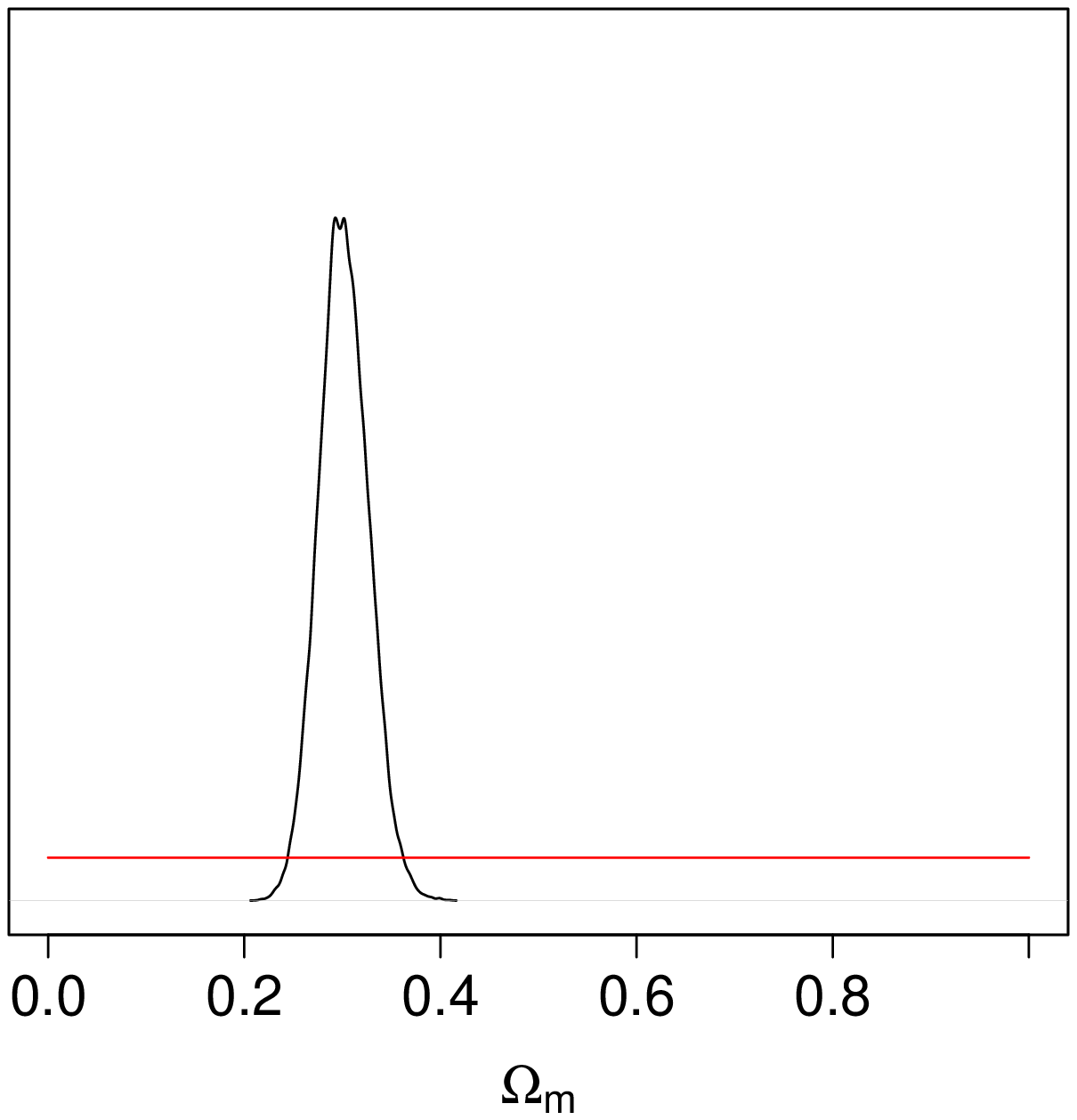}
  \hspace{-1.2cm}\includegraphics[width=2.1in]{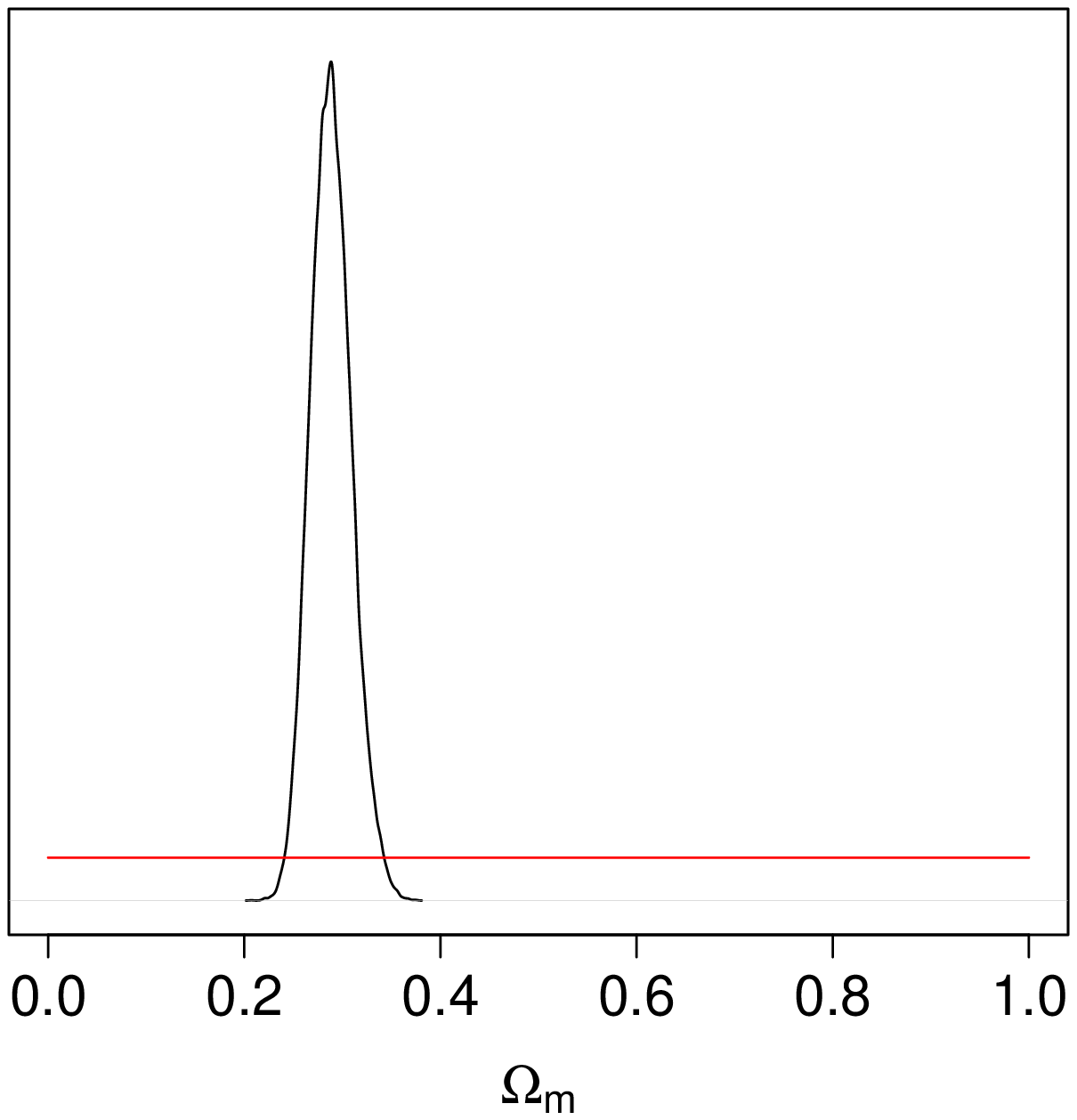}} 
  \caption{\label{data}Results for reconstructing $w(z)$ from
    currently available data. The top row shows reconstruction results
    for $w(z)$ (red line; the black dashed line shows $w=-1$) for an
    exponential covariance function, i.e. $\alpha=1$, including
    different cosmological probes, the second row shows the
    corresponding posterior for $\Omega_m$ (red lines: priors, black
    lines: posteriors).  The first column shows the results from
    supernova data only, the second column includes CMB measurements,
    the third column uses supernova and BAO data, and the fourth
    column shows the results for a combined supernova-BAO-CMB
    analysis. The light blue contours show the 95\% confidence level,
    the dark blue contours the 68\% confidence level. As is to be
    expected, the error bars shrink somewhat if more data sources are
    included, though the effect is small due to the limited number of
    extra data points. As found previously by others
    (e.g.~\cite{Wang:2009sn}), current data are consistent with a
    cosmological constant. }
\end{figure*}

The prior for $\Omega_m$ for the analysis of supernova data alone is
informed by the 7-year WMAP analysis~\cite{komatsu} for a $w$CDM model
combining CMB, BAO, and $H_0$ measurement. Once a second cosmological
probe is included in the analysis, the assumption on this prior can be
relaxed and we choose a uniform prior for the analysis of the combined
data sets. We also allow for an uncertainty in the overall calibration
of the supernova data, $\Delta_\mu$, and choose a wide, uniform prior
for $\Delta_\mu$.

\begin{table*}
     \caption{Union 2 Data set- 95\% PIs, the last two columns are
       results from Ref.~\cite{amanullah}, Table 11 for comparison.}
     \centering \scriptsize
         \begin{tabular}{ ccccc cc|||cc }
             \hline
Data Type & $\Omega_{m}$ & $\Delta_{\mu}$  & $\sigma_2$ & $\rho$ & $\kappa^2$ &$\vartheta$ &$\Omega_m$~\cite{amanullah} & $w$~\cite{amanullah}\\ [0.9ex]
\hline
SNe &$ 0.279 ^{+ 0.070 }_{ -0.074 }$&$ -0.003 ^{+ 0.028 }_{ -0.028 }$&$ 0.985 ^{+ 0.124 }_{ -0.110 }$&$ 0.870^{+ 0.127 }_{ -0.303 }$&$ 0.353 ^{+ 0.393 }_{ -0.192 }$ &-1.00 & $0.270^{+0.021}_{-0.021}$ & -1 (fixed) \\
SNe+BAO &$ 0.302 ^{+ 0.051 }_{ -0.048 }$&$ -0.004 ^{+ 0.028 }_{ -0.027 }$&$ 0.981 ^{+ 0.122 }_{ -0.109 }$&$ 0.864 ^{+ 0.132 }_{ -0.322 }$&$ 0.363 ^{+ 0.437 }_{ -0.202 }$ &-1.07 & $0.309^{+0.032}_{-0.032}$ & $-1.114^{+0.098}_{-0.112}$\\
SNe+CMB &$ 0.274 ^{+ 0.049 }_{ -0.041 }$&$ -0.002 ^{+ 0.028 }_{ -0.027 }$&$ 0.982 ^{+ 0.123 }_{ -0.109 }$&$ 0.865 ^{+ 0.132 }_{ -0.333 }$&$ 0.358 ^{+ 0.447 }_{ -0.199 }$ &-1.00 & $0.268^{+0.019}_{-0.017}$ & $-0.997^{+0.050}_{-0.055}$\\
SNe+BAO+CMB &$ 0.289 ^{+ 0.044 }_{ -0.038 }$&$ -0.005 ^{+ 0.027 }_{ -0.027 }$&$ 0.981 ^{+ 0.122 }_{ -0.109 }$&$ 0.869 ^{+ 0.127 }_{ -0.302 }$&$ 0.366 ^{+ 0.436 }_{ -0.201 }$& -1.01 & $0.277^{+0.014}_{-0.014}$ & $-1.009^{+0.050}_{-0.054}$\\
\hline
         \end{tabular}
     \label{table:newmuu2}  
\end{table*}

Next we discuss the likelihoods for the different probes.  We assume
that the SNe, CMB, and BAO measurements are independent of each other
which allows us to derive a likelihood for each probe separately.  The
likelihood for the supernova data is given by:
\begin{equation}\label{L}
L^{\rm SN}(\sigma,\theta)\propto
\left(\frac{1}{\tau_i\sigma}\right)^n
\exp\left(-\frac{1}{2}\sum_{i=1}^n
  \left(\frac{\mu_i-\mu(z_i,\theta)}{\tau_i\sigma}\right)^2\right),    
\end{equation}
where $\theta$ encapsulates the cosmological parameters as well as the
hyperparameters, i.e., $\{\Delta_\mu,\Omega_m,
\kappa,\rho\}$ and $\sigma^2$ is the associated variance, expected to
be close to unity.  For the CMB data we have an equivalent expression:
\begin{equation}
  L^{\rm CMB}(\theta) \propto \frac{1}{\tau_{z^*}}\exp\left({ -\frac{1}{2} {\left(  \frac{y^*- R(z^*,\theta)}{\tau_{z^*}}  \right) } ^2}\right).
\end{equation}
Since we only have one data point, we cannot assign a variance
parameter.  The likelihood for the BAO data is slightly more
complicated. For each BAO point we have two observed distance
measures. These measurements $(y_{1i},y_{2i})$ are correlated and we
assume that they have a correlated and bivariate Normal distribution,
given by:
\begin{equation} \label{eq52}
  \begin{bmatrix}
    y_{1i} \\
    y_{2i} \\
  \end{bmatrix}
 \sim MVN
\begin{bmatrix}
   \begin{bmatrix}  D_A(z_i)/r_s \\  H(z_i)*r_s  \\
   \end{bmatrix},  \sigma_B^2{\bf K}
\end{bmatrix},
\end{equation}
where
\begin{equation}
{\bf K}=\begin{bmatrix}
      \sigma^2_{y_{1i}}& r_{12i}\sigma_{y_{1i}}\sigma_{y_{2i}}\\
      r_{21i}\sigma_{y_{1i}}\sigma_{y_{2i}}& \sigma^2_{y_{2i}}\\
   \end{bmatrix},
\end{equation}
and $\sigma_B$ is the associated variance parameter.
This leads to the following likelihood for the BAO data:
\begin{equation}\label{eq54}
L^{\rm BAO} (\sigma_{B},\theta)
\propto \frac{1}{|{\sigma_B^2\bf K}|^{m/2}}  
\exp\left({-\frac{1}{2\sigma^2_B}\sum_{i=1}^m\left( {\bf D}'{\bf K^{-1}}{\bf
D}\right)}\right),
\end{equation}
with
\begin{equation}
{\bf D}=\left(\begin{array}{c}
y_{1i}-D_a(z)/r_s\\
y_{2i}-H(z)*r_s
\end{array}\right).
\end{equation}
We can find the combined likelihood simply by multiplying the
likelihoods since we assume the different probes are uncorrelated:
\begin{equation} 
L^{\rm total}=L^{\rm SNe}*L^{\rm CMB}*L^{\rm BAO}.
\end{equation}

\section{Results for Current Observations}
\label{realdat}

We begin our analysis by reconstructing $w(z)$ from currently
available data. We use the supernova data set recently released by
Amanullah et al.~\cite{amanullah}. This so-called Union-2 compilation
(extending the Union compilation from Ref.~\cite{hicken09}) consists
of 557 supernovae between redshift $z=0.015$ and $z=1.4$.  The
magnitude errors in the data set range between 0.08 and 1.02, with an
average error of $\sim 0.2$.

In addition to the supernova data we include the most recent BAO
measurements from the Two-degree-Field Galaxy Redshift Survey (2dFGRS)
at $z=0.2$ and the Sloan Digital Sky Survey (SDSS)~\cite{2dF} at
$z=0.35$ given by
\begin{eqnarray}
r_s(z_d)/d_V(z=0.2)&=&0.1905\pm 0.0061,\\
r_s(z_d)/d_V(z=0.35)&=&0.1097\pm 0.0036,
\end{eqnarray}
where $d_V(z)=[(1+z)^2d_A^2 cz/H(z)]^{1/3}$. 
For the CMB analysis, we use the most recent measurement of the shift
parameter $R$ from WMAP-7~\cite{komatsu}, given by
\begin{equation}
R(z_\star)=1.719\pm 0.019.
\end{equation}
In order to have a complete description of the problem we have to
specify some additional cosmological parameters that are expected to
have little or no effect on dark energy. These parameters -- fixed at
the best-fit WMAP-7 values from their $\Lambda$CDM analysis -- are:
$\Omega_r=4.897\cdot 10^{-5}$, $z_d=1020.3$, $z_\star=1090.79$, and
$\Omega_b/\Omega_r=914.54$.

We carry out four different analyses: supernova data by themselves
with a Gaussian prior for $\Omega_m$ given in Eq.~(\ref{om}), and
combined analyses for supernova data and CMB, supernova data and BAO,
and for all three probes. For the combined data sets we can relax the
prior assumptions on $\Omega_m$ and use a wide uniform prior, given in
Eq.~(\ref{om_u}).  The results are summarized in
Table~\ref{table:newmuu2} and Fig.~\ref{data}.

All results are consistent with a cosmological constant, i.e. $w=-1$,
as can be seen in the first row in Fig.~\ref{data}. The supernova data
by themselves have no constraining power on $\Omega_m$ and therefore
force us to choose a rather strong prior. The lower panels in
Fig.~\ref{data} show the prior (red line) and posterior (black line)
for $\Omega_m$ demonstrating this point clearly. If we include either
CMB or BAO or both, the constraints on $\Omega_m$ get much better. As
can been seen in Table~\ref{table:newmuu2}, the error estimates for
$\Omega_m$ shrink by almost a factor of two if all probes are
combined.  Overall, the supernova data by themselves lead to a
slightly higher value of $\Omega_m$, while the combination with CMB
data leads to a lower value. The inclusion of the BAO points shifts up
the value for $\Omega_m$ considerably, by more than 10\% compared to
the supernova--CMB analysis. Nevertheless, within the error bars, all
values for $\Omega_m$ are consistent and agree well with the best-fit
WMAP-7 values including different probes. The value for the shift
parameter $\Delta_{\mu}$ is very close to zero in all cases.

A brief comparison with Ref.~\cite{amanullah} also shows very good
agreement. For ease of comparison, we quote their results in the last
two columns of Table~\ref{table:newmuu2} for the case of a flat
Universe and $w=const$. The trends in the best-fit value for
$\Omega_m$ are exactly the same as we find, the value is lowest for
the case of supernova+CMB data and highest for supernova+BAO data.
They also find that for the supernova+BAO analysis, $w$ is slightly
below $w=-1$ while for all other cases it is very close to $w=-1$.
Their analysis is also consistent with a cosmological constant. It is
very interesting to note that our error estimates for $w(z)$ are also
very similar to the findings of Amanullah et al., even though their
assumption of $w=const.$ is very restrictive. This is very encouraging
since it shows that our method leads to tight error bounds without
loss of flexibility in allowing for time variations in $w(z)$. In
contrast, a $w_0-w_a$ fit would have increased the error bars
considerably.

\section{Results for Simulated Data}
\label{simdat}
In this section we investigate how well our method works for
reconstructing $w(z)$ with future high-quality data. Current
limitiations -- uncertainties in the data and limited statistics --
prevent us from extracting possible time variations in $w(z)$
reliably. The error bands are still rather large and results are in
complete agreement with a cosmological constant. Future measurements
will hopefully change this: if there is a small time variation in
$w(z)$ we should be able to detect it. In our previous
paper~\cite{holsclaw1} we generated a supernova data set, assuming
high-quality measurements from a WFIRST-like mission. We showed that a
set of $\sim$ 2300 supernovae out to a redshift of $z=1.7$ and perfect
knowledge of $\Omega_m$ allows us to confidently extract time
variations in the dark energy equation of state. We also showed that
larger uncertainties in $\Omega_m$ degraded this result due to
well-known degeneracies between $w$ and $\Omega_m$. These degeneracies
can be broken by including different data sources. We show in the
following that a combination of accurate supernova data with results
from a BAO survey such as BigBOSS will provide sufficient information
to enable a reliable and interesting reconstruction of $w(z)$. The
combination of these different data sources eliminates the degeneracy
problem and provides reliable constraints on the time variation of
$w(z)$ without requiring ``perfect'' knowledge of $\Omega_m$.

\subsection{The Simulated Data}

\begin{figure}[b]
\centerline{
  \includegraphics[width=2.1in]{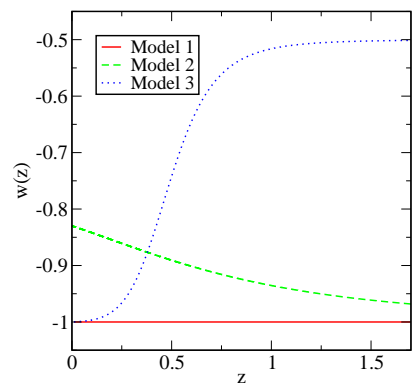}} 
  \caption{\label{eossim}Dark energy equation of state $w(z)$ for our
    three simulated models.} 
  \end{figure}

  We generate simulated data for all three probes (supernovae, BAO,
  CMB) and three different cosmological models. The models used are
  the same as in our previous work~(see Ref.~\cite{holsclaw1} for more
  details). Model 1 has a constant dark energy equation of state
  $w=-1$, Model 2 is based on a quintessence model with a minimally
  coupled scalar field and a dark energy equation of state
  $w(z)=(\dot\phi/2-V_0\phi^2)/(\dot\phi^2/2+V_0\phi^2)$, and for
  Model 3 we choose a slightly more extreme quintessence model with
  $w(z)=-1.0006+308472/(\exp[20/(1+z)+617439])$. The resulting
  equations of state are shown in Figure~\ref{eossim}. For each model
  we choose $\Omega_m=0.27$ and fix $H_0=70.4$~km/s/Mpc,
  $\omega_b=0.0226$, $\omega_r=2.469\cdot10^{-5}$, $z_\star=1090.89$,
  and $z_d=1020.5$. While Model 3 is already ruled out observationally
  it provides a good example for a rather sharp transition in $w(z)$.
  For each model we create two data sets: (i) We assume the
  best-possible scenario, a space mission to obtain supernova
  measurements out to redshift $z=1.7$ and in addition a BigBOSS-like
  BAO survey, and CMB data; (ii) good ground based supernova
  measurements in combination with BigBOSS and CMB measurements. In
  the following we provide some details on the assumptions for the
  different data sets.

\subsubsection{Supernova Measurements}

As mentioned above we investigate two different sets of simulated
supernova measurements. The first one is the same as we used in
Ref.~\cite{holsclaw1}. It contains 2298 data points distributed over a
redshift range of $0<z<1.7$ with larger concentration of supernovae in
the midrange redshift bins ($0.4<z<1.1$) and at low redshift
($z<0.1$). The exact distribution is shown in Ref.~\cite{holsclaw1} in
Fig.1. For the distance modulus we assume an error of
$\tau_i=0.13$. The measurements are presented in the following form:
\begin{equation}
\tilde\mu_i=\alpha_i+\epsilon_i.
\end{equation}
In this notation, the observations $\tilde\mu_i$ follow a normal
distribution with mean $\alpha(z_i)$, the standard deviation being set
by the distribution of the error, $\epsilon_i$, representing a
mean-zero normal distribution with standard deviation,
$\tau_i\sigma$. Here, $\tau_i$ is the observed error and $\sigma$
accounts for a possible rescaling. In addition, we assume that the
errors are independent.

\begin{figure}[t]	
\centerline{
  \includegraphics[width=2.1in]{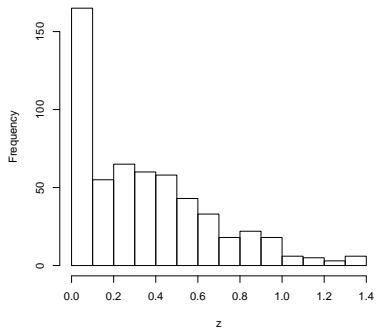}} 
  \caption{\label{snedist}Redshift distribution of the small supernova
    data set. The simulated data has exactly the same distribution as
    the real data. }
\end{figure}

\begin{figure}[t]	
\centerline{
  \includegraphics[width=2.1in]{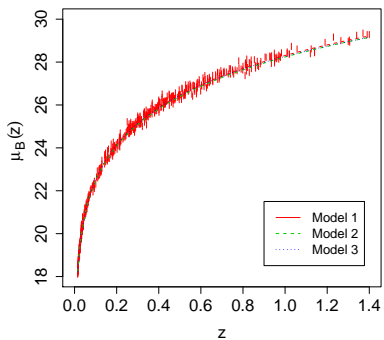}} 
  \caption{\label{sne}Small supernova data set with error bars and the 3 simulated models. }
\end{figure}

For the second set of simulated supernova data we consider the same
number of data points as currently available from ground-based surveys
(557 measurements). The redshift distribution is shown in
Fig.~\ref{snedist}. The distribution extends to $z=1.4$ with a maximum
at low redshift and around $z=0.3$. Only a handful of supernovae are
available at higher redshifts. Since we assume that the measurements
are taken from the ground, we increase the errors on the distance
modulus to $\tau_i=0.15$.  Figure~\ref{sne} shows the distance modulus
redshift relation of these measurements for Model 1 with error
bars. The exact relations for Model 2 and 3 are shown in addition. The
differences between the three models are very small, pointing to the
challenge of the reconstruction task.

\subsubsection{CMB Measurements}

For the CMB points we use the following realizations (the exact values
for $R$ for each model are given in parentheses):
\begin{eqnarray}
{\rm Model~1:} & R(z_\star)=1.736\pm 0.019 & (R^{\rm ex}=1.723),\\
{\rm Model~2:} & R(z_\star)=1.716\pm 0.019 & (R^{\rm ex}=1.702),\\
{\rm Model~3:} & R(z_\star)=1.683\pm 0.019 & (R^{\rm ex}=1.670).
\end{eqnarray}

\begin{figure}[t]	
\centerline{
  \includegraphics[width=1.9in]{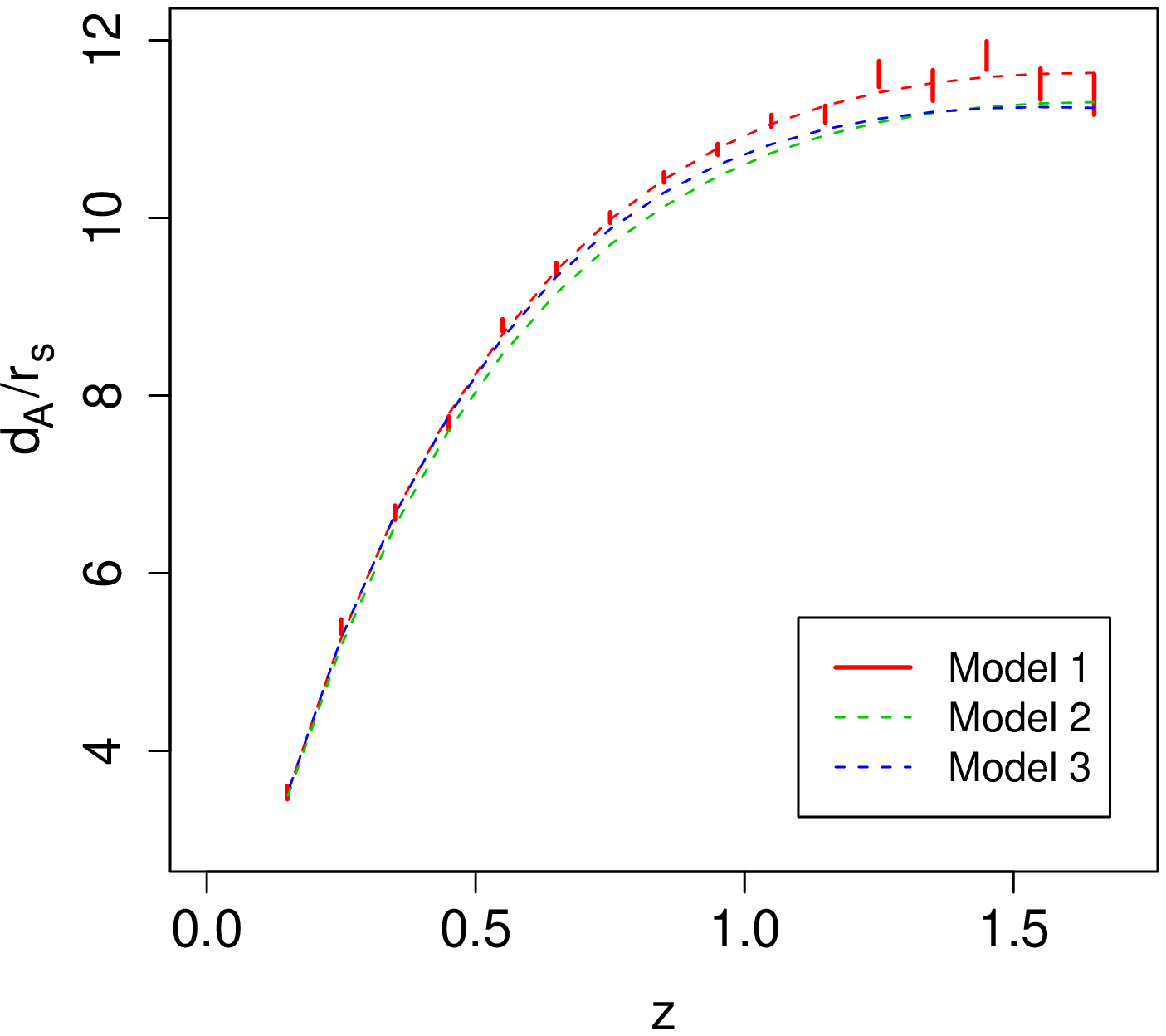}
  \hspace{-0.3cm}\includegraphics[width=1.9in]{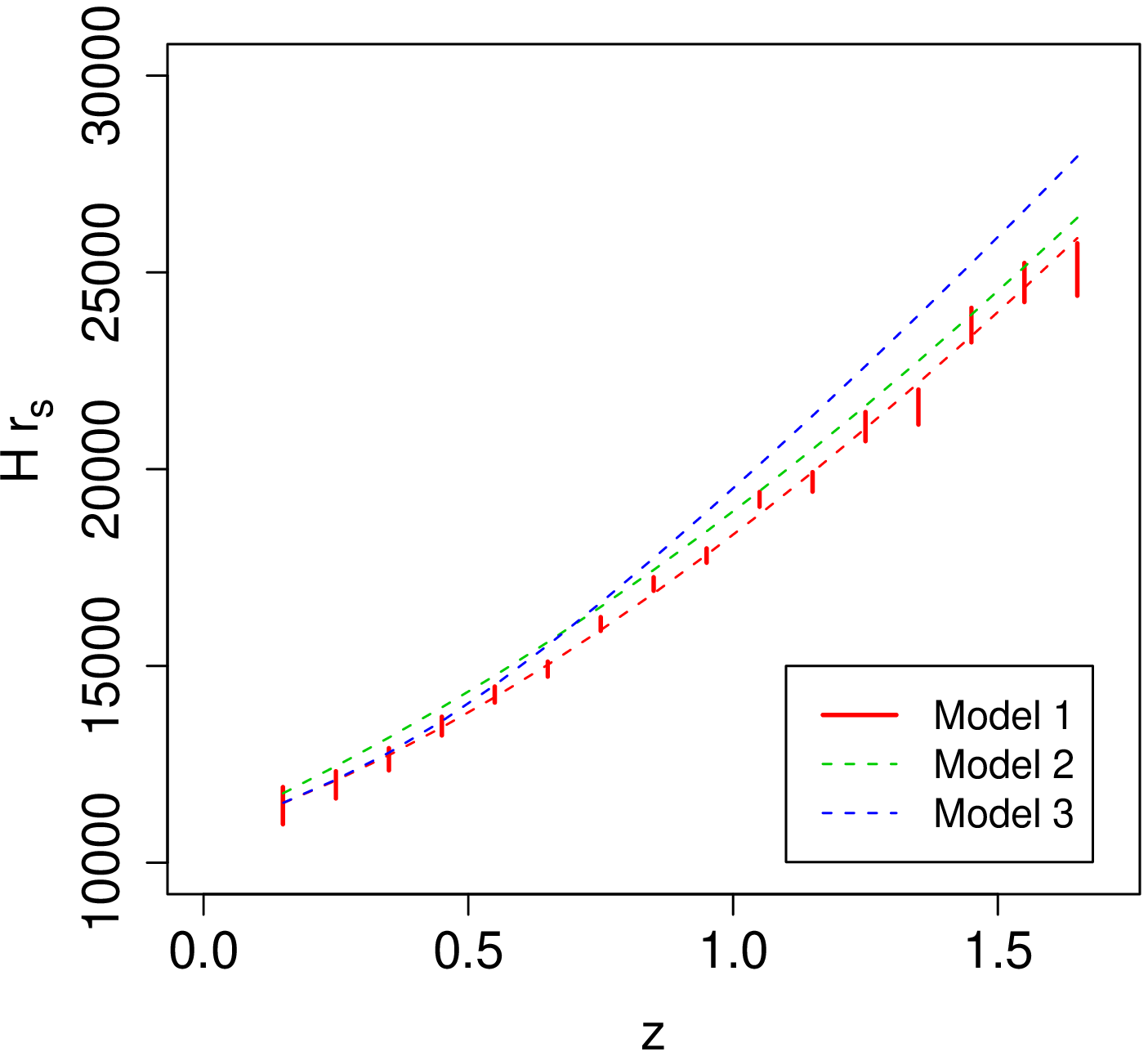}} 

 \vspace{-.5cm} 

  \centerline{
   \includegraphics[width=1.9in]{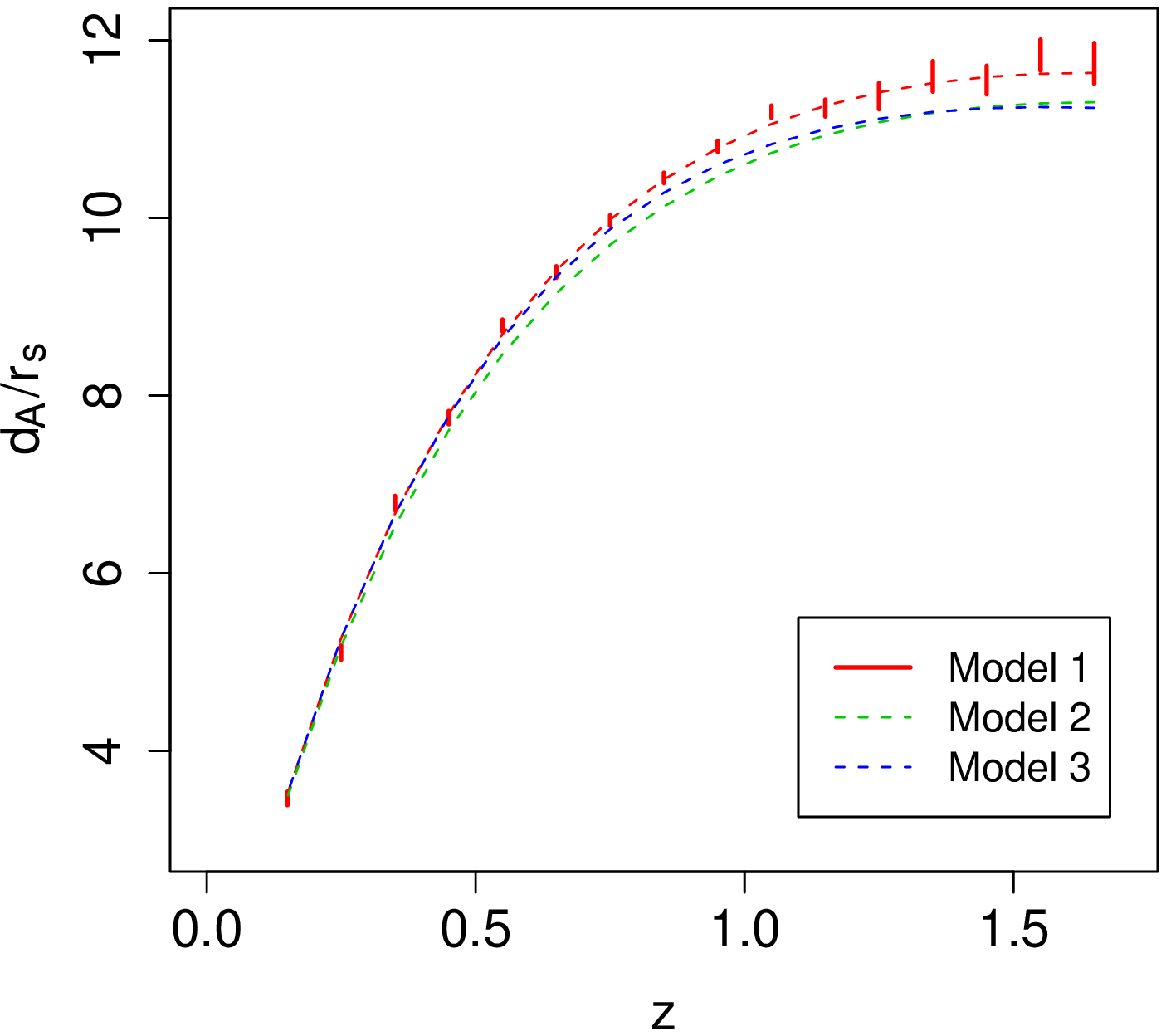}
  \hspace{-0.3cm}\includegraphics[width=1.9in]{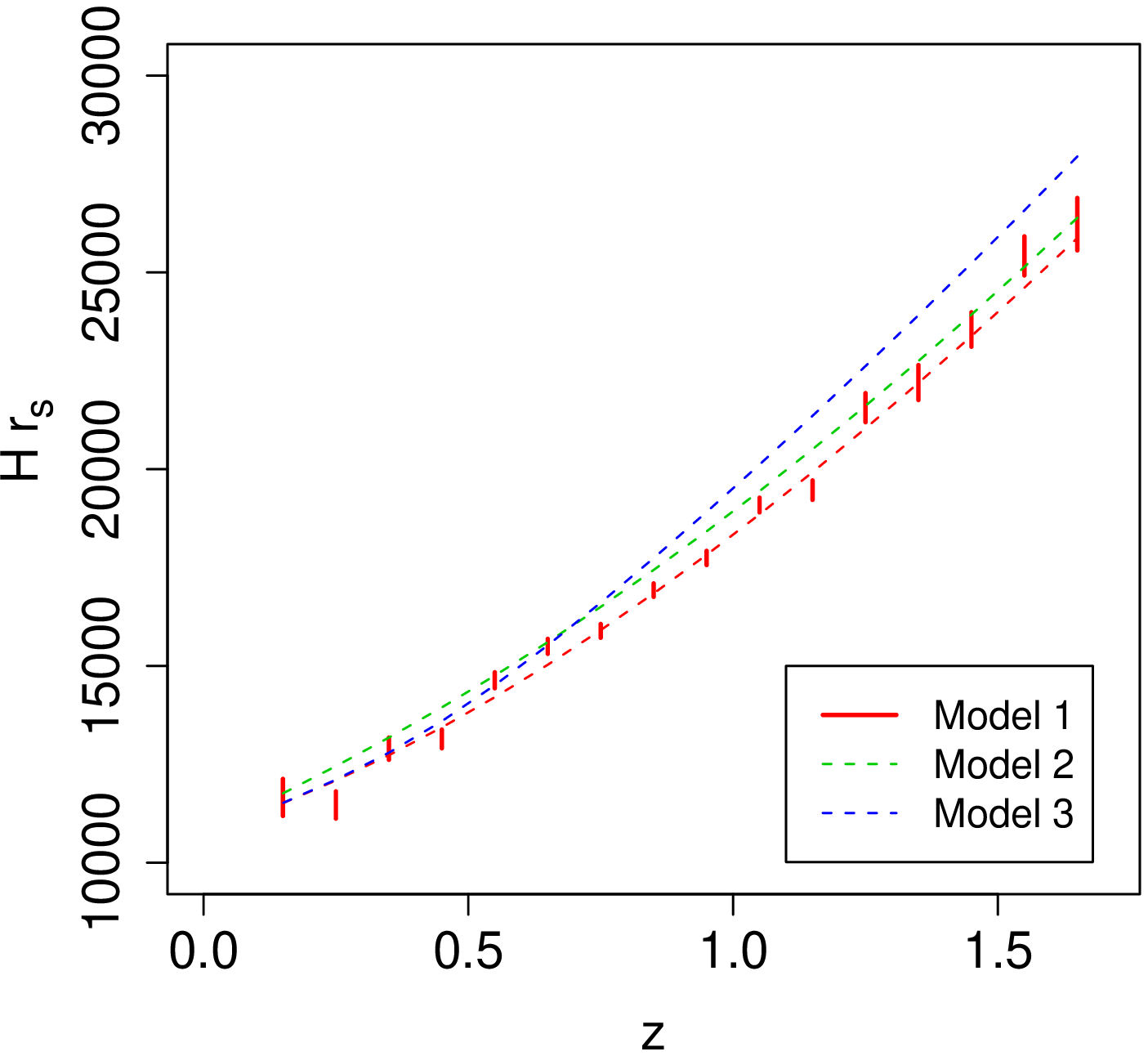}} 
  
   \vspace{-.5cm} 

  \caption{\label{bao}Two realizations of 20 simulated BAO points for a
    BigBOSS-like survey. Left column: angular distance diameter, right
    column: Hubble parameter both in terms of the sound
    horizon. Model 1 ($w=const.$) is shown with error bars with one
    standard deviation. For Model 2 (green dashed) and Model 3 (blue
    dotted) we show the exact predictions.}
\end{figure}

\subsubsection{BAO Measurements}
Future BAO surveys such as BigBOSS will obtain measurements of the
angular diameter distance, $d_A(z)$, as well as the Hubble parameter
$H(z)$, in terms of the sound horizon at the epoch of baryon drag,
$r_s(z_d)$. For our simulated BAO data sets, we follow the
specifications for a BigBOSS survey as outlined in
http://bigboss.lbl.gov/docs/BigBOSS\_NOAO\_public.pdf. We assume a
survey area of 24000 deg$^2$ (covering northern and southern skies)
and adopt the galaxy density distribution estimated in the BigBOSS
proposal (Table 2.3 in the aforementioned document). In this proposal,
measurements from luminous red galaxies and emission-line galaxies are
combined. The resulting distribution accounts for several sources of
inefficiency (discussed in the BigBOSS proposal) leading to a
degradation of the galaxy number density at high redshift. Often, a
constant galaxy density over the whole redshift range is assumed. We
studied this case as well and found that the eventual results in both
cases are very similar. In order to derive estimates for the errors of
the simulated measurements, we use a publicly available code
introduced in Ref.~\cite{bao2}. The formula used to obtain BAO errors
in this code is a 2D approximation of the full Fisher matrix
formalism. In \cite{bao2}, the results for the full Fisher matrix
calculation and this method are shown to match well. Although these
results are for $\Lambda$CDM, they should hold for other cosmologies.

The input parameters for the code are: $\sigma_8$ at the present
epoch, $\Sigma_{\perp} = \Sigma_0 G = $ transverse {\em rms}
Lagrangian displacement, with $G =$ growth factor normalized such that
$G = (1+z)^{-1}$ at high redshift, $\Sigma_0 = 12.4 h^{-1}$ Mpc for a
cosmology with $\sigma_8 = 0.9$ at present and scaling linearly with
$\sigma_8$; $\Sigma_{\parallel} = \Sigma_0 G (1+f) =$ line of sight
{\em rms} Lagrangian displacement, with $f = d({\rm ln}G)/d({\rm ln}
a), G, \Sigma_0$ as before; and the number density $= 3 \times 10^{-4}
h^3/{\rm Mpc}^3$ (\cite{bigboss}). $G, f, \sigma_8$ are input
correctly for each model. The biggest possible source of error are the
formulae used for $\Sigma_{\perp}, \Sigma_{\parallel}$; these were
shown to be reasonable fits to the true values in \cite{bao1}. The
value of $\Sigma_0$ given is also for the cosmology used in
\cite{bao1}. For a different cosmology, $\Sigma_0$ would obviously be
different, and the simplest way to deal with this, as suggested in the
paper, is to scale it linearly with $\sigma_8$. This may not be
completely accurate as we use very different cosmological models but
should yield a reliable estimate.

\begin{figure}[b]
  \centerline{
   \includegraphics[width=1.9in]{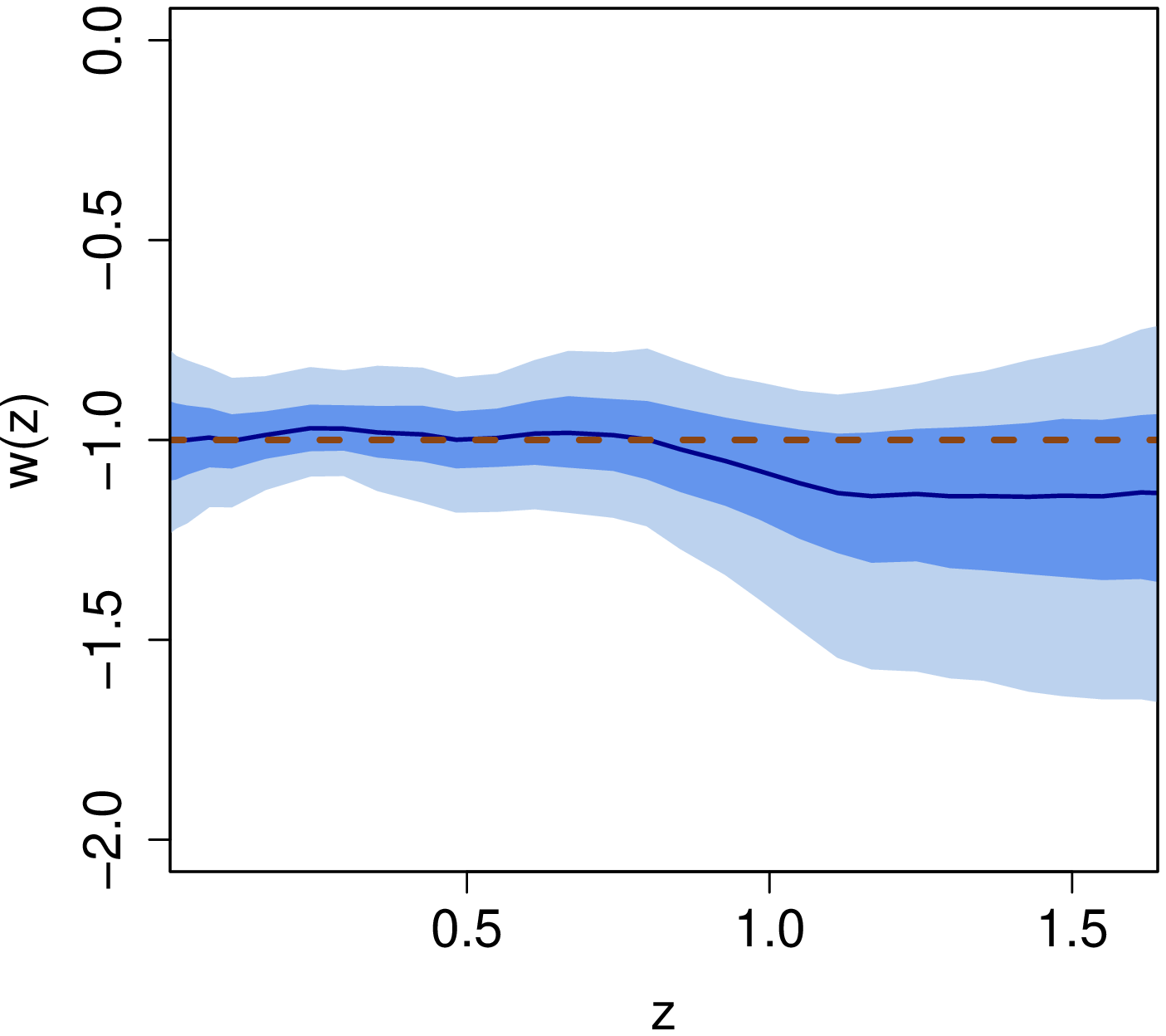}
  \hspace{-0.3cm}\includegraphics[width=1.9in]{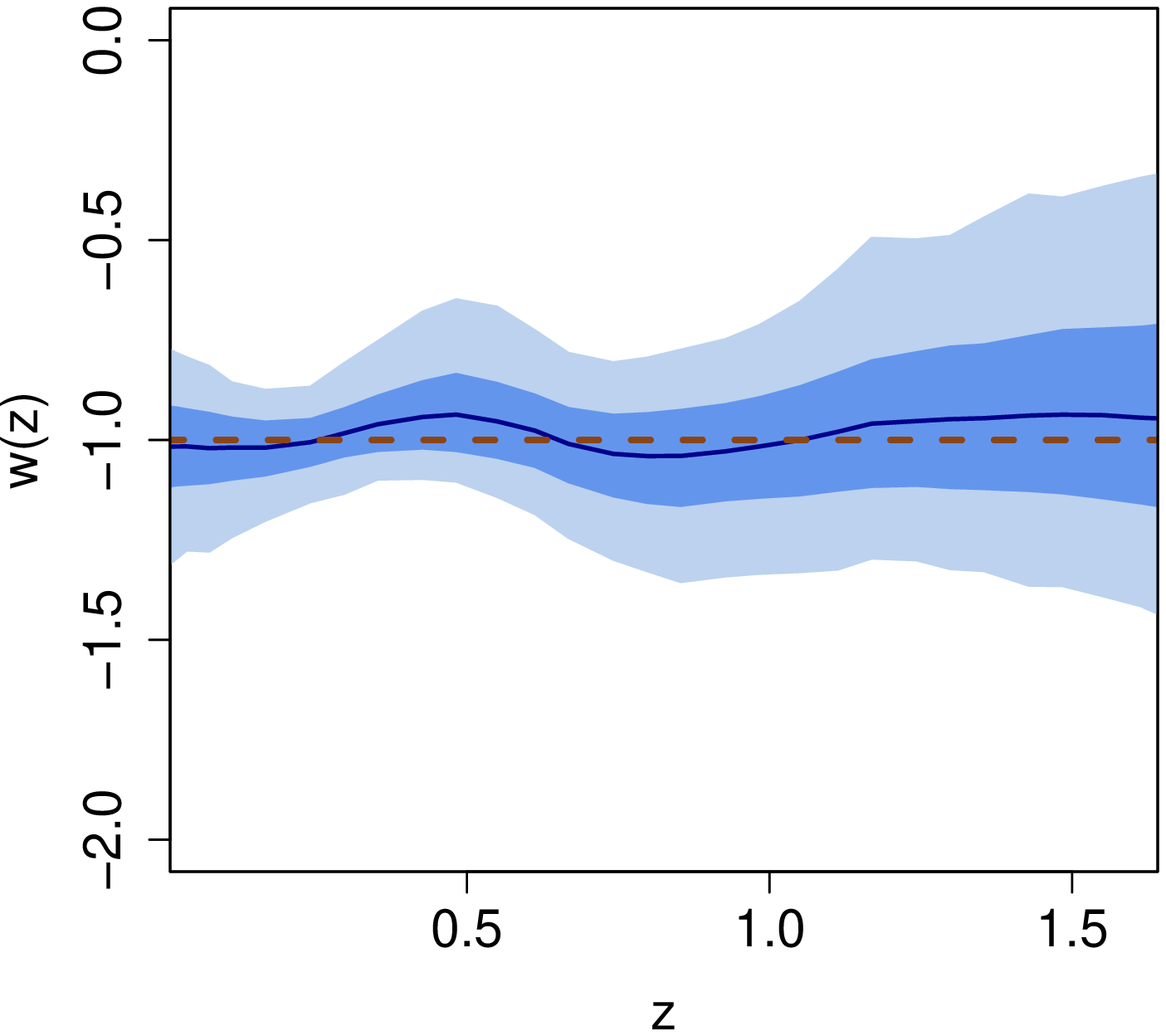}} 
  
\vspace{-.5cm} 

  \centerline{
   \includegraphics[width=1.9in]{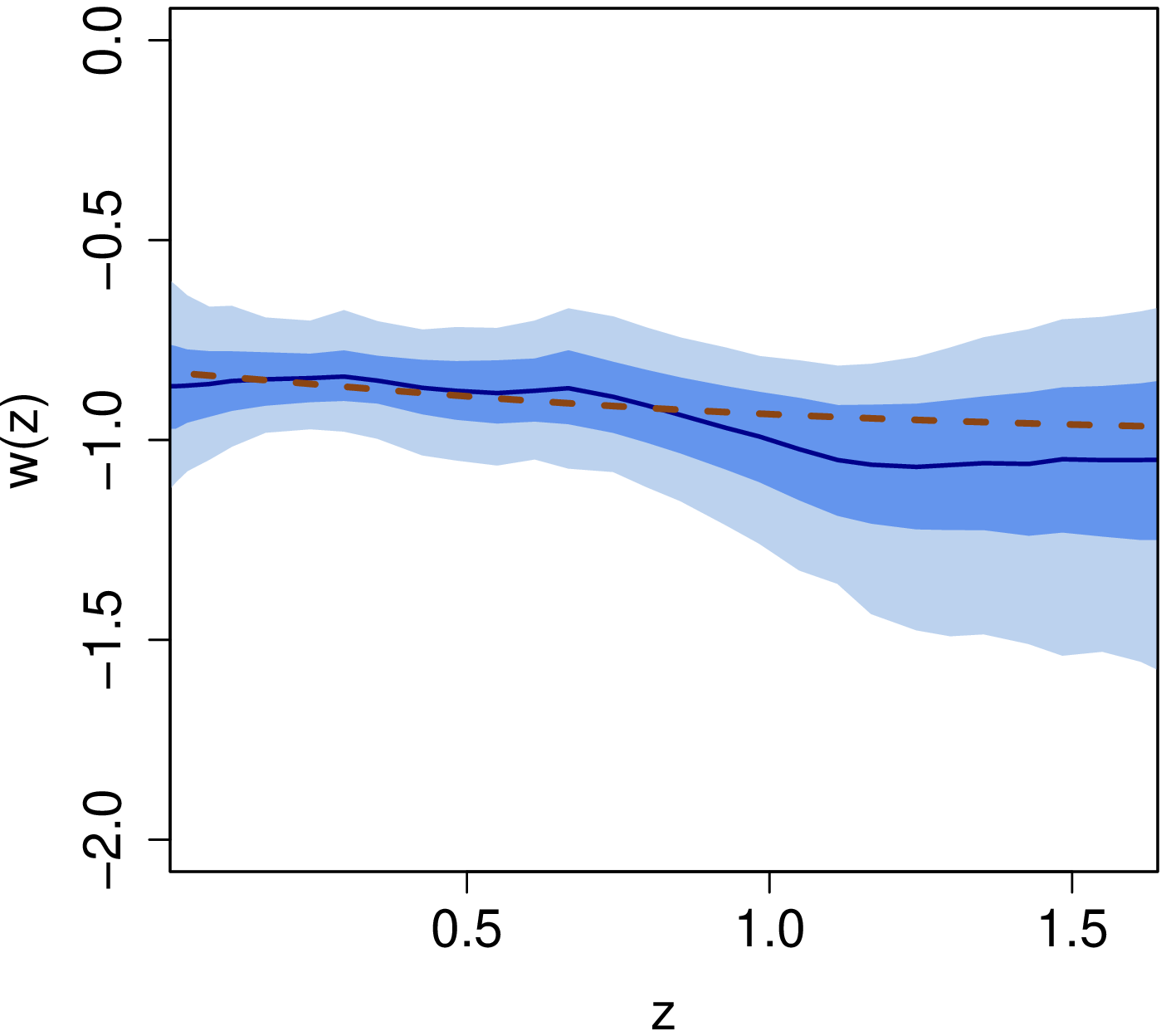}
  \hspace{-0.3cm}\includegraphics[width=1.9in]{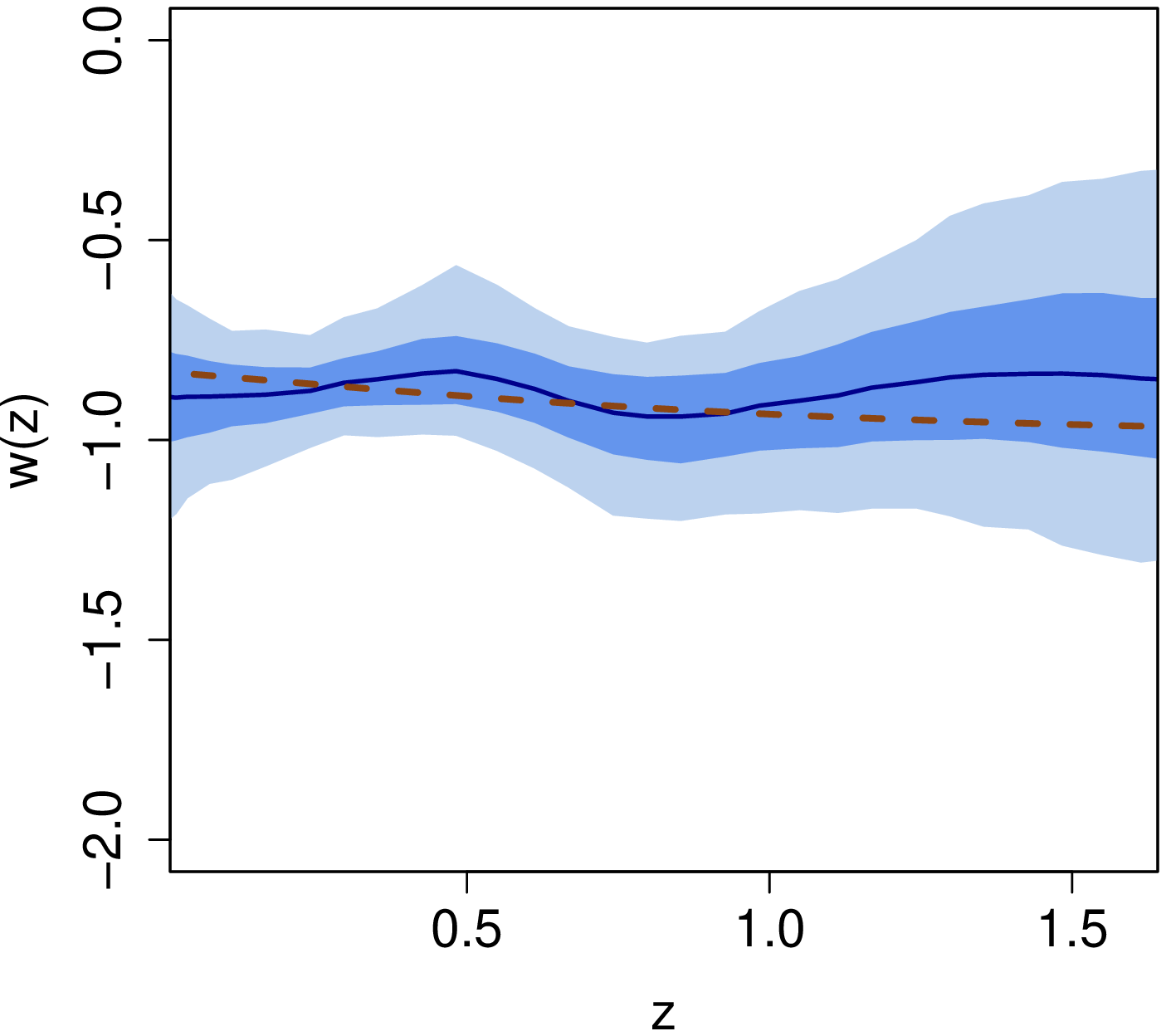}} 
  
  \vspace{-.5cm} 

  \centerline{
  \includegraphics[width=1.9in]{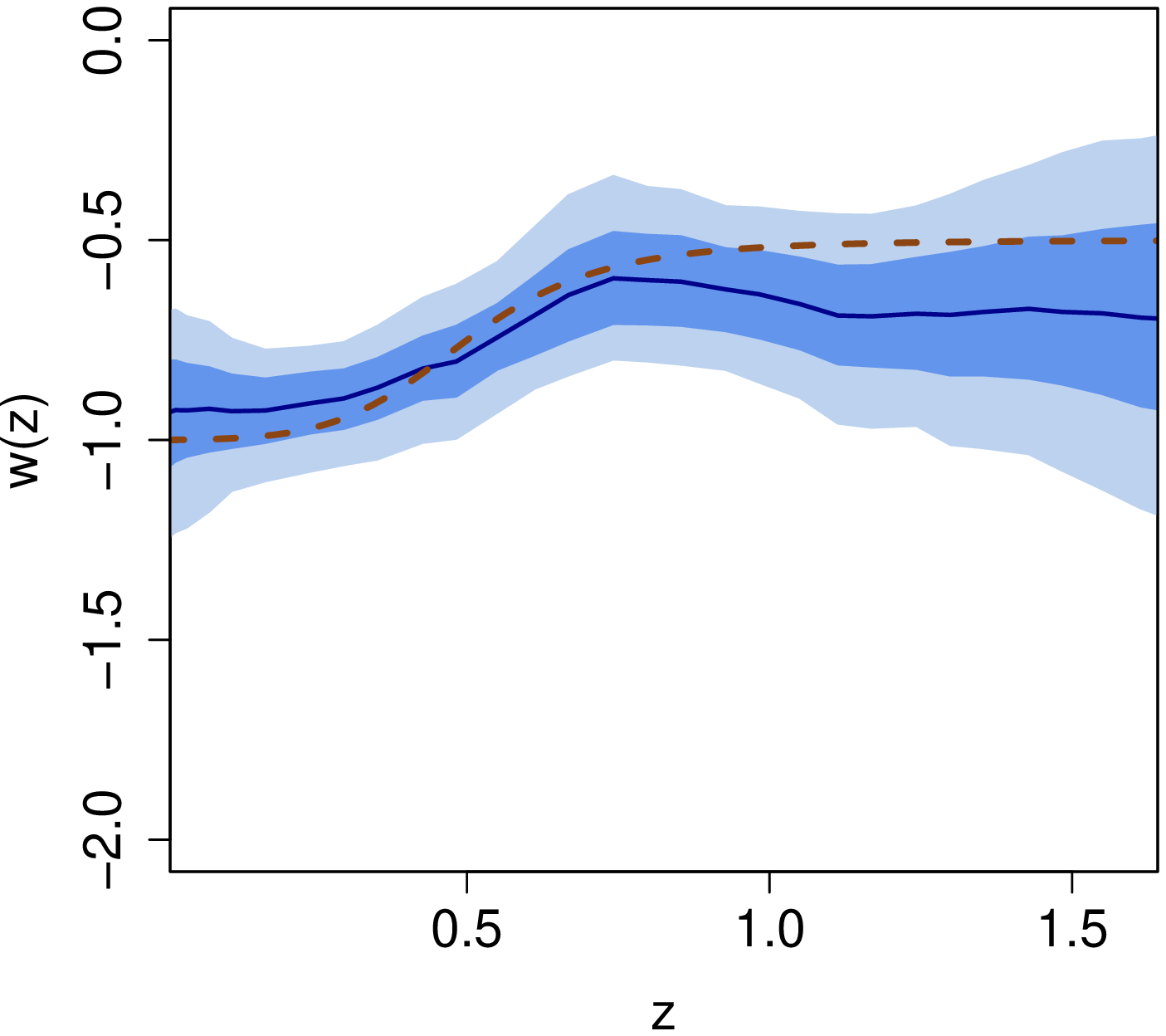}
  \hspace{-0.3cm}\includegraphics[width=1.9in]{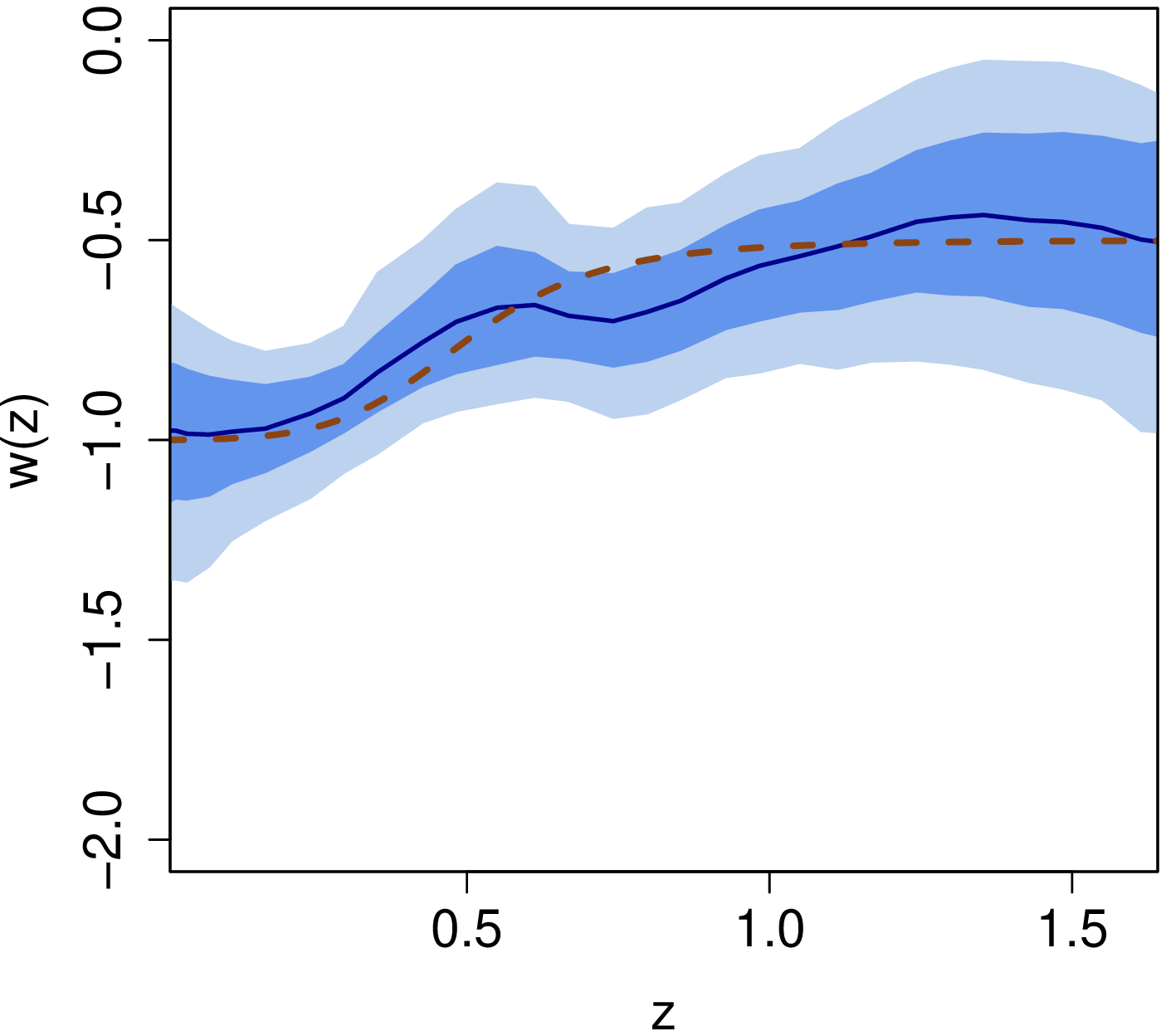} } 
\caption{\label{BAOonly}Left column: reconstruction 
result for $w(z)$ for realization I of the BAO data later used in
combination with the small supernova sample. Right column: results for
realization II later used in combination with the large supernova
sample. Top to bottom: results for Models 1 - 3. The dashed line shows
the underlying theoretical model, the dark blue region shows the 68\%
confidence level, the light blue region the 95\% confidence level, the
dark blue line shows the mean reconstructed history. In all cases, the
reconstruction results capture the ``truth" within the error bands
reliably.}
\end{figure}

\begin{figure*}[t]	
\centerline{
  \includegraphics[width=2.1in]{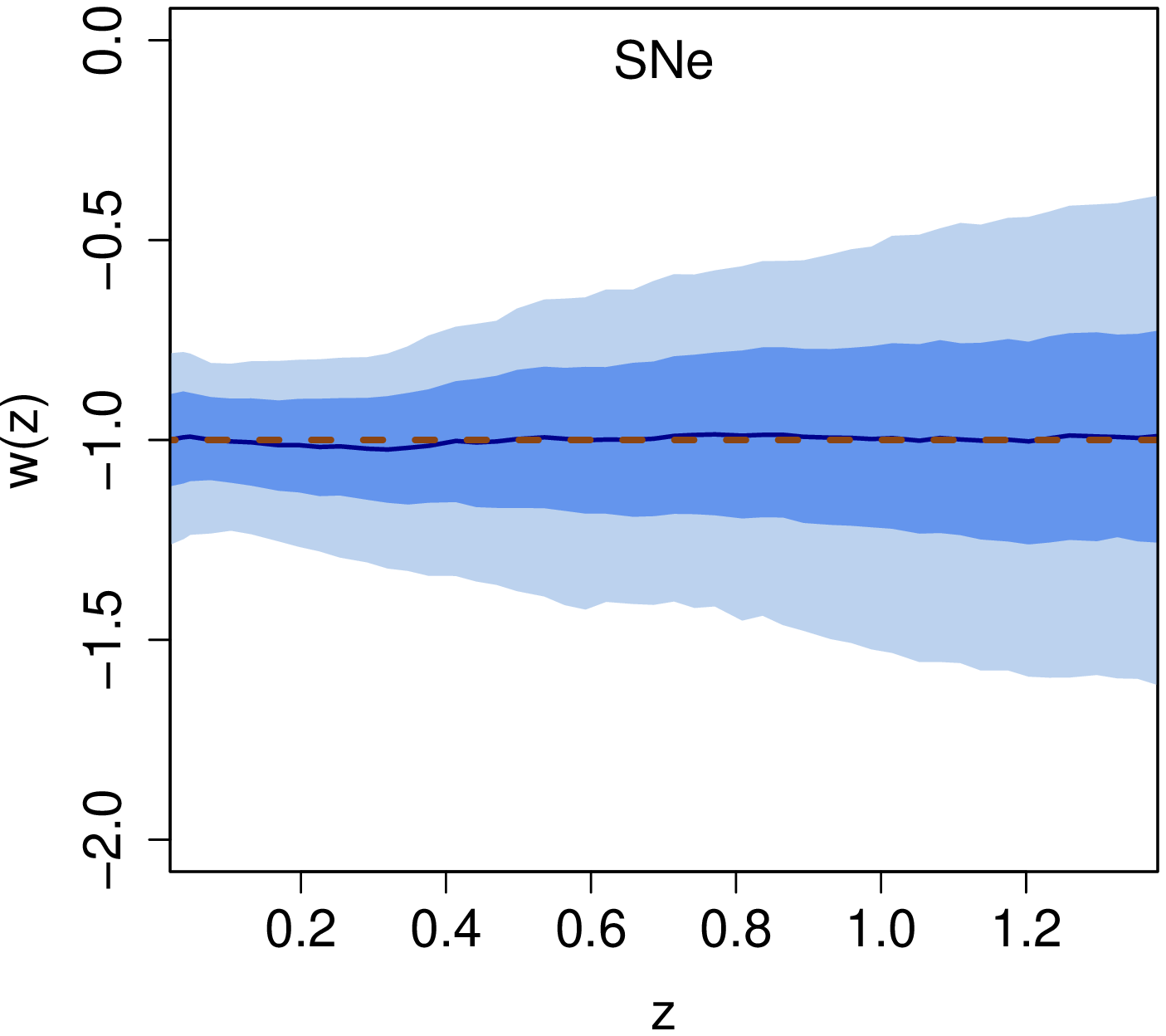}
  \hspace{-1.2cm}\includegraphics[width=2.1in]{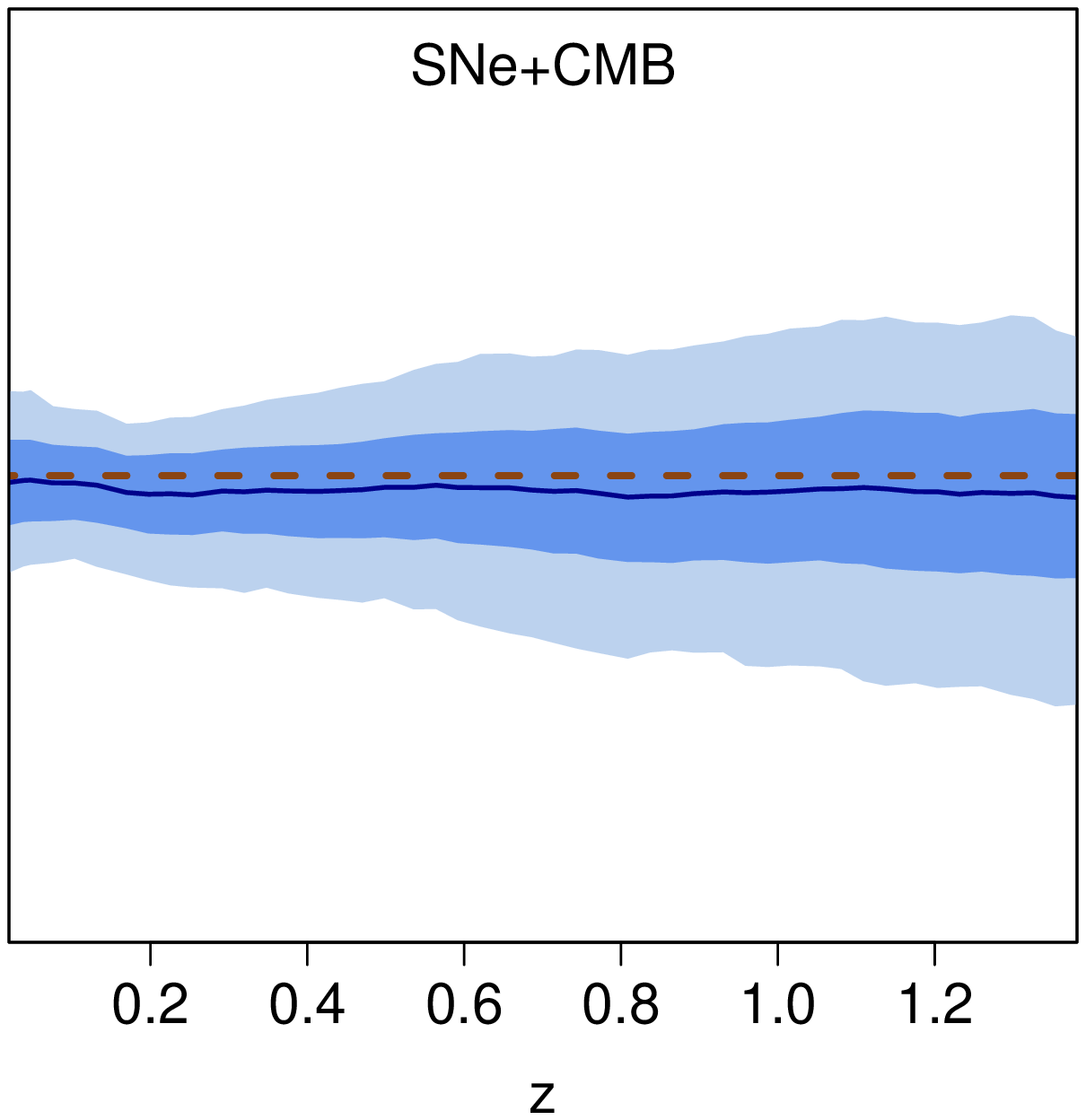}
  \hspace{-1.2cm}\includegraphics[width=2.1in]{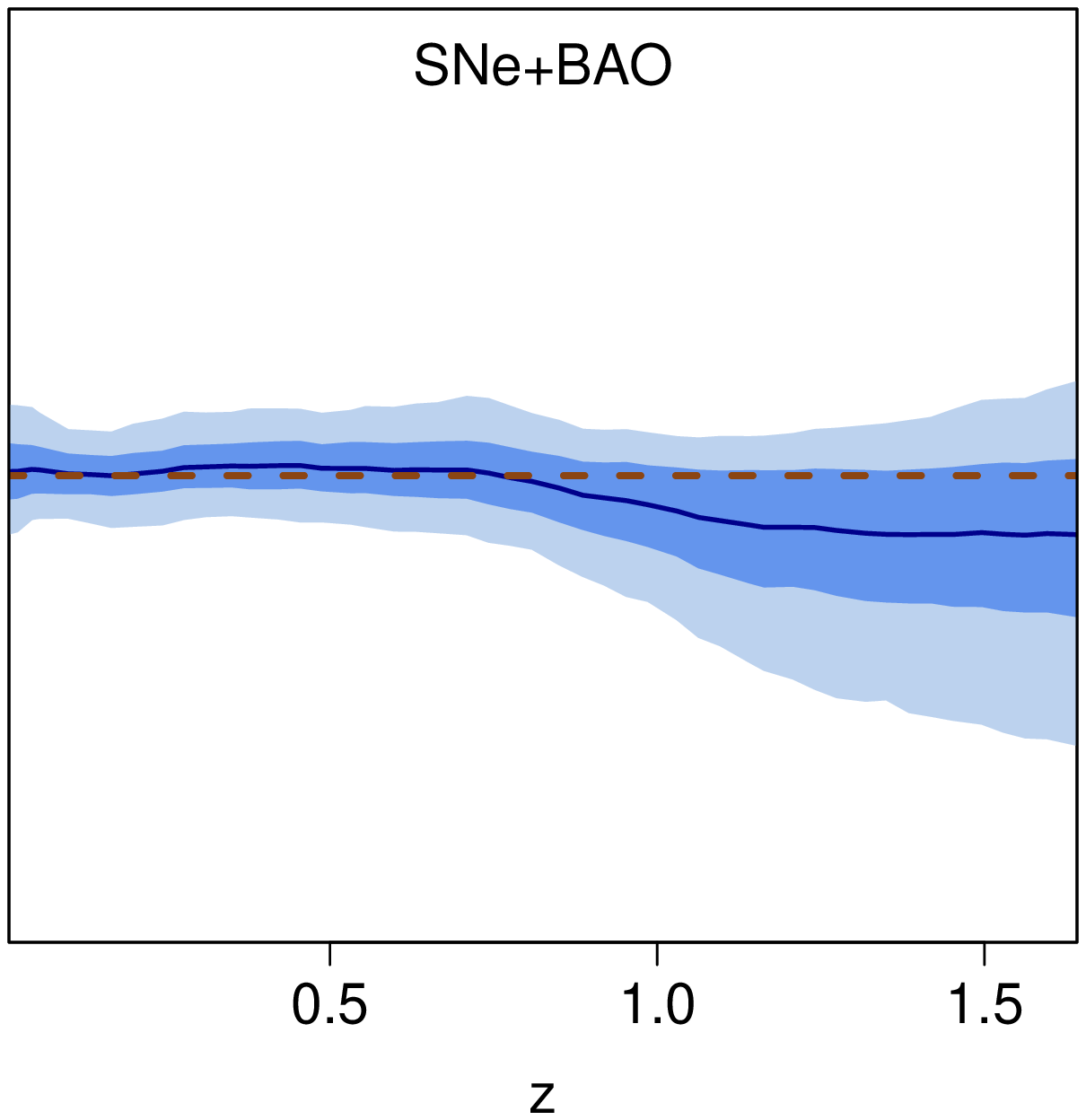}
  \hspace{-1.2cm}\includegraphics[width=2.1in]{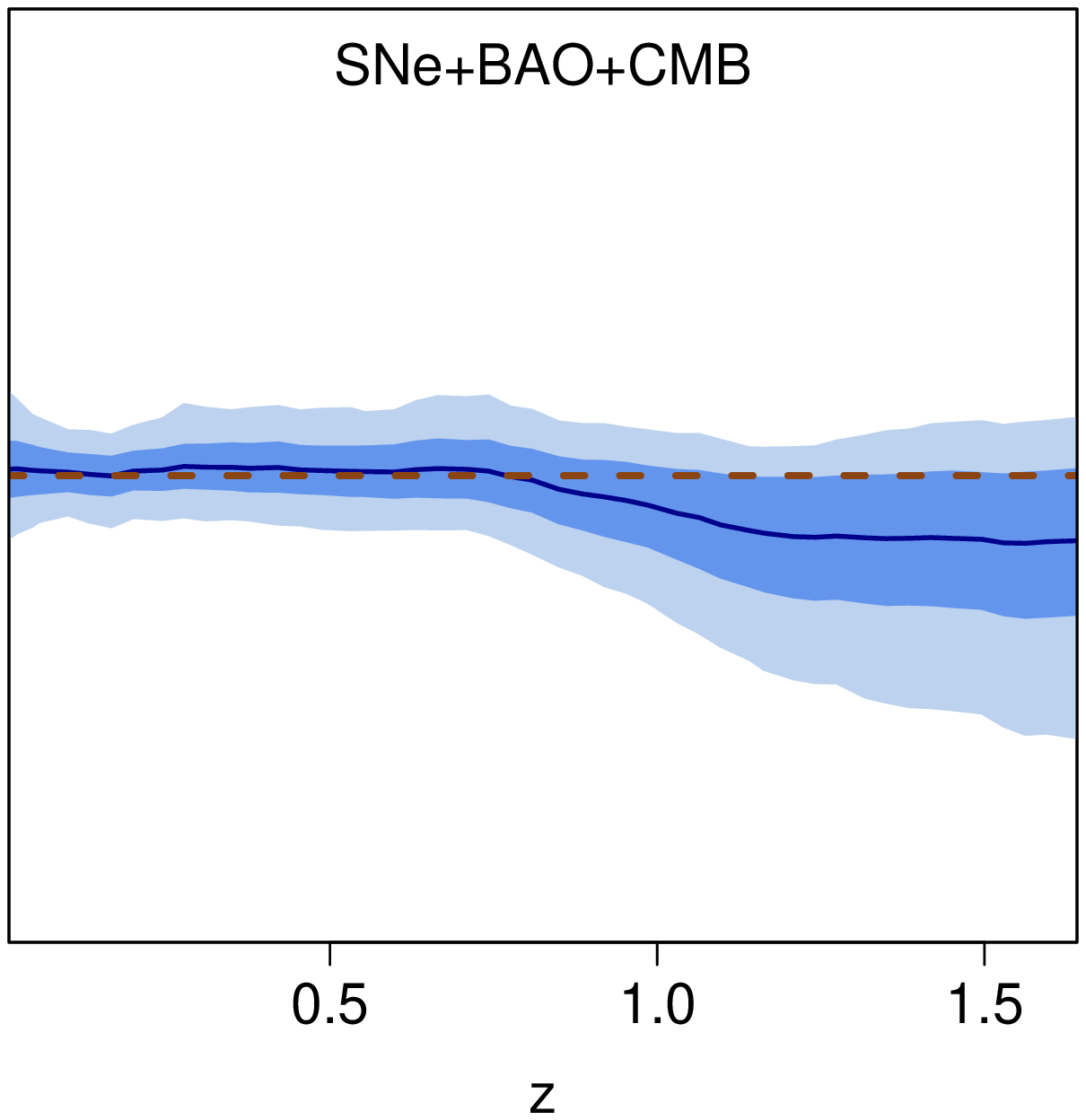}}

  \vspace{-0.53cm} 

  \centerline{ \includegraphics[width=2.1in]{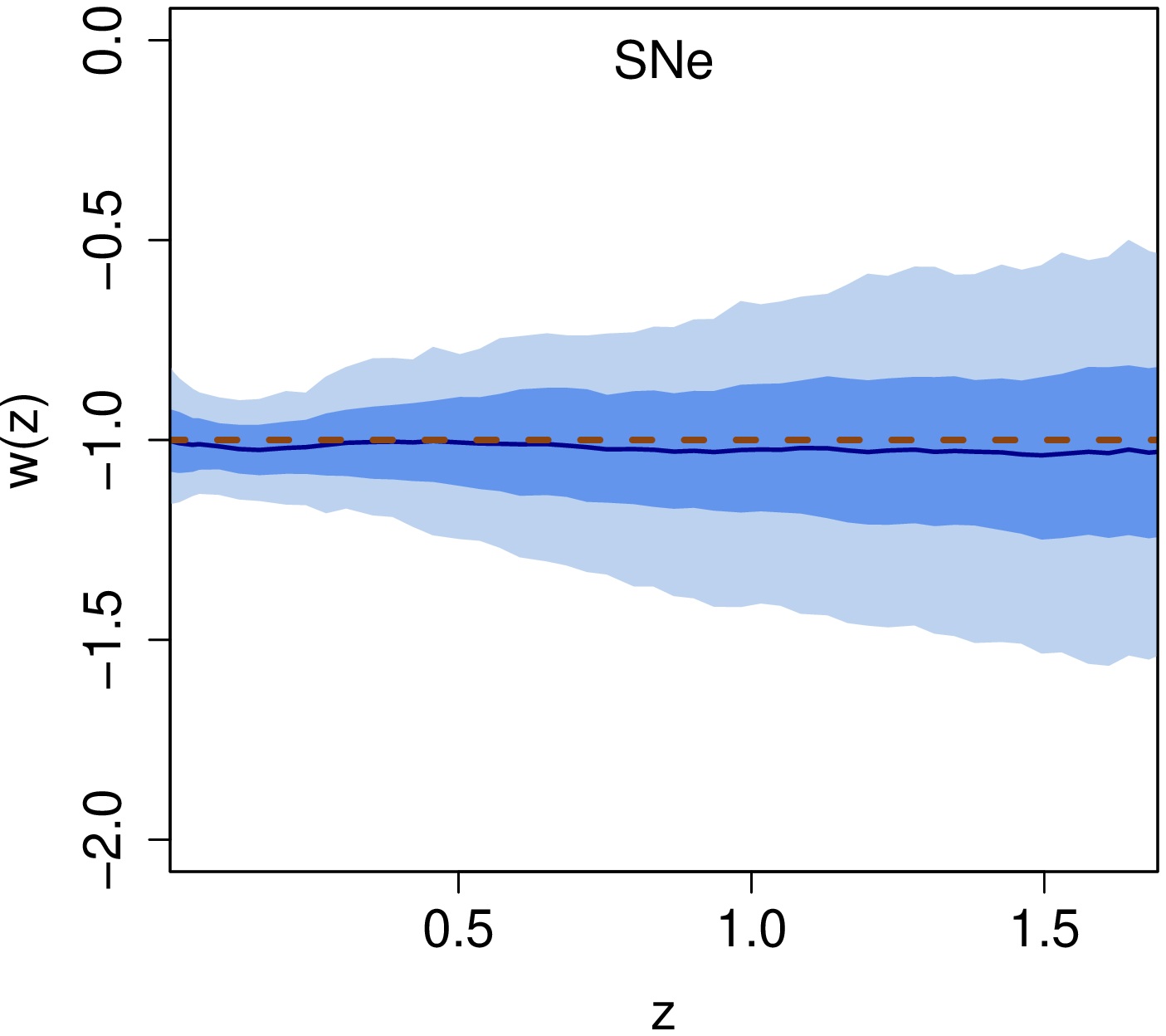}
  \hspace{-1.2cm}\includegraphics[width=2.1in]{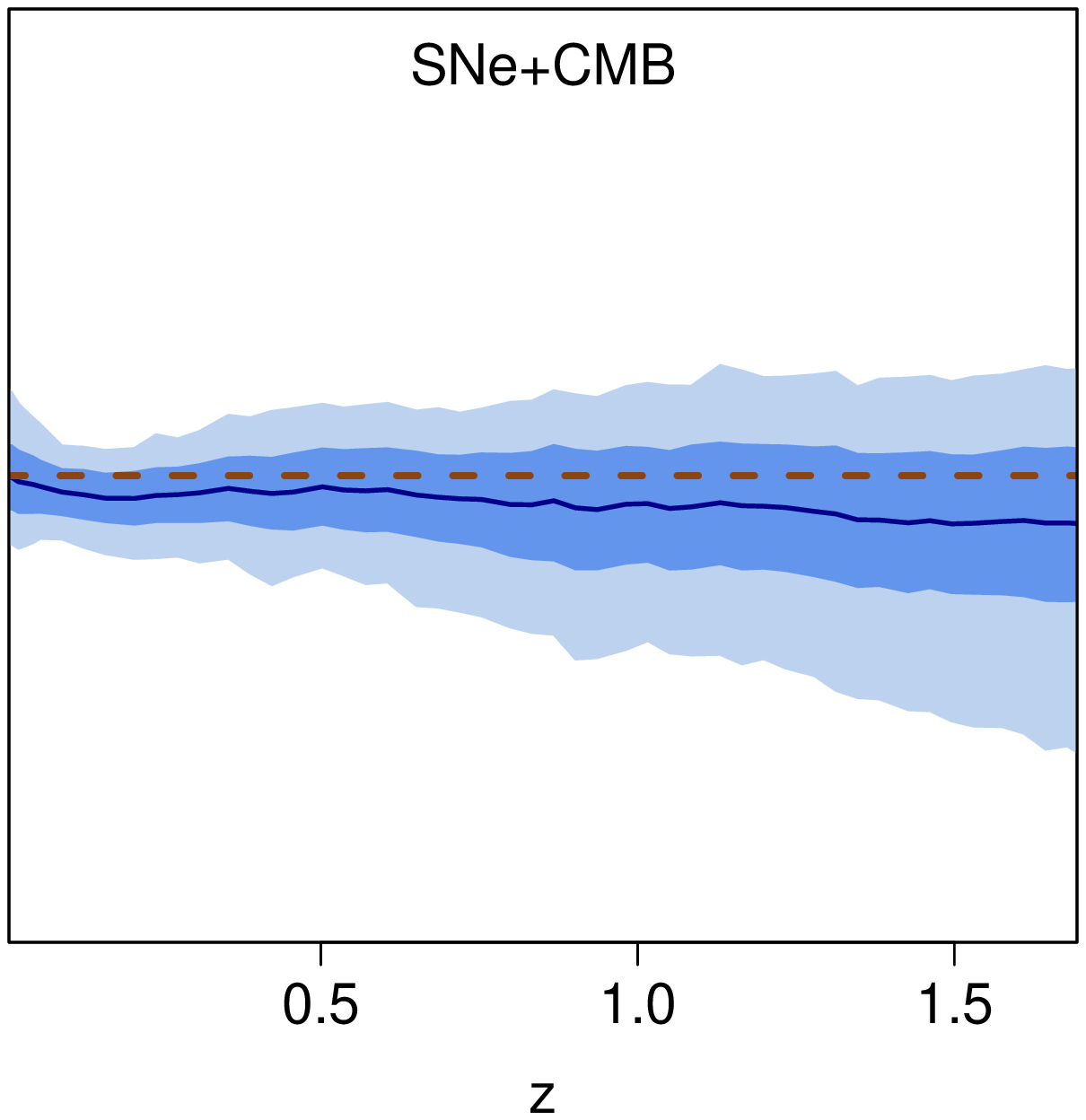}
  \hspace{-1.2cm}\includegraphics[width=2.1in]{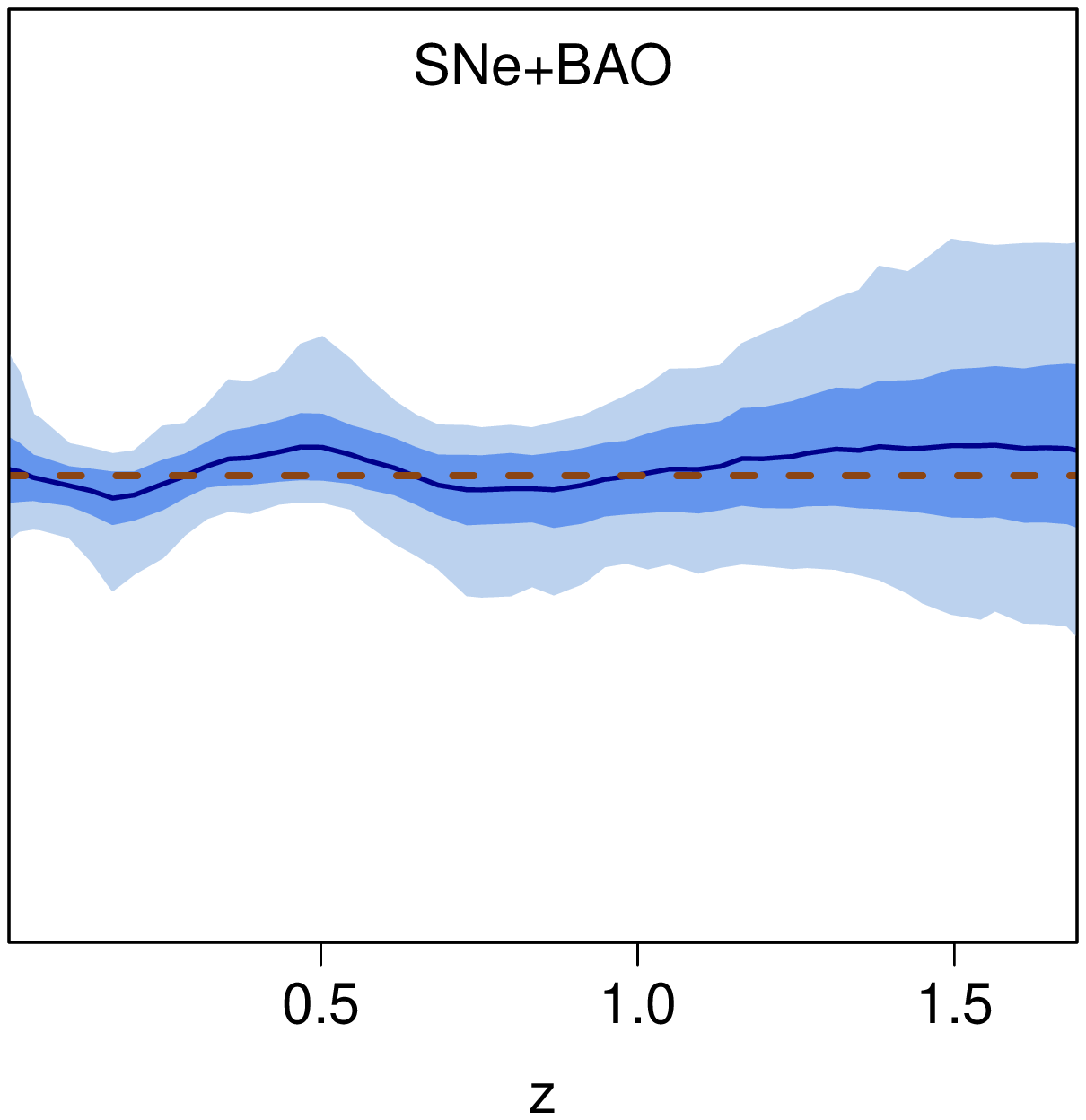}
  \hspace{-1.2cm}\includegraphics[width=2.1in]{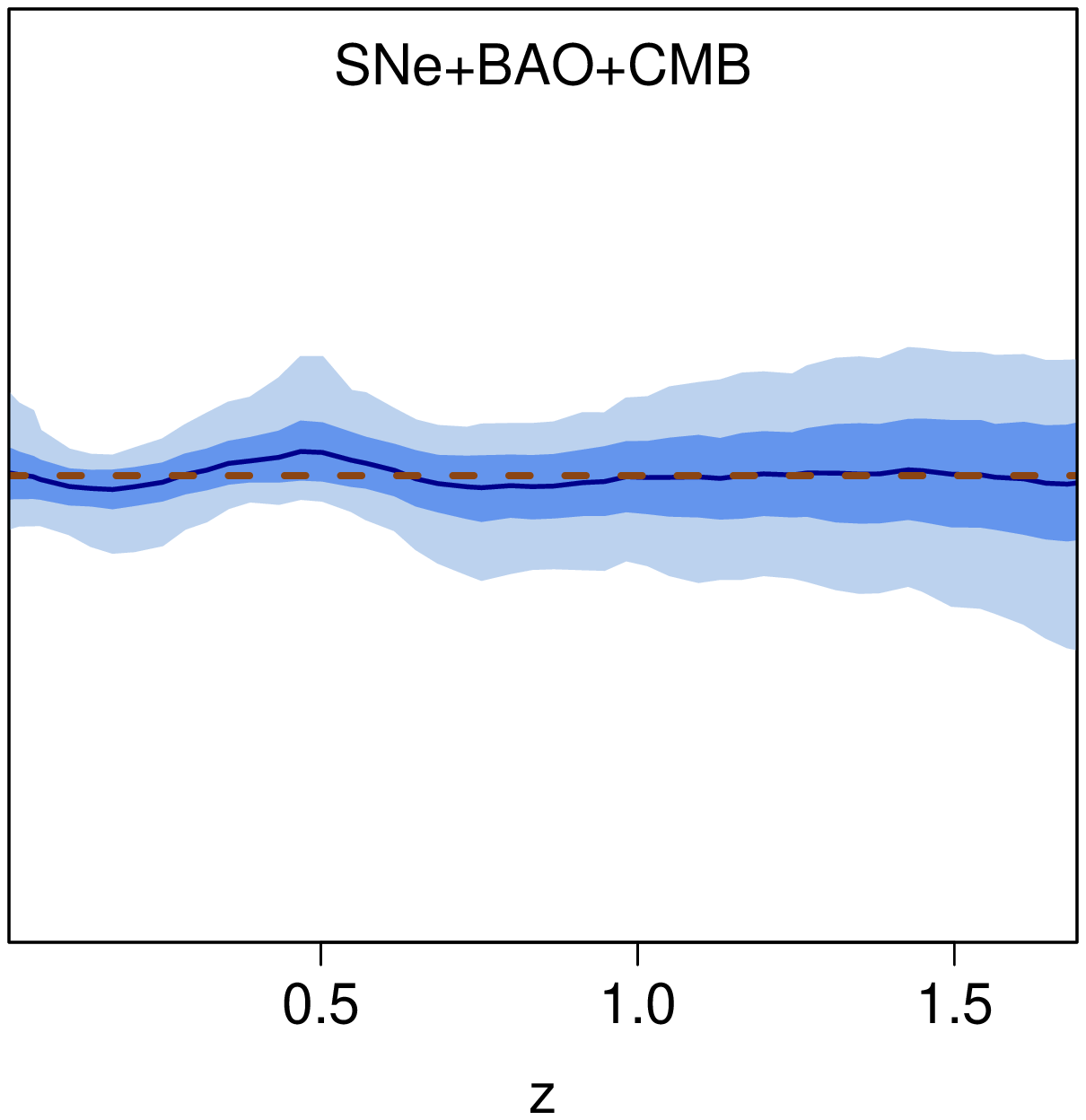}}
  \caption{\label{simdata1}Reconstruction results for the model with
  $w=-1$. From left to right different probes are considered, SNe,
  SNe+CMB, SNe+BAO, and a combination of all three measurements. The
  red dashed line shows the truth, the blue solid line the mean result
  for the reconstruction. The blue shaded region shows the 68\%
  confidence level, the light blue shaded region the 95\% confidence
  level. The upper row shows the results for the small supernova data
  set (557 supernovae with $\tau_i=0.15$ out to $z=1.4$) while the
  lower row shows potential space-based supernova measurements (2298
  supernovae with $\tau_i=0.13$ out to $z=1.7$). The CMB data point is
  the same in all cases where it is included, the BAO data are of same
  quality but two different realizations out to $z=2$. Note that the
  redshift range varies in the different panels depending on which
  probes are included.  }
\end{figure*}

\begin{figure*}	
\centerline{
  \includegraphics[width=2.1in,angle=0]{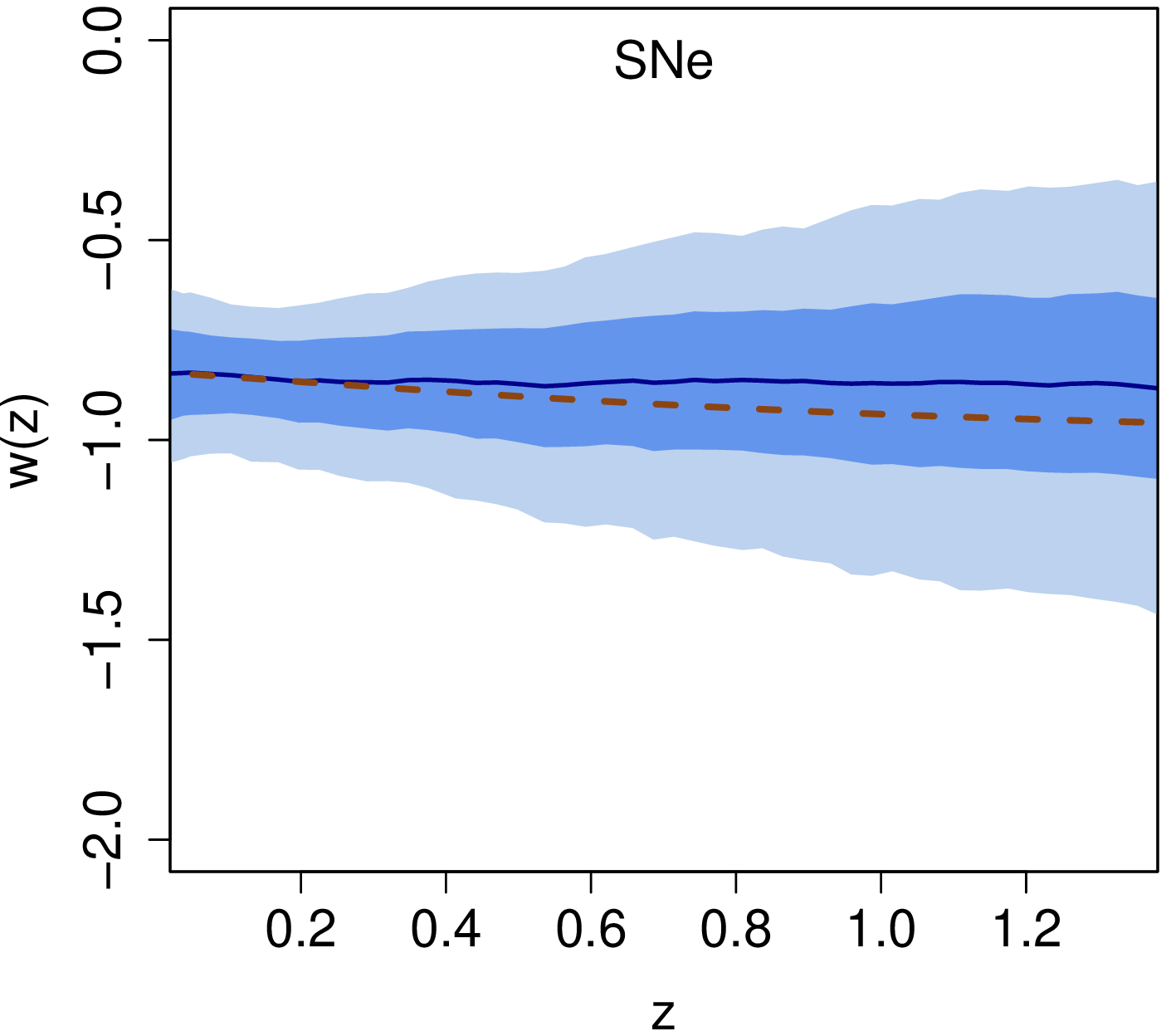}
  \hspace{-1.2cm}\includegraphics[width=2.1in,angle=0]{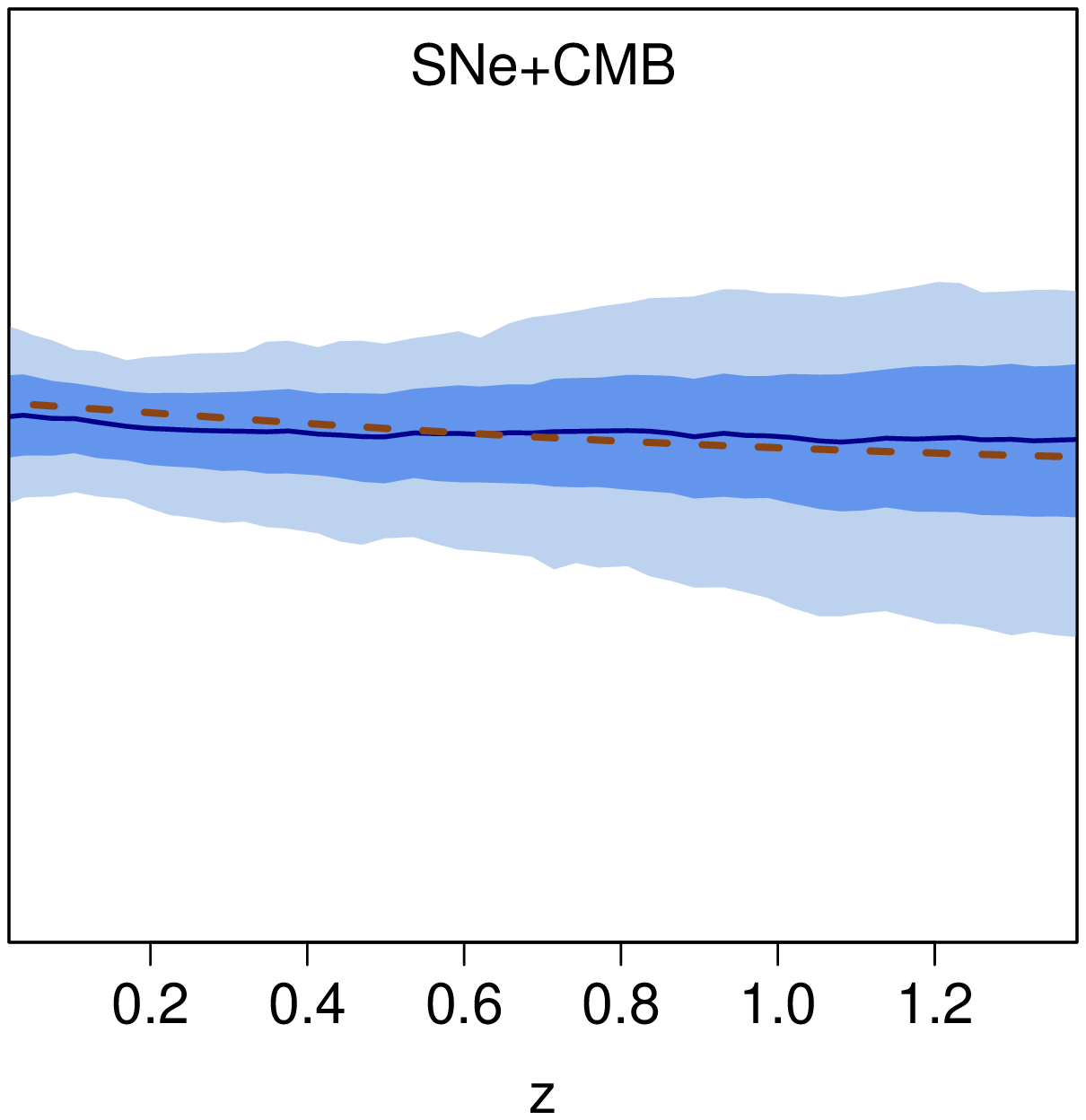}
  \hspace{-1.2cm}\includegraphics[width=2.1in,angle=0]{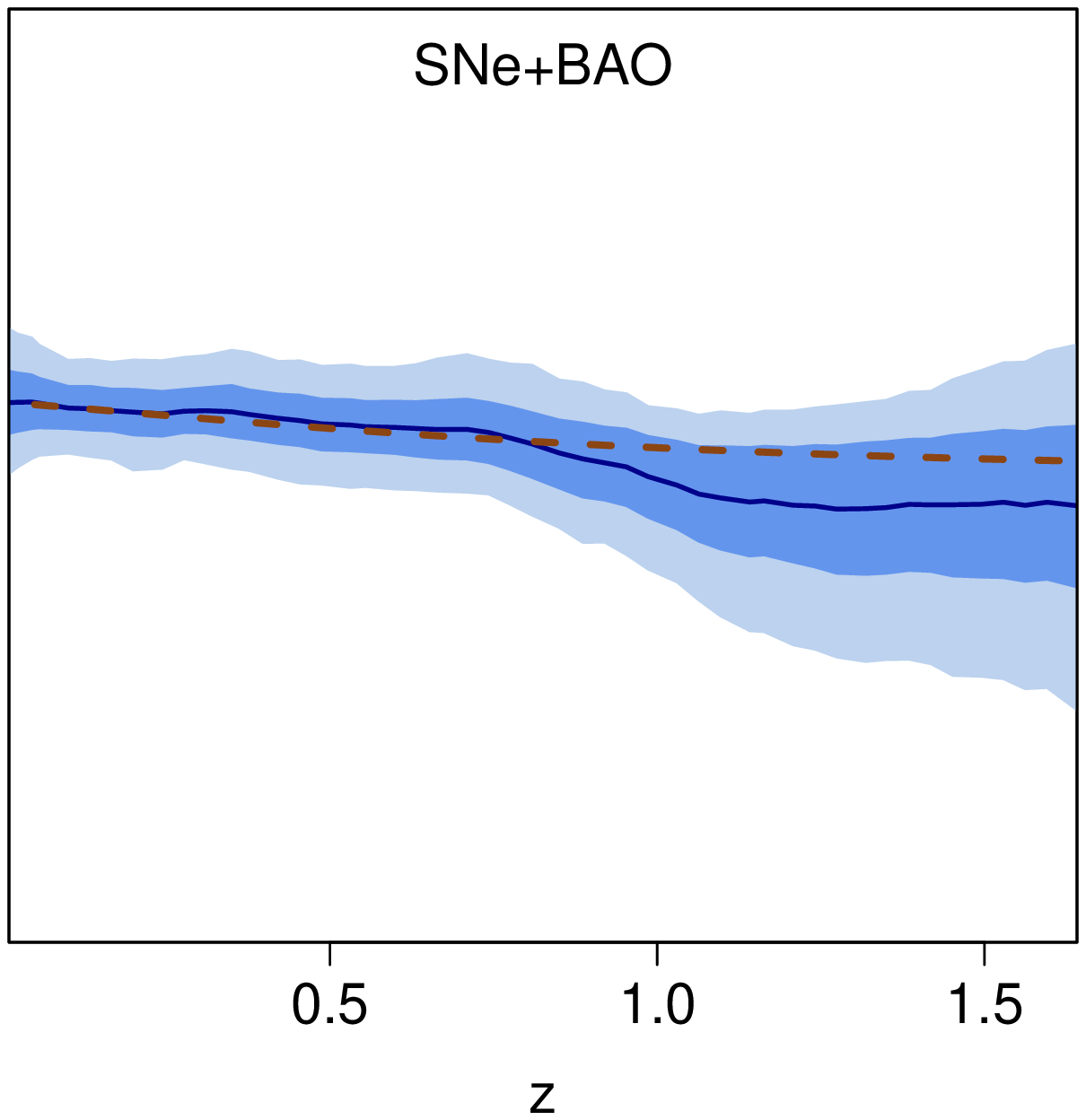}
  \hspace{-1.2cm}\includegraphics[width=2.1in,angle=0]{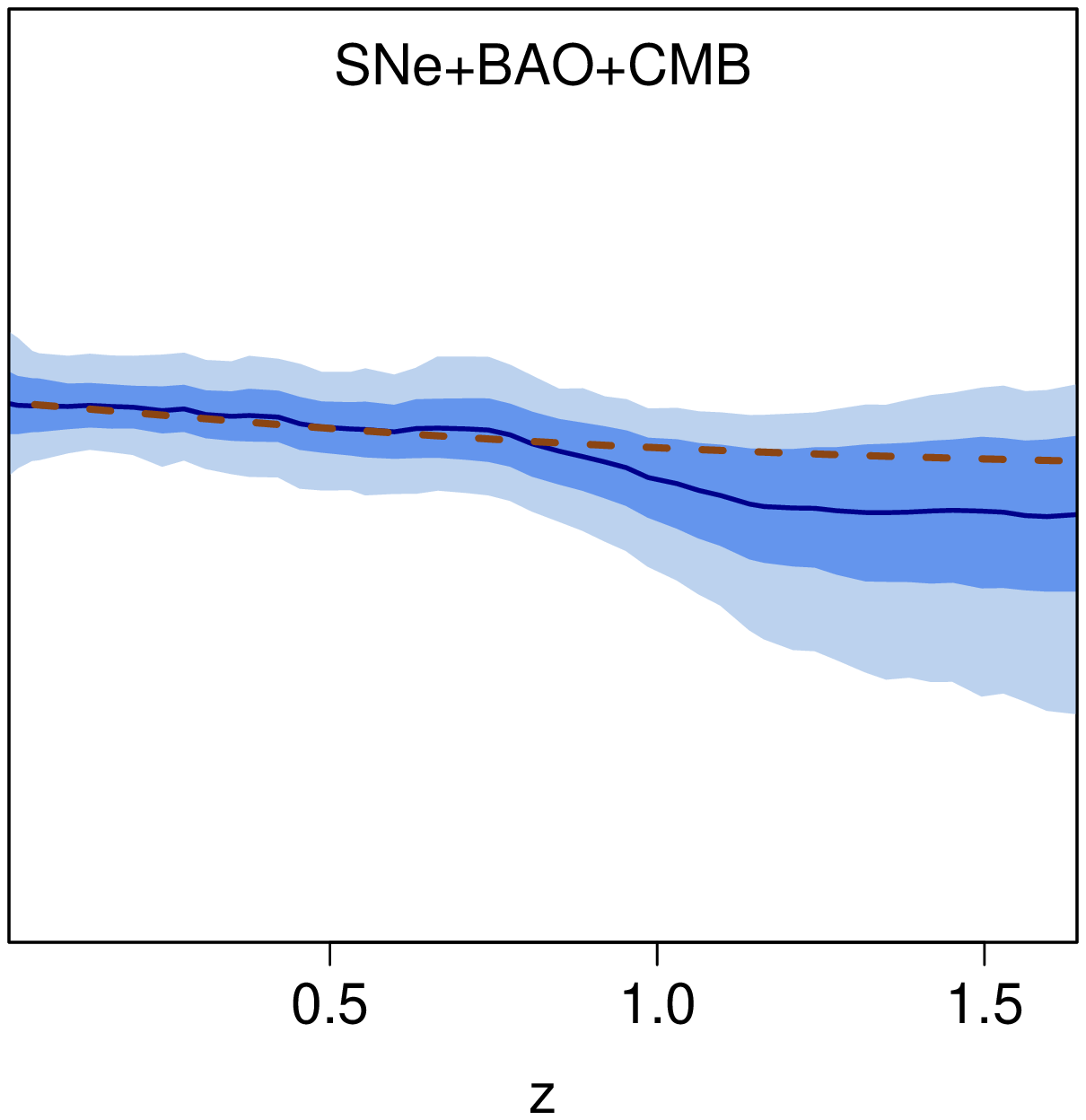}} 

  \vspace{-0.53cm} 

  \centerline{
   \includegraphics[width=2.1in,angle=0]{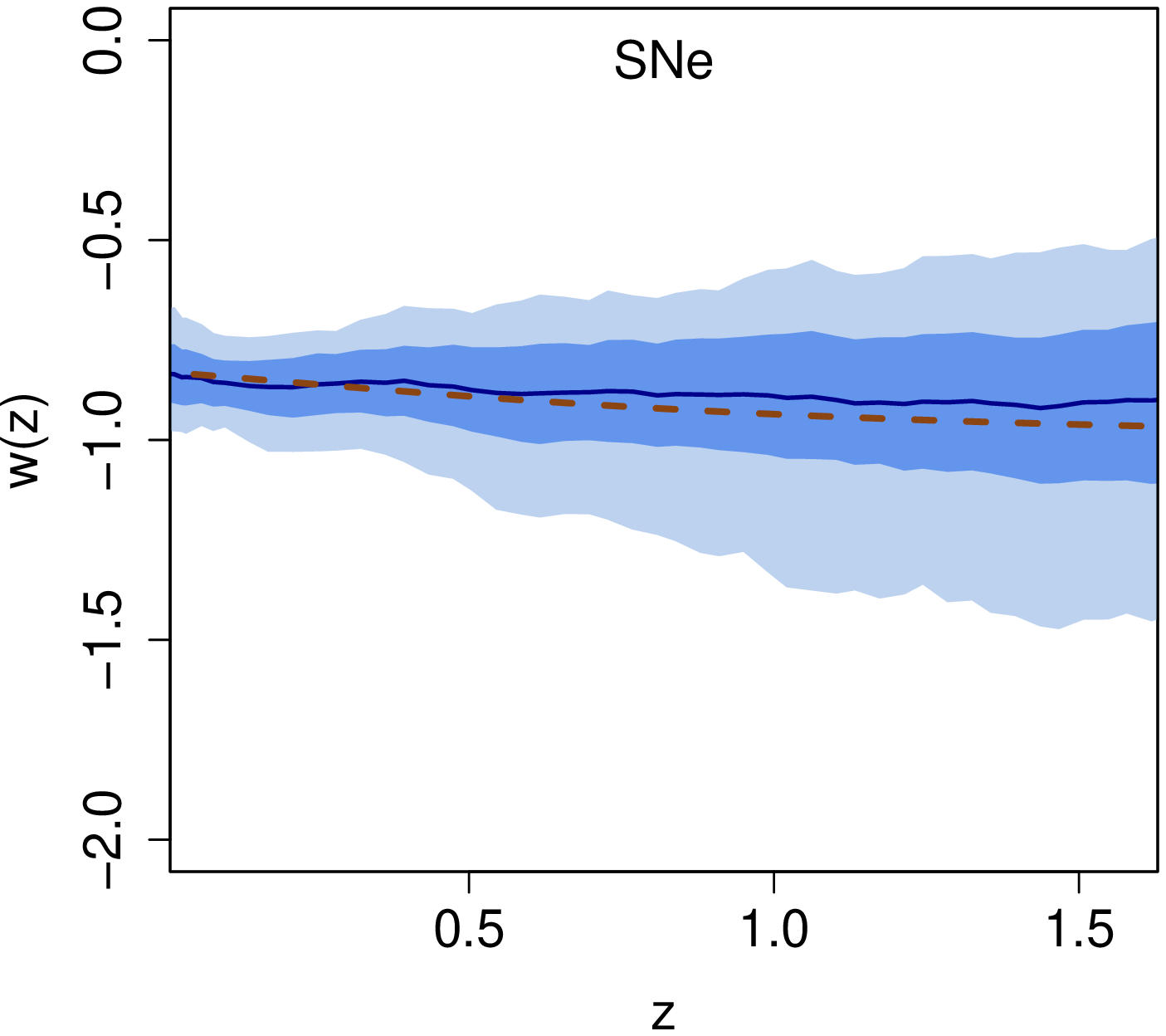}
  \hspace{-1.2cm}\includegraphics[width=2.1in,angle=0]{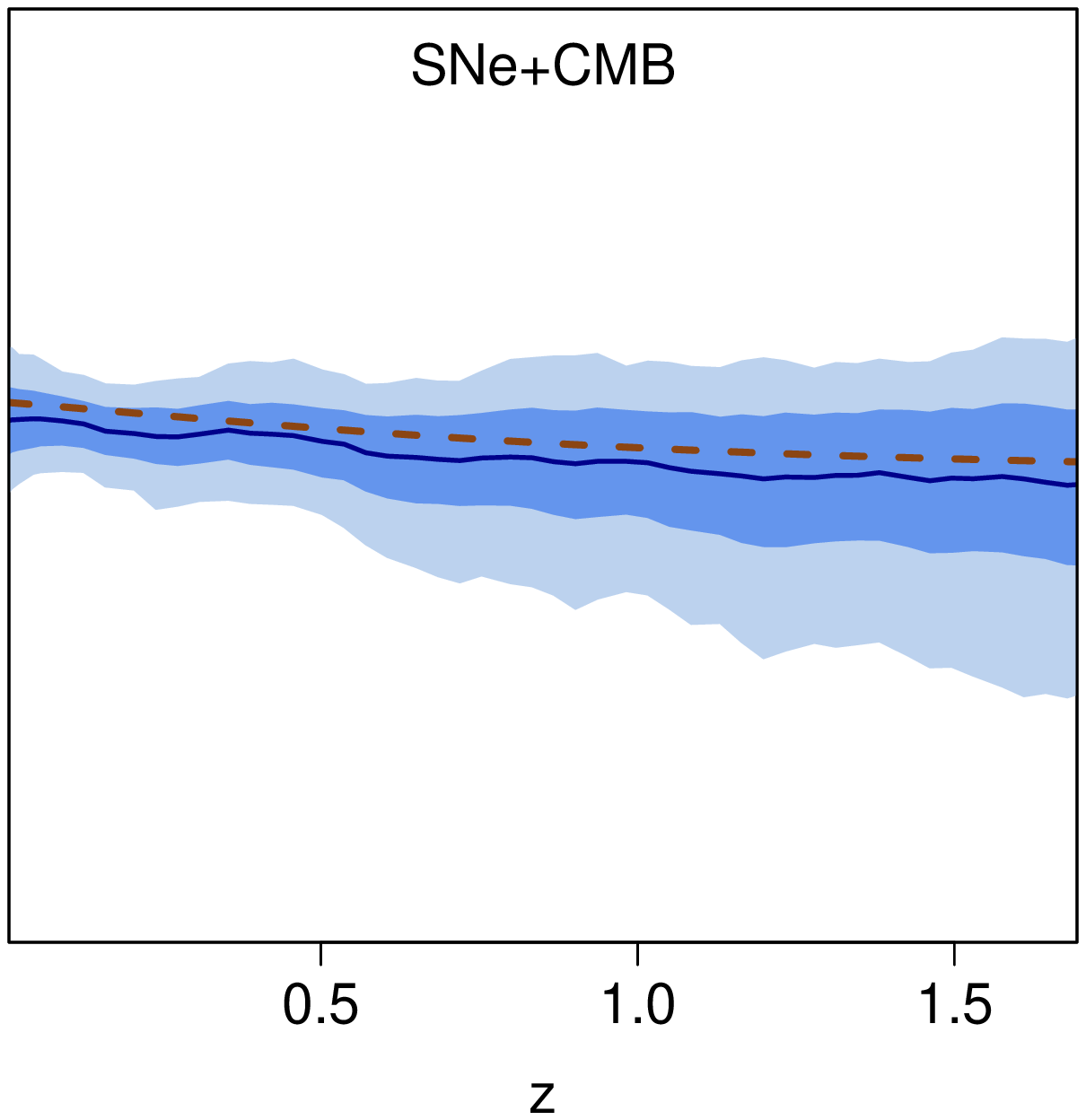}
  \hspace{-1.2cm}\includegraphics[width=2.1in,angle=0]{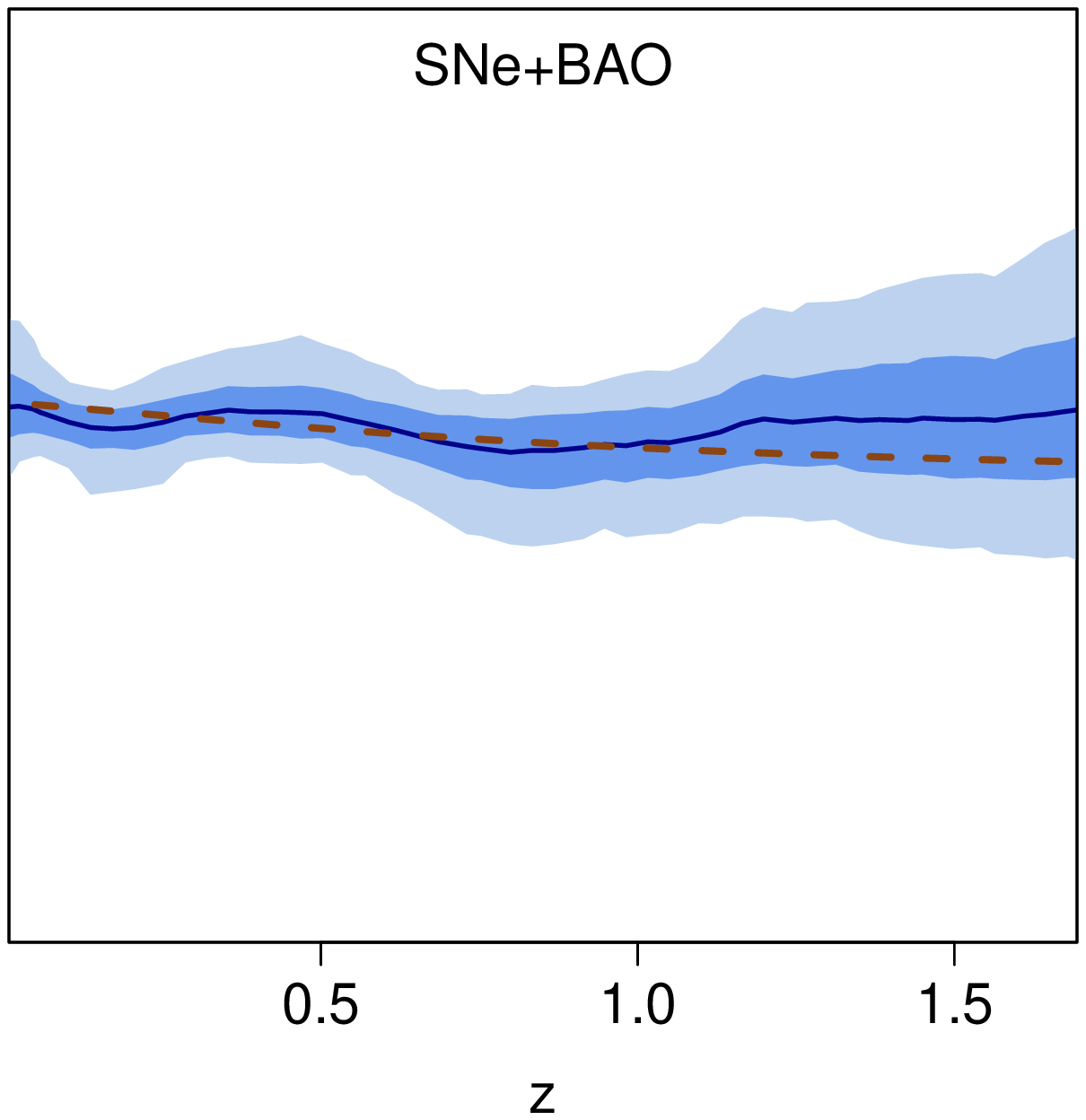}
  \hspace{-1.2cm}\includegraphics[width=2.1in,angle=0]{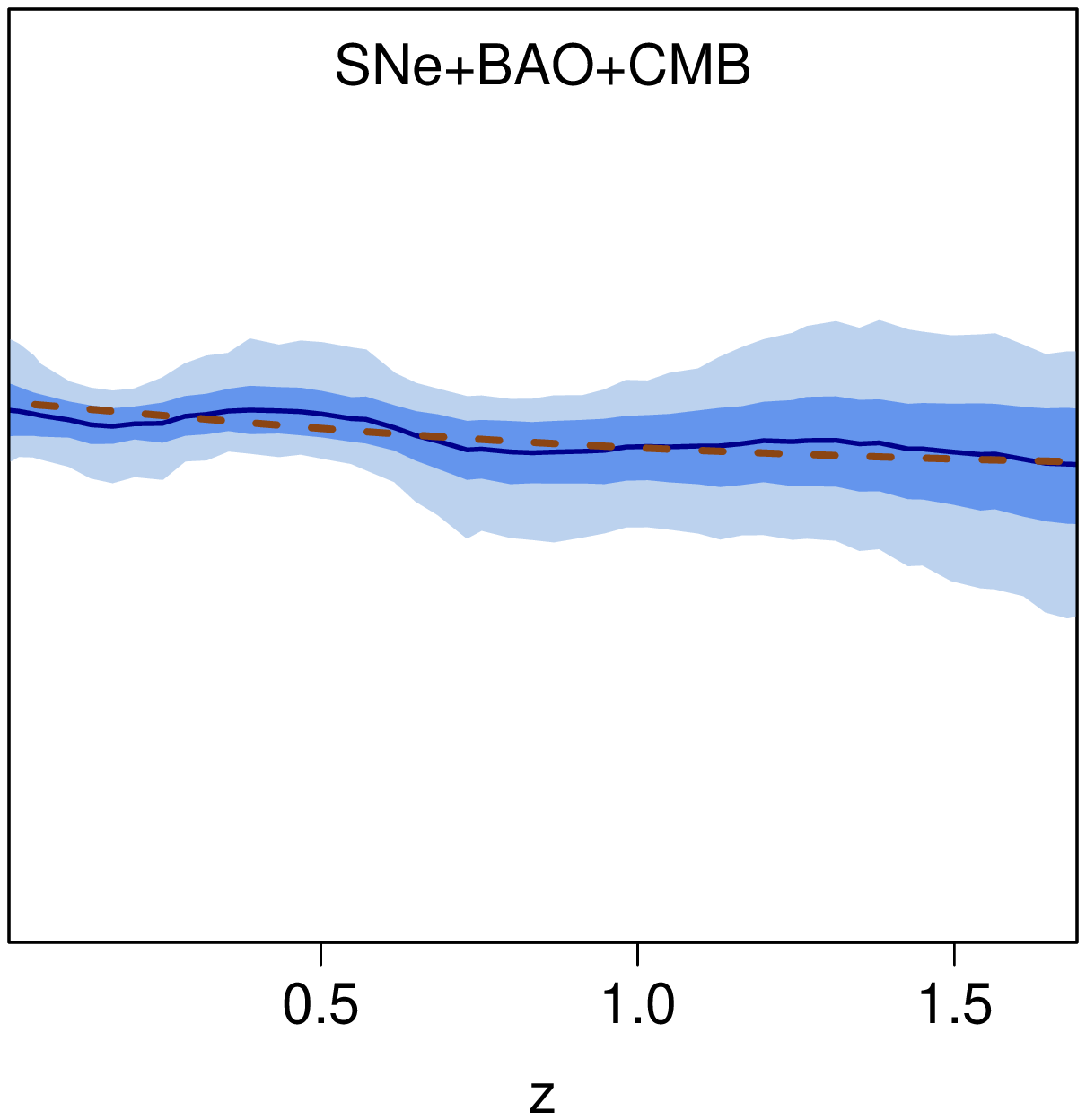}} 
  \caption{\label{simdata2}Same as in Fig.~\ref{simdata1} but for Model 2, the quintessence model.  }
\end{figure*}

Figure~\ref{bao} shows two realizations for a $\Lambda$CDM model
(Model 1) for the angular distance diameter $D_A(z)/r_s$ in the left
column and for the Hubble parameter $H(z)r_s$ in the right column. In
addition, we show the exact predictions for Model 2 and 3. We use two
realizations for the BAO data to demonstrate the dependence of the
reconstruction as a function of realization. Because observations
represent only one realization, this imposes an an irreducible
limitation on the reconstruction program, whether non-parametric or
not. We will return to this issue in future work.

\subsection{Results}
\subsubsection{Prelude}

Before we present our results for the combined analysis of different
cosmological probes we show the constraints we obtain from the
simulated BAO data alone on $w(z)$. The results are already remarkably
good. We choose a flat prior for $\Omega_m$ for this analysis.

Figure~\ref{BAOonly} shows the results for both realizations presented
in Figure~\ref{bao}, Table~\ref{table550} provides the best fit values
for $\Omega_m$ all three models for the left column (Realization I)
and Table~\ref{table2000} for the right column (Realization II). For
Model 1 (first row) the predictions are slightly low for the first
realization but overall the results are consistent with the input
model., $w=-1$. We verified that this result does not change
considerably if the tighten the prior on $\Omega_m$. Similar trends
can be seen for Model 2 and 3. We will come back to these trends later
in the discussion on the results for combined data sets. The value for
$\Omega_m$ for realization I (Table~\ref{table550}) is slightly high
in all cases -- adding CMB measurements decreases the error on
$\Omega_m$ but in fact shifts the best fit values even higher. The
second realization leads to values for $\Omega_m$ very close to the
input value for BAO measurements only, the CMB point again shifts it
up slightly. The reconstruction from the BAO data only works
remarkably well -- in all cases the underlying model is captured
within the error bars reliably.

\subsubsection{Combining Different Data Sets}
\begin{figure*}[t]	
\centerline{
  \includegraphics[width=2.1in,angle=0]{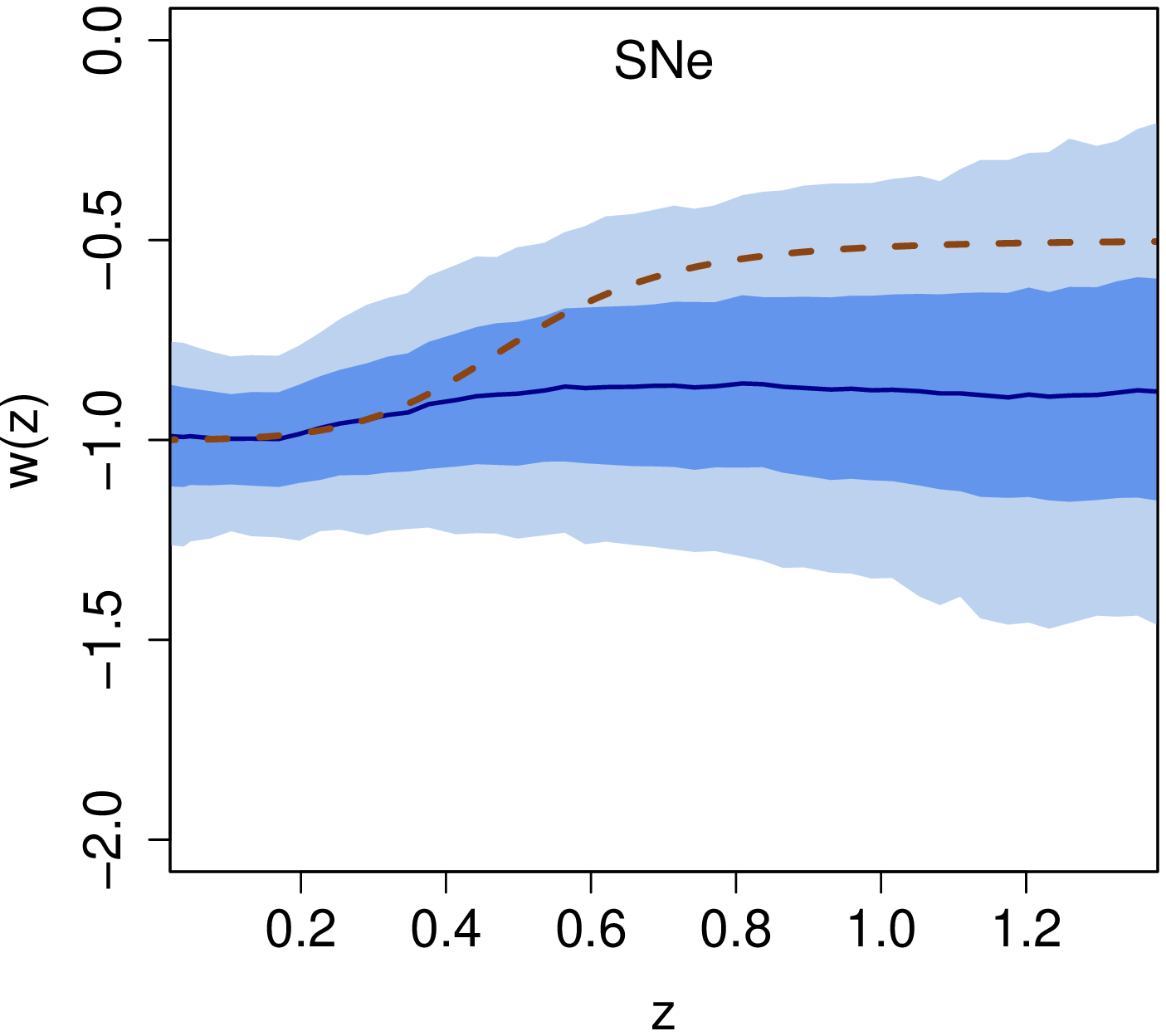}
  \hspace{-1.2cm}\includegraphics[width=2.1in,angle=0]{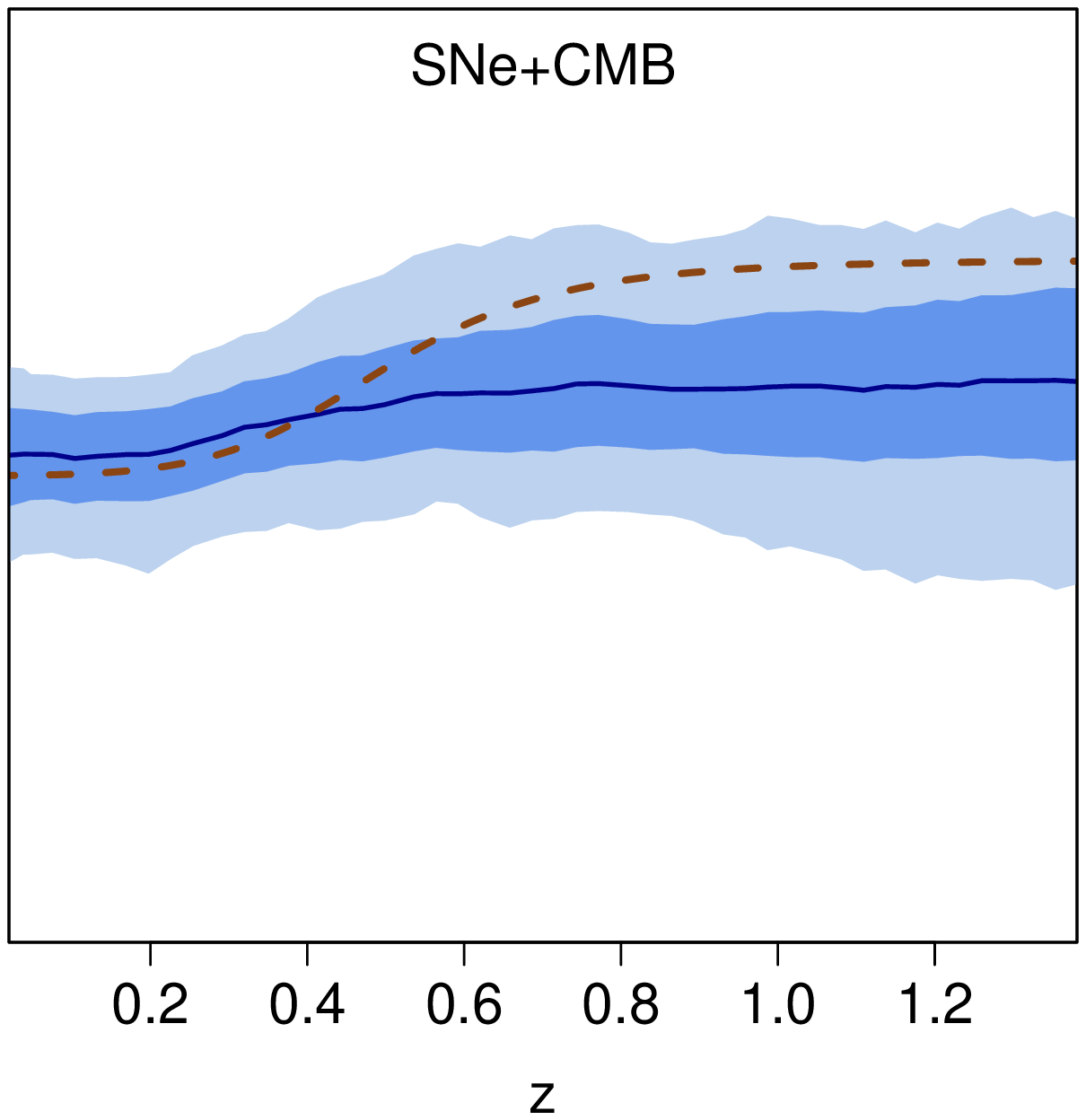}
  \hspace{-1.2cm}\includegraphics[width=2.1in,angle=0]{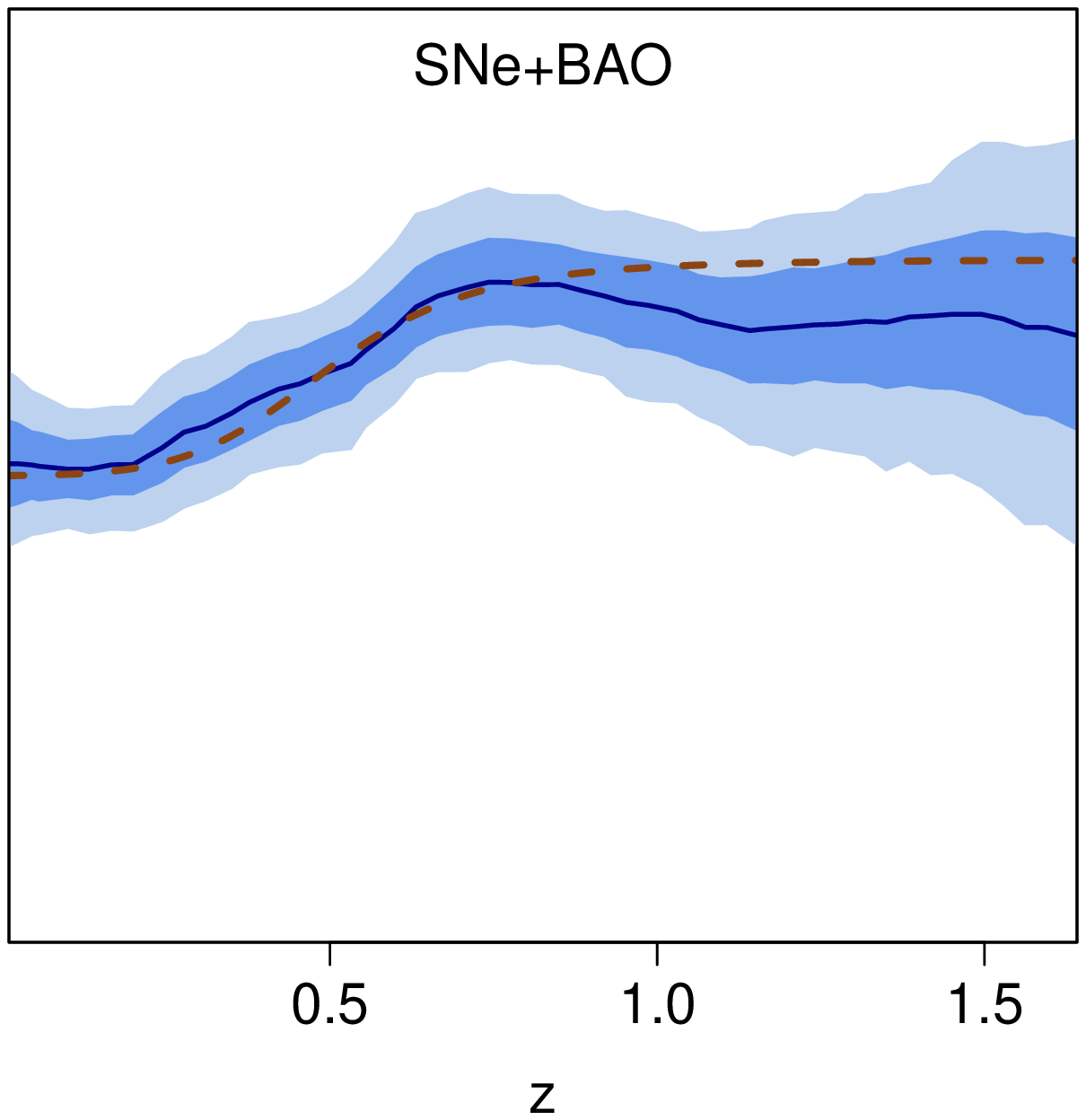}
  \hspace{-1.2cm}\includegraphics[width=2.1in,angle=0]{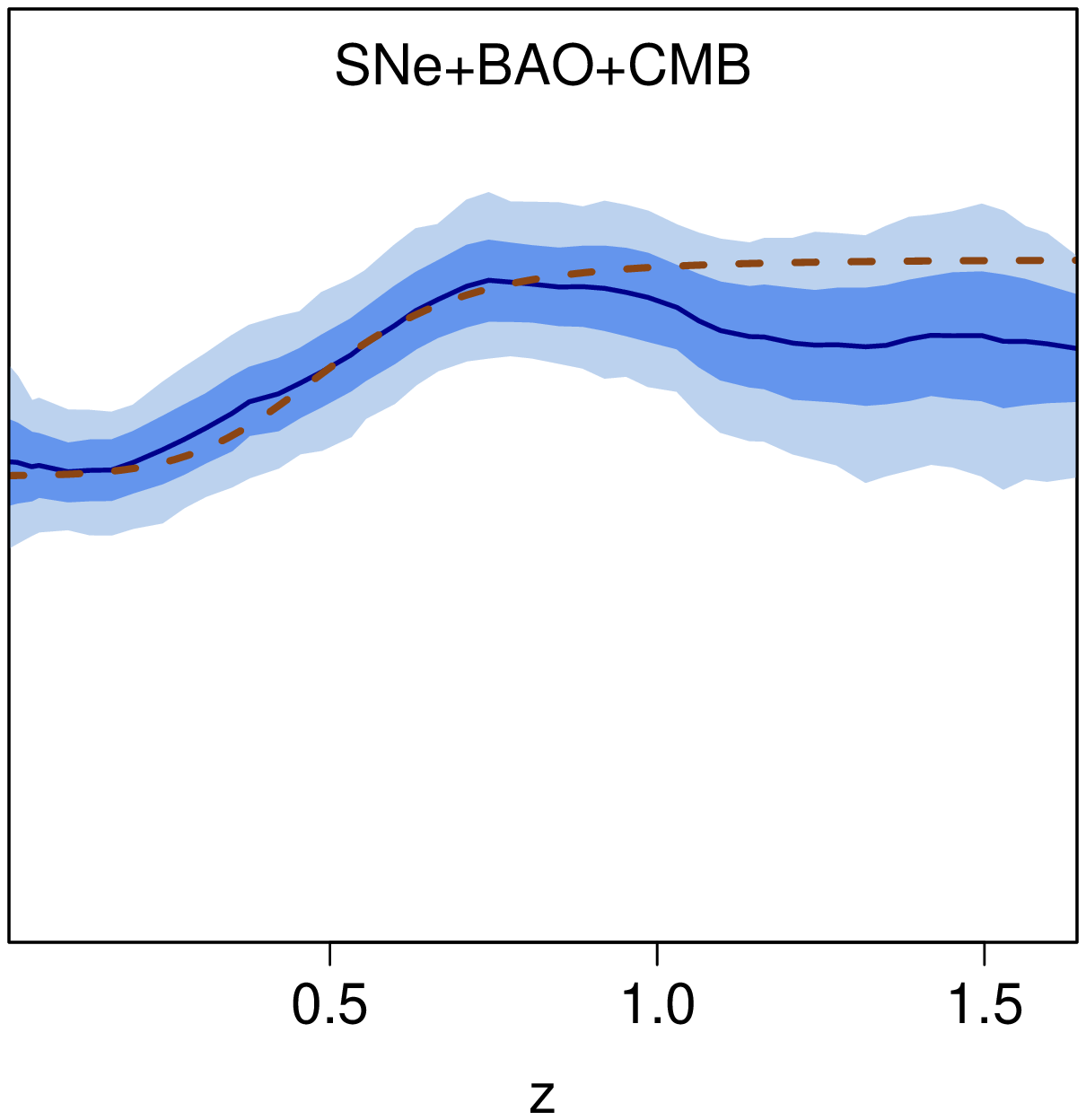}} 

  \vspace{-0.53cm} 

  \centerline{
   \includegraphics[width=2.1in,angle=0]{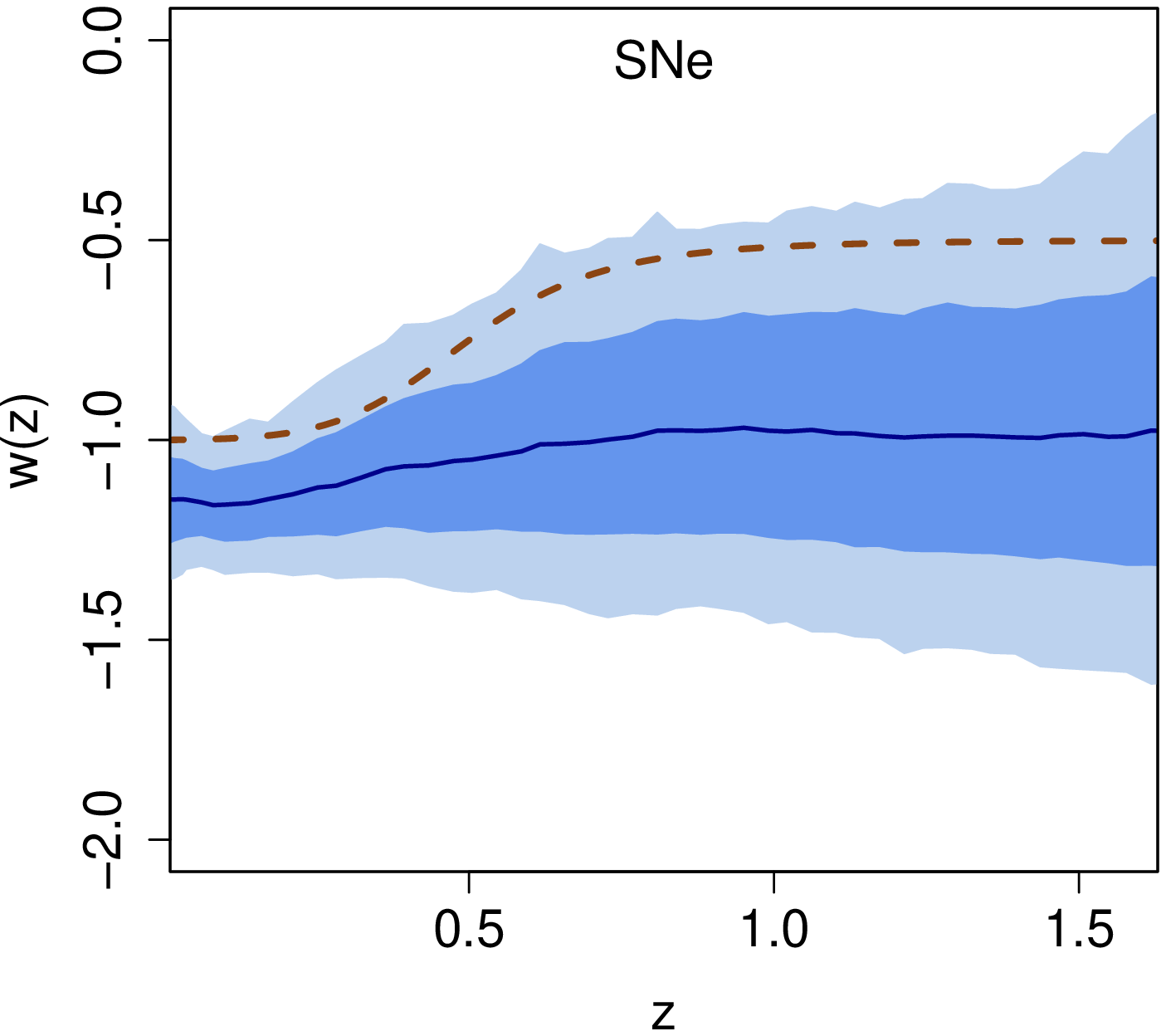}
  \hspace{-1.2cm}\includegraphics[width=2.1in,angle=0]{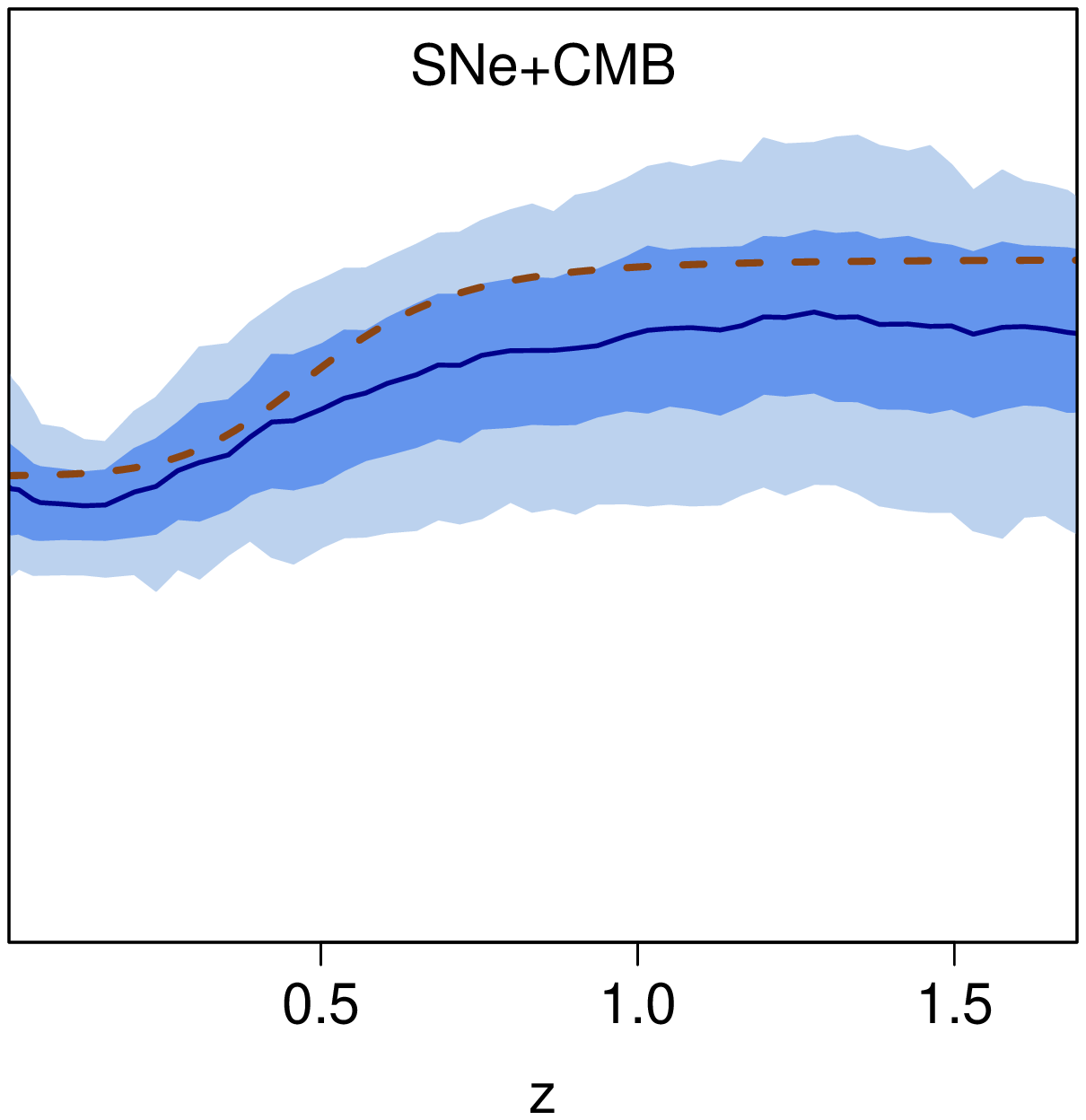}
  \hspace{-1.2cm}\includegraphics[width=2.1in,angle=0]{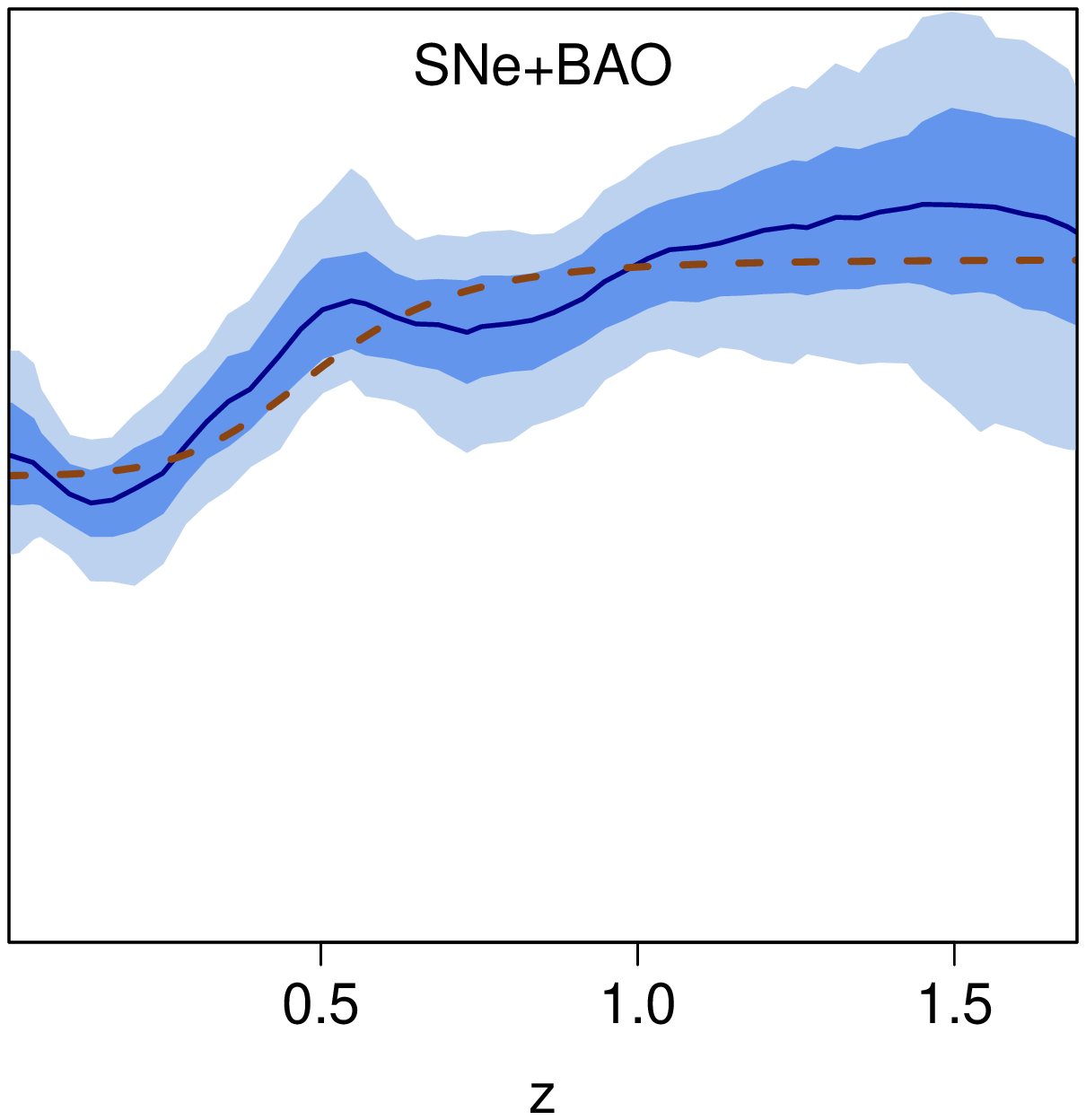}
  \hspace{-1.2cm}\includegraphics[width=2.1in,angle=0]{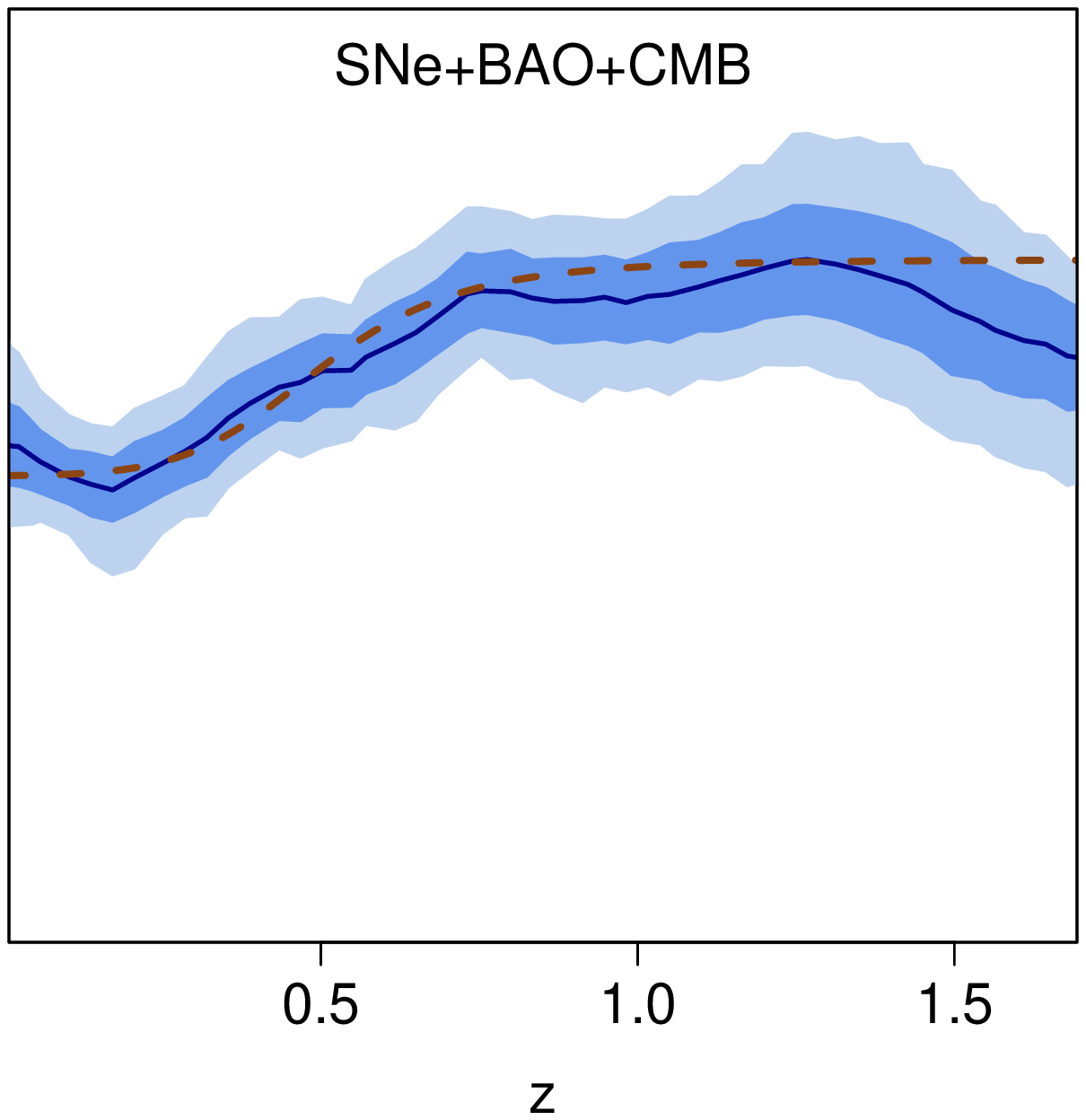}} 
  \caption{\label{simdata3}Same as in Fig.~\ref{simdata1} but for Model 3. }
\end{figure*}

Next we present the results for Model 1 - 3 for several
different combinations of data as discussed above:
\medskip

\noindent
{\it Ground-based supernova mission
  (Figs.~\ref{simdata1}-\ref{simdata3}, upper rows;
  Table~\ref{table550}):}
\begin{itemize}
\item 557 supernovae out to $z=1.4$, $\tau_i=0.15$
\item supernovae + CMB measurement
\item supernovae + 20 BAO points (realization I)
\item supernova + BAO + CMB measurements
\end{itemize}
{\it Space-based supernova mission
  (Figs.~\ref{simdata1}-\ref{simdata3}, lower rows;
  Table~\ref{table2000}):}
\begin{itemize}
\item 2298 supernovae out to $z=1.7$, $\tau_i=0.13$
\item supernovae + CMB measurement 
\item supernovae + 20 BAO points (realization II)
\item supernova + BAO + CMB measurements
\end{itemize}

\begin{table*}[htp]
        \caption{Posterior 95\% PIs, 557 SNe points}
        \centering
                \begin{tabular}{ ccc ccc c cc}
                        \hline
Data Type & Data & $\Omega_{m}$ & $\Delta_{\mu}$ & $\sigma^2$ & $\sigma_B^2$ & $\rho$ & $\kappa^2$ &$\vartheta$\\
[0.9ex]
\hline
SNe         & $\mu_1$ &$ 0.282 ^{+ 0.064 }_{ -0.069 }$&$ 0.003 ^{+ 0.026 }_{ -0.027 }$&$ 1.05 ^{+ 0.13 }_{ -0.12 }$&n/a                         &$ 0.87 ^{+ 0.12 }_{ -0.30 }$&$ 0.35 ^{+ 0.41 }_{ -0.19 }$& -1.00\\
            & $\mu_2$ &$ 0.277 ^{+ 0.070 }_{ -0.075 }$&$ 0.002 ^{+ 0.026 }_{ -0.026 }$&$ 1.05 ^{+ 0.13 }_{ -0.12 }$&n/a                         &$ 0.88 ^{+ 0.12 }_{ -0.30 }$&$ 0.35 ^{+ 0.40 }_{ -0.19 }$& -0.87\\
            & $\mu_3$ &$ 0.291 ^{+ 0.071 }_{ -0.082 }$&$ 0.007 ^{+ 0.027 }_{ -0.027 }$&$ 1.05 ^{+ 0.13 }_{ -0.12 }$&n/a                         &$ 0.85 ^{+ 0.14 }_{ -0.30 }$&$ 0.37 ^{+ 0.44 }_{ -0.21 }$& -1.00\\
SNe+CMB     & $\mu_1$ &$ 0.293 ^{+ 0.043 }_{ -0.038 }$&$ 0.004 ^{+ 0.026 }_{ -0.026 }$&$ 1.05 ^{+ 0.13 }_{ -0.12 }$&n/a                         &$ 0.87 ^{+ 0.13 }_{ -0.30 }$&$ 0.36 ^{+ 0.41 }_{ -0.20 }$& -1.07\\
            & $\mu_2$ &$ 0.297 ^{+ 0.046 }_{ -0.042 }$&$ 0.002 ^{+ 0.026 }_{ -0.025 }$&$ 1.05 ^{+ 0.13 }_{ -0.12 }$&n/a                         &$ 0.87 ^{+ 0.12 }_{ -0.31 }$&$ 0.36 ^{+ 0.45 }_{ -0.20 }$& -0.94\\
            & $\mu_3$ &$ 0.277 ^{+ 0.065 }_{ -0.050 }$&$ 0.009 ^{+ 0.026 }_{ -0.026 }$&$ 1.05 ^{+ 0.13 }_{ -0.12 }$&n/a                         &$ 0.85 ^{+ 0.14 }_{ -0.31 }$&$ 0.37 ^{+ 0.46 }_{ -0.21 }$& -0.78\\
SNe+BAO     & $\mu_1$ &$ 0.280 ^{+ 0.016 }_{ -0.015 }$&$ 0.004 ^{+ 0.023 }_{ -0.022 }$&$ 1.05 ^{+ 0.13 }_{ -0.12 }$&$ 0.89 ^{+ 0.60 }_{ -0.37 }$&$ 0.90 ^{+ 0.10 }_{ -0.26 }$&$ 0.34 ^{+ 0.40 }_{ -0.18 }$& -1.04 \\
            & $\mu_2$ &$ 0.280 ^{+ 0.017 }_{ -0.016 }$&$ 0.002 ^{+ 0.024 }_{ -0.023 }$&$ 1.05 ^{+ 0.13 }_{ -0.12 }$&$ 0.88 ^{+ 0.60 }_{ -0.36 }$&$ 0.88 ^{+ 0.11 }_{ -0.29 }$&$ 0.34 ^{+ 0.40 }_{ -0.19 }$& -0.94 \\
            & $\mu_3$ &$ 0.283 ^{+ 0.018 }_{ -0.019 }$&$ 0.008 ^{+ 0.026 }_{ -0.026 }$&$ 1.05 ^{+ 0.13 }_{ -0.12 }$&$ 0.88 ^{+ 0.61 }_{ -0.37 }$&$ 0.81 ^{+ 0.14 }_{ -0.26 }$&$ 0.39 ^{+ 0.48 }_{ -0.22 }$& -0.73 \\
SNe+BAO+CMB & $\mu_1$ &$ 0.280 ^{+ 0.015 }_{ -0.015 }$&$ 0.004 ^{+ 0.023 }_{ -0.023 }$&$ 1.05 ^{+ 0.13 }_{ -0.12 }$&$ 0.89 ^{+ 0.60 }_{ -0.36 }$&$ 0.90 ^{+ 0.10 }_{ -0.26 }$&$ 0.34 ^{+ 0.39 }_{ -0.18 }$& -1.05 \\
            & $\mu_2$ &$ 0.280 ^{+ 0.016 }_{ -0.015 }$&$ 0.002 ^{+ 0.024 }_{ -0.023 }$&$ 1.05 ^{+ 0.13 }_{ -0.12 }$&$ 0.88 ^{+ 0.60 }_{ -0.36 }$&$ 0.88 ^{+ 0.11 }_{ -0.26 }$&$ 0.35 ^{+ 0.40 }_{ -0.19 }$& -0.95 \\
            & $\mu_3$ &$ 0.284 ^{+ 0.017 }_{ -0.017 }$&$ 0.008 ^{+ 0.026 }_{ -0.025 }$&$ 1.05 ^{+ 0.13 }_{ -0.12 }$&$ 0.87 ^{+ 0.60 }_{ -0.36 }$&$ 0.80 ^{+ 0.15 }_{ -0.26 }$&$ 0.37 ^{+ 0.42 }_{ -0.20 }$& -0.73 \\
BAO         & $\mu_1$ &$ 0.280 ^{+ 0.030 }_{ -0.028 }$&   n/a                         &   n/a                      &$ 0.90 ^{+ 0.61 }_{ -0.37 }$&$ 0.88 ^{+ 0.12 }_{ -0.30 }$&$ 0.36 ^{+ 0.41 }_{ -0.19 }$& -1.05 \\
            & $\mu_2$ &$ 0.276 ^{+ 0.035 }_{ -0.032 }$&   n/a                         &   n/a                      &$ 0.89 ^{+ 0.62 }_{ -0.37 }$&$ 0.86 ^{+ 0.13 }_{ -0.31 }$&$ 0.34 ^{+ 0.38 }_{ -0.18 }$& -0.92 \\
            & $\mu_3$ &$ 0.297 ^{+ 0.037 }_{ -0.044 }$&   n/a                         &   n/a                      &$ 0.92 ^{+ 0.66 }_{ -0.39 }$&$ 0.82 ^{+ 0.16 }_{ -0.28 }$&$ 0.37 ^{+ 0.42 }_{ -0.20 }$& -0.75 \\
BAO+CMB     & $\mu_1$ &$ 0.281 ^{+ 0.025 }_{ -0.024 }$&   n/a                         &   n/a                      &$ 0.89 ^{+ 0.61 }_{ -0.37 }$&$ 0.88 ^{+ 0.11 }_{ -0.28 }$&$ 0.35 ^{+ 0.39 }_{ -0.19 }$& -1.07 \\
            & $\mu_2$ &$ 0.279 ^{+ 0.029 }_{ -0.025 }$&   n/a                         &   n/a                      &$ 0.89 ^{+ 0.61 }_{ -0.37 }$&$ 0.88 ^{+ 0.12 }_{ -0.29 }$&$ 0.35 ^{+ 0.41 }_{ -0.19 }$& -0.95 \\
            & $\mu_3$ &$ 0.296 ^{+ 0.035 }_{ -0.035 }$&   n/a                         &   n/a                      &$ 0.92 ^{+ 0.66 }_{ -0.39 }$&$ 0.81 ^{+ 0.16 }_{ -0.31 }$&$ 0.37 ^{+ 0.41 }_{ -0.20 }$& -0.74 \\ 
\hline
                \end{tabular}
        \label{table550} 
\end{table*}

\begin{table*}[htp]
        \caption{Posterior 95\% PIs, 2298 supernova}
        \centering
                \begin{tabular}{ ccc cccc c c}
                        \hline
Data Type & Data & $\Omega_{m}$ & $\Delta_{\mu}$  & $\sigma^2$ &$\sigma_B^2$ & $\rho$ &
$\kappa^2$ & $\vartheta$\\
[0.9ex]
\hline
SNe         & $\mu_1$ &$ 0.270 ^{+ 0.032 }_{ -0.043 }$&$ -0.003 ^{+ 0.019 }_{ -0.018
}$& $0.97 ^{+ 0.06 }_{ -0.05 }$& n/a                        &$ 0.90 ^{+ 0.10 }_{
-0.27 }$&$ 0.34 ^{+ 0.37 }_{ -0.18 }$& -1.00\\
            & $\mu_2$ &$ 0.263 ^{+ 0.046 }_{ -0.051 }$&$ -0.004 ^{+ 0.018 }_{ -0.018
}$& $0.97 ^{+ 0.06 }_{ -0.06 }$& n/a                        &$ 0.90 ^{+
0.10 }_{ -0.27 }$&$ 0.34 ^{+ 0.40 }_{ -0.18 }$& -0.87\\
            & $\mu_3$ &$ 0.327 ^{+ 0.040 }_{ -0.070 }$&$ -0.007 ^{+ 0.019 }_{ -0.019
}$& $0.97 ^{+ 0.06 }_{ -0.06 }$& n/a                        &$ 0.85 ^{+
0.14 }_{ -0.32 }$&$ 0.35 ^{+ 0.40 }_{ -0.19 }$& -0.92\\
SNe+CMB     & $\mu_1$ &$ 0.278 ^{+ 0.024 }_{ -0.024 }$&$ -0.003 ^{+ 0.020 }_{ -0.019
}$& $0.97 ^{+ 0.06 }_{ -0.06 }$& n/a                        &$ 0.89 ^{+ 0.11 }_{
-0.32 }$&$ 0.34 ^{+ 0.39 }_{ -0.18 }$& -1.04\\
            & $\mu_2$ &$ 0.279 ^{+ 0.027 }_{ -0.026 }$&$ -0.006 ^{+ 0.019 }_{ -0.018
}$& $0.97 ^{+ 0.06 }_{ -0.06 }$& n/a                        &$ 0.90 ^{+
0.10 }_{ -0.29 }$&$ 0.34 ^{+ 0.38 }_{ -0.19 }$& -0.90\\
            & $\mu_3$ &$ 0.292 ^{+ 0.050 }_{ -0.043 }$&$ -0.002 ^{+ 0.021 }_{ -0.020
}$& $0.97 ^{+ 0.06 }_{ -0.06 }$& n/a                        &$ 0.81 ^{+
0.16 }_{ -0.29 }$&$ 0.40 ^{+ 0.48 }_{ -0.24 }$& -0.82\\  
SNe+BAO     & $\mu_1$ &$ 0.269 ^{+ 0.011 }_{ -0.011 }$&$ -0.002 ^{+ 0.021 }_{ -0.019
}$&$ 0.97 ^{+ 0.06 }_{ -0.06 }$&$ 1.29 ^{+ 0.87 }_{ -0.54 }$&$ 0.88 ^{+ 0.12 }_{
-0.36 }$&$ 0.35 ^{+ 0.45 }_{ -0.20 }$& -0.97 \\
            & $\mu_2$ &$ 0.269 ^{+ 0.011 }_{ -0.010 }$&$ -0.005 ^{+ 0.020 }_{ -0.018
}$&$ 0.97 ^{+ 0.06 }_{ -0.06 }$&$ 1.30 ^{+ 0.87 }_{ -0.54 }$&$ 0.89 ^{+
0.10 }_{ -0.28 }$&$ 0.35 ^{+ 0.40 }_{ -0.19 }$& -0.88 \\
            & $\mu_3$ &$ 0.268 ^{+ 0.013 }_{ -0.015 }$&$ -0.002 ^{+ 0.021 }_{ -0.022
}$&$ 0.97 ^{+ 0.06 }_{ -0.06 }$&$ 1.22 ^{+ 0.85 }_{ -0.52 }$&$ 0.75 ^{+
0.21 }_{ -0.32 }$&$ 0.40 ^{+ 0.47 }_{ -0.22 }$& -0.63 \\
SNe+BAO+CMB & $\mu_1$ &$ 0.269 ^{+ 0.010 }_{ -0.010 }$&$ -0.001 ^{+ 0.018 }_{ -0.017
}$&$ 0.97 ^{+ 0.06 }_{ -0.06 }$&$ 1.31 ^{+ 0.88 }_{ -0.54 }$&$ 0.90 ^{+ 0.10 }_{
-0.30 }$&$ 0.35 ^{+ 0.42 }_{ -0.19 }$& -1.00 \\
            & $\mu_2$ &$ 0.270 ^{+ 0.010 }_{ -0.010 }$&$ -0.004 ^{+ 0.018 }_{ -0.017
}$&$ 0.97 ^{+ 0.06 }_{ -0.06 }$&$ 1.31 ^{+ 0.88 }_{ -0.54 }$&$ 0.91 ^{+
0.09 }_{ -0.29 }$&$ 0.34 ^{+ 0.40 }_{ -0.19 }$& -0.90 \\
            & $\mu_3$ &$ 0.269 ^{+ 0.011 }_{ -0.011 }$&$  0.002 ^{+ 0.020 }_{ -0.021
}$&$ 0.97 ^{+ 0.06 }_{ -0.06 }$&$ 0.66 ^{+ 0.46 }_{ -0.28 }$&$ 0.76 ^{+
0.18 }_{ -0.31 }$&$ 0.39 ^{+ 0.48 }_{ -0.21 }$& -0.71 \\
BAO         & $\mu_1$ &$ 0.270 ^{+ 0.036 }_{ -0.049 }$&   n/a                       
  &   n/a                      &$ 1.33 ^{+ 0.92 }_{ -0.56 }$&$ 0.86 ^{+ 0.14 }_{
-0.31 }$&$ 0.35 ^{+ 0.41 }_{ -0.20 }$& -1.00 \\
            & $\mu_2$ &$ 0.264 ^{+ 0.037 }_{ -0.041 }$&   n/a                       
  &   n/a                      &$ 1.33 ^{+ 0.92 }_{ -0.56 }$&$ 0.87 ^{+
0.13 }_{ -0.31 }$&$ 0.35 ^{+ 0.40 }_{ -0.19 }$& -0.88 \\
            & $\mu_3$ &$ 0.278 ^{+ 0.052 }_{ -0.059 }$&   n/a                       
  &   n/a                      &$ 1.35 ^{+ 1.03 }_{ -0.60 }$&$ 0.77 ^{+
0.20 }_{ -0.30 }$&$ 0.37 ^{+ 0.45 }_{ -0.20 }$& -0.68 \\
BAO+CMB     & $\mu_1$ &$ 0.279 ^{+ 0.031 }_{ -0.025 }$&   n/a                       
  &   n/a                      &$ 1.35 ^{+ 0.92 }_{ -0.56 }$&$ 0.88 ^{+ 0.11 }_{
-0.31 }$&$ 0.35 ^{+ 0.41 }_{ -0.19 }$& -1.03 \\
            & $\mu_2$ &$ 0.275 ^{+ 0.032 }_{ -0.027 }$&   n/a                       
  &   n/a                      &$ 1.35 ^{+ 0.93 }_{ -0.56 }$&$ 0.88 ^{+
0.11 }_{ -0.32 }$&$ 0.36 ^{+ 0.43 }_{ -0.20 }$& -0.92 \\
            & $\mu_3$ &$ 0.285 ^{+ 0.039 }_{ -0.044 }$&   n/a                       
  &   n/a                      &$ 1.38 ^{+ 1.01 }_{ -0.62 }$&$ 0.78 ^{+
0.20 }_{ -0.32 }$&$ 0.40 ^{+ 0.47 }_{ -0.22 }$& -0.70 \\
\hline
                \end{tabular}
        \label{table2000} 
\end{table*}

As for the real data, we choose a stronger prior for $\Omega_m$ in the
case of analyzing supernova data only while we use a flat prior for
any combination of data. Figure~\ref{simdata1} shows the results for
Model 1. The reconstruction from supernova data only works very well
-- the additional data points (comparing the upper and lower panel)
help reduce the error bands (note that the redshift range in the lower
row showing the results for 2298 supernovae extends out further) and
also lead to a better estimate for $\Omega_m$ with tighter error
bounds, given in Tables~\ref{table550}, \ref{table2000}. The addition
of the CMB point (second column in Figure~\ref{simdata1}) allows us to
choose a much less strict prior on $\Omega_m$, i.e. a flat prior.
Overall, the reconstruction works well with the combination of
supernova and CMB measurements, the error bands on $w(z)$ shrink
considerably. The estimate for $\Omega_m$ is slightly too high leading
to a small overall underestimation of $w(z)$ [we remind the reader of
the degeneracy of $\Omega_m$ and $w(z)$]. In the third column we show
the supernova+BAO analysis. In this case, both results extend to
$z=1.7$ due to the BAO data at those redshifts. In the upper row, the
supernova data only covers a redshift range out to $z=1.4$, the
overall result is similar to the result from the BAO data only
(Figure~\ref{BAOonly}) though the error bands shrink considerably.
Combining all three data sets leads to even narrower error bands
(fourth column). In the lower row the small downward trend from the
CMB point is compensated by the small upward trend from the BAO
measurements at high redshifts, leading to an almost perfect
reconstruction result. In the upper row, both CMB and BAO realization
have a small downward trend in $w(z)$ which surveys in the final
result. Overall, the ``truth'' is captured well in all cases and lies
well within the error bounds. We would like to emphasize that the dark
blue line in the figures only represents the mean of the
reconstruction result; much more significant are the error bands
themselves -- these must capture the true underlying model to
establish a valid approach.

The results for Model 2 and 3 are similar, shown in
Figures~\ref{simdata2} and \ref{simdata3}. Model 2 exhibits a small
time variation which could be extracted from future data. The powerful
combination of all three probes can be gauged by the relatively small
error bands shown in the fourth column in Figure~\ref{simdata2}. At
low to intermediate redshifts (out to $z\sim 0.6$) a cosmological
constant is clearly disfavored. The supernova data alone would not
have had enough information to disfavor $w=-1$ at any redshift, as the
error bands in this case clearly include a cosmological constant. The
inclusion of high redshift supernova data improves the results
somewhat, the overall reconstruction shown in the lower left corner of
Figure~\ref{simdata2} is excellent with narrow error bands. In this
case, the constraints for $\Omega_m$ are also very close to the input
value for the theoretical model with tight error bands.

Model~3 has a rather strong variation in $w(z)$. While this model is
observationally ruled out already, it provides a good test bed for our
new approach to demonstrate that more complicated dark energy equation
of states can be reconstructed. As we discussed in detail in
Ref.~\cite{holsclaw1} the degeneracy between $\Omega_m$ and $w(z)$
makes the reconstruction task rather difficult -- the left panels in
Figure~\ref{simdata3} show the constructed $w(z)$ from supernova data
only with a Gaussian prior on $\Omega_m$. The error bars are rather
wide and include a cosmological constant comfortably. The addition of
the CMB point already improves the result considerably, in this case we
choose a flat prior on $\Omega_m$. The best-fit value for $\Omega_m$
is very close to the input value of 0.27 compared to the case where we
analyze supernova data only. The inclusion of the BAO data (third and
fourth column) in both cases (557 and 2298 supernova data points)
improves the results even more. The time dependence is well captured
and the estimate for $\Omega_m$ is also very good.

Some final remarks on the content of Tables~\ref{table550},
\ref{table2000}: in addition to the results discussed above, we
provide some information on the results for the combination of BAO and
CMB measurements. Overall, the extra information from the CMB
measurement does not help very much to improve the results, contrary
to what we find when we add this information to the supernova data. In
addition to the constraints on the cosmological parameters and error
behavior of the data (given by $\sigma$ for the supernova data and
$\sigma_B$ for the BAO data) we list the final hyperparameters for the
GP model in the last three columns. Perhaps the most interesting
parameter here is the adjusted mean value for $w(z)$ given by
$\vartheta$ in the last column. As we described in
Ref.~\cite{holsclaw1} in detail, we start the GP model with some value
for $\vartheta$ (in the case of Model~1, $\vartheta=-1$ is the natural
choice for example) and run the reconstruction program for some time.
The results then have information about an improved value for the mean
of the GP model and the analysis framework can be adjusted
accordingly. As can be seen in the Tables, the final values for
$\vartheta$ are close to the mean value of the underlying truth.
Because the adjustment scheme works extremely well, we started
basically all reconstruction evaluations at $\vartheta=-1$, the GP
model automatically suggesting better mean values if the choice was
non-optimal. Overall, the reconstruction of $w(z)$ works very well
when multiple sources are included.

\section{Conclusion}
\label{conclusion}

In this paper we have introduced a new non-parametric reconstruction
scheme for the dark energy equation of state $w(z)$ combining multiple
cosmological probes. The reconstruction scheme is based on a GP
modeling approach and provides very good constraints on $w(z)$ with
reliable error bars. The basic method was introduced in
Ref.~\cite{holsclaw1} for supernova data only. Here we extend the
methodology to include BAO and CMB measurements. We have carried out
an analysis of currently available data and found excellent agreement
with a cosmological constant consistent with a large number of recent
publications, including
Refs.~\cite{hicken09,amanullah,2dF,komatsu,Wang:2009sn}. We have also
demonstrated our method on simulated data for different cosmological
models. In all cases, the GP model approach performed very well.

An important aspect of our new approach (as stressed in
Refs.~\cite{holsclaw1,holsclaw2}) is the simultaneous constraint of
the cosmological parameters as well as the hyperparameters of the GP
model from the data. In comparison to parametric approaches, our new
method is more flexible and can therefore capture even subtle time
variations in $w(z)$ if the data quality is good enough. It produces
narrow error bands over the full redshift ranges considered. For a
more detailed comparison with parametrized methods, see
Ref.~\cite{holsclaw1}.

The combination of different data probes mitigates the problem of
degeneracies between $w(z)$ and $\Omega_m$ as to be expected. An
encouraging observation is that even the BAO data alone (of high
quality from a BigBOSS-like survey) can deliver good constraints on
the time dependence of the dark energy equation of state, clearly
competitive with space based supernova observations.

Our new non-parametric reconstruction approach lends itself to
analysis of the promise of future dark energy probes in a reliable
way. For example, possible tension in the data due to e.g.
insufficient understanding of systematic errors would lead to an
increase in the error bands when combining different probes (a
different attempt to solve this problem with parametric methods is
discussed in e.g. Ref.~\cite{tension}). The GP based approach can
therefore help to optimize future dark energy missions.

\begin{acknowledgments}

We would like to thank the Institute for Scalable Scientific Data
Management for supporting this work.  Part of this research was
supported by the DOE under contract W-7405-ENG-36.  UA, SH, KH, and DH
acknowledge support from the LDRD program at Los Alamos National
Laboratory.  KH was supported in part by NASA. SH and KH acknowledges
the Aspen Center for Physics, where part of this work was carried
out. We would like to thank Andreas Albrecht, Eric Linder, Adrian
Pope, Martin White, and Michael Wood-Vasey for useful discussions.
\end{acknowledgments}

\end{document}